\documentclass[fleqn,usenatbib]{mnras}

\usepackage{newtxtext,newtxmath}

\usepackage[T1]{fontenc}
\usepackage{ae,aecompl}

\sfcode`A=1000 \sfcode`B=1000 \sfcode`C=1000 \sfcode`D=1000
\sfcode`E=1000 \sfcode`F=1000 \sfcode`G=1000 \sfcode`H=1000
\sfcode`I=1000 \sfcode`J=1000 \sfcode`K=1000 \sfcode`L=1000
\sfcode`M=1000 \sfcode`N=1000 \sfcode`O=1000 \sfcode`P=1000
\sfcode`Q=1000 \sfcode`R=1000 \sfcode`S=1000 \sfcode`T=1000
\sfcode`U=1000 \sfcode`V=1000 \sfcode`W=1000 \sfcode`X=1000
\sfcode`Y=1000 \sfcode`Z=1000

\usepackage{graphicx}	
\usepackage{amsmath}	
\usepackage{amssymb}	
\usepackage{threeparttable}
\usepackage[caption = false]{subfig}

\newcommand{\IRAS}{\textbf{IRAS}}
\newcommand{\Lsol}{{\hbox{L$_\odot$}}}
\newcommand{\Msol}{{\hbox{M$_\odot$}}}
\newcommand{\Sp}{{\it Spitzer}}
\newcommand{\Her}{{\it Herschel}}

\title[AGN Fraction in Merging Galaxies]{The AGN Luminosity Fraction in Merging Galaxies}

\author[J. Dietrich et al.]{
Jeremy Dietrich,$^{1, 2}$\thanks{E-mail: jdietrich1@email.arizona.edu}
Aaron S.~Weiner,$^{1, 3}$
Matthew L.N.~Ashby,$^{1}$
\newauthor
Christopher C.~Hayward,$^{4}$
Juan Rafael Mart\'inez-Galarza,$^{1}$
Andr\'es F.~Ramos Padilla,$^{1, 5}$
\newauthor
Lee Rosenthal,$^{6}$
Howard A.~Smith,$^{1}$
S.~P.~Willner,$^{1}$
and Andreas Zezas$^{1, 7, 8}$
\\
$^{1}$Harvard-Smithsonian Center for Astrophysics, 60 Garden St, Cambridge, MA 02138\\
$^{2}$Department of Astronomy and Steward Observatory, University of Arizona, 933 N Cherry Ave, Tucson, AZ 85719\\
$^{3}$Department of Physics, Applied Physics, and Astronomy, Rensselaer Polytechnic Institute, 110 8th St, Troy, NY 12180\\
$^{4}$Center for Computational Astrophysics, Flatiron Institute, 162 Fifth Avenue, New York, NY 10010\\
$^{5}$Leiden Observatory, Leiden University, P.O.  Box 9513, 2300RA Leiden, The Netherlands\\
$^{6}$Department of Astronomy, California Institute of Technology, Pasadena, CA 91125\\
$^{7}$Physics Department \& Institute of Theoretical \& Computational Physics, University of Crete, 71003 Heraklion, Crete, Greece\\
$^{8}$Foundation for Research and Technology-Hellas, 71110 Heraklion, Crete, Greece\\
}

\date{Accepted XXX. Received YYY; in original form ZZZ}

\pubyear{2018}

\begin{document}
\label{firstpage}
\pagerange{\pageref{firstpage}--\pageref{lastpage}}
\maketitle

\begin{abstract}
Galaxy mergers are key events in galaxy evolution, often causing massive starbursts and fueling active galactic nuclei (AGN).  In these highly dynamic systems, it is not yet precisely known how much starbursts and AGN respectively contribute to the total luminosity, at what interaction stages they occur, and how long they persist.  Here we estimate the fraction of the bolometric infrared (IR) luminosity that can be attributed to AGN by measuring and modeling the full ultraviolet to far-infrared spectral energy distributions (SEDs) in up to 33 broad bands for 24 merging galaxies with the Code for Investigating Galaxy Emission.  In addition to a sample of 12 confirmed AGN in late-stage mergers, found in the {\it Infrared Astronomical Satellite} Revised Faint Source Catalog, our sample includes a comparison sample of 12 galaxy mergers from the \textit{Spitzer} Interacting Galaxies Survey, mostly early stage.  We also SED modeling of merger simulations to validate our methods, and we supplement the SEDs with mid-IR spectra of diagnostic lines obtained with \textit{Spitzer}'s InfraRed Spectrograph.  The estimated AGN contributions to the IR luminosities vary from system to system from 0\% up to $\sim$91\% but are significantly greater in the later-stage, more luminous mergers, consistent with what is known about galaxy evolution and AGN triggering.  

\end{abstract}




\section{Introduction}

Galaxy interactions have long been known to influence galaxy evolution, and a large majority of galaxies in the universe show signs of previous interactions (e.g., Struck 1999).  Mergers trigger enhanced star formation (SF) and generate or fuel active galactic nuclei (\cite{san88}; \citeauthor{hon15} \citeyear{hon15}, \citeauthor{bra15} \citeyear{bra15} and references therein).  In addition, mergers produce disturbed morphological features (\citeauthor{too72} \citeyear{too72}; \citeauthor{lan13} \citeyear{lan13} [hereafter L13] and references therein).  Tidal tails and nuclear disruptions are the most obvious indications that two or more galaxies are interacting or merging.

The enhanced SF seen in galaxy mergers is, in most cases, the main power source for (ultra) luminous infrared galaxies ([U]LIRGs) in the local universe \citep{san96}.  \citet{vei02} have shown that many of these galaxies contain clear morphological indicators of past galaxy interactions.  However, not all galaxy mergers have enhanced IR emission.  The star formation rate (SFR) depends on the merger stage, as SF and AGN activity are enhanced in the later stages of mergers \citep{lac14}.

In efforts to address when and how SF and AGN activity proceed, (U)LIRGs and other luminous late-stage mergers have become prime targets for  space-based telescopes such as the \textit{Spitzer Space Telescope} \citep{wer04}, the \textit{Herschel Space Observatory} \citep{pil10}, the \textit{Galaxy Evolution Explorer (GALEX)} \citep{mar05}, and \textit{Swift} \citep{geh04} among others.  This suite of instruments provide highly reliable photometry by virtue of their privileged vantage point above Earth's atmosphere, and they are sensitive to the multiple processes contributing to galaxies' spectral energy distributions (SEDs):  \textit{Spitzer}'s infrared cameras measure the enhanced mid-infrared (MIR) emission from both AGN and SF, \textit{Herschel} views emission principally from the diffuse galactic dust, and the UV satellites are sensitive to the emission arising from young, hot stars.

SEDs that span the range from UV to FIR wavelengths reflect all significant energetic processes occurring in galaxies.  They are therefore indispensable for inferring galaxies' underlying physical properties, including but by no means limited to SFRs, masses and luminosities of the galactic dust and the effects of AGN \citep{hay15}.  For galaxy mergers, the MIR emission arises from dust heated by both SF and AGN  \citep{gru08}.  However, the relative proportions each process contributes are not well understood and vary enormously over time and from one system to another. Moreover, the ability to detect high-redshift galaxies is increasing, but the SEDs of these distant sources are much easier to obtain than spatial morphology and are therefore our best keys to understand the physical processes underway. A clear understanding of SEDs in the local universe is a prerequisite for drawing reliable conclusions about distant sources.

Many studies are being performed to calculate the fraction of luminosity that comes from the AGN in merging galaxies. Studies utilize wide wavelength ranges (from total IR to UV/X-ray) and span redshifts from the local universe ($z < 0.06$) out to the distant universe ($1 < z < 4$) (e.g., \citeauthor{cie15} \citeyear{cie15}; \citeauthor{dro16} \citeyear{dro16}; \citeauthor{fer16} \citeyear{fer16}; \citeauthor{vad16} \citeyear{vad16}; \citeauthor{vil17} \citeyear{vil17}).  Other recent studies have also characterized the SEDs out to 24~\micron\ of radio-loud AGN \citep{wil17} or specifically focused on the far-IR SED shape, where dust and AGN emission dominate \citep{saf16, cow17}. Accurate constraints on the AGN contribution to the total luminosity are necessary for precise estimates of other energetic processes such as SFR. In galaxies at cosmological distances for which primarily photometric data are available, we can ascertain which physical processes are providing the luminosity for the galaxy.

This work presents an analysis of 24 merging galaxies organized into two samples. First we re-measure in a uniform and self-consistent manner and then analyze the full SEDs of 12 late-stage merging (U)LIRGs and estimate their fractional AGN contributions across the entire IR range from 1--1000\micron\ (hereafter referred to as `total IR').  Our SEDs cover UV to far-IR/submillimeter wavelengths, providing a comprehensive view into the processes at work in merging galaxies.  We compare the results for late-stage mergers to those for 12 previously studied early-stage mergers.   

This paper is organized as follows.  Section~2 presents the galaxy samples, the observations, and the data reduction.  Section~3 describes the SED analysis.  Section~4 describes the same type of SED analysis of merger simulations. Section~5  discusses results, and Section~6 summarizes conclusions.

\section{Observations}

\subsection{Galaxy samples}

We chose the late-stage merger sample to represent strongly interacting, infrared-luminous systems.  Candidate systems were identified by the \textit{Infrared Astronomical Satellite} (\textit{IRAS}, \citeauthor{neu84} \citeyear{neu84}) Revised Faint Source Catalogue (FSC, \citeauthor{wan14} \citeyear{wan14}).  We selected interacting systems by cross-referencing \IRAS\ sources with Version 2 of the Galaxy Zoo public galaxy classification program \citep{wil13} to yield 453 systems.  Of these, 85 have far infrared luminosities at least in the `luminous infrared galaxy' (LIRG: $L_{\rm IR}>10^{11}$~\Lsol) range, and 7 are in the ULIRG ($L_{\rm IR}>10^{12}$~\Lsol) range. We classified all these systems by merger stage \citep{wei18} finding 38 with morphological evidence for strong interaction such as long tidal tails or heavily disturbed morphology.  These systems are designated as ``Stage 4'' or later by \citet{wei18}.  Of these 38 galaxies, only 12 had available photometry from all of \Sp/IRAC, \Sp/MIPS 24~\micron, and \textit{Herschel}/SPIRE at 250, 350, and 500~\micron.  These 12 constitute the late-stage merger sample listed in Table~\ref{tab1}.  By construction the sample is pure for strong interactions and high luminosities--indeed 11 of the 12 are in the top 20 luminosities of all 453 in the Galaxy Zoo sample. However, the sample is far from complete because of our requirement to have full data sets, especially  \Her\ data.  This may have introduced a bias toward `interesting' systems and therefore ones with extreme properties, but any bias is probably related to obvious properties such as morphology and luminosity rather than parameters that can be revealed only by detailed analysis.

For a control set to compare with the late-stage mergers, a `Reference Sample' was drawn from the SIGS galaxy sample \citep{bra15}.  SIGS consists of 103 galaxies in 48 systems selected by a combination of galaxy proximity on the sky and morphological disturbance. SIGS therefore includes all merger stages from non-interacting systems to early approach to late stages (\citealt{bra15}, L13). From the SIGS sample, we selected 12 galaxies with UV--submm photometry comparable to what was available for the Late-Stage Merger Sample.  We adopted the merger stage classifications from L13 for these objects.  Nearly all of them are Stages~2--3 implying at least mild but at most moderate distortions and galaxies still separated from each other \citep{wei18}. The Reference Sample members are listed in  Table~\ref{tab1}.  The sample is deliberately heterogeneous but contains a range of systems that are merging but have not yet reached the final merger stage.  The requirement for many-band photometry introduces a bias toward well-studied systems, which are likely if anything to be those with especially strong merger signatures, i.e., the Reference Sample probably resembles the Late-Stage Merger Sample more closely than the full SIGS sample would.

One difference between the samples is their redshift distributions. The  Reference Sample galaxies all lie within $z\le0.012$, but the Late-Stage Merger Sample galaxies are in the range $0.01\le z\le0.06$.  This reflects the fact that late-stage mergers are less common than early-stage ones---the early stages last longer than late stages---and it's necessary to search larger distances to find equal numbers of late-stage systems.  None of the galaxies requires a significant k-correction, and all inhabit the present-day universe.

\begin{table*}
\centering
\footnotesize
\caption{Basic data for the two galaxy samples}
\label{tab1}
\begin{threeparttable}
\begin{tabular}{c c c c c c c}
\hline
\hline
Galaxy Name & Redshift (\it{z}) & $D_L$ (Mpc)$^a$ & log $L_{IR}$ ($M_{\odot})^a$ & Stage$^b$ & AGN$^c$ & Ref$^d$\\
\hline
\multicolumn{5}{c}{Late-Stage Merger Sample}\\
\hline
IRAS~08572+3915 & 0.05835 & 265 & 12.08 & 4 & LINER & (1)\\
IRAS~15250+3609 & 0.05516 & 249 & 12.04 & 5 & LINER & (1)\\
Mrk~231 & 0.04217 & 188 & 12.51 & 6 & Seyfert 1 & (4)\\
Mrk~273 & 0.03778 & 168	& 12.05 & 5 & Seyfert 2 & (2)\\
Mrk~463 & 0.05035 & 227 & 11.73 & 4.5 & Seyfert 2 & (3)\\
NGC~2623 & 0.01851 & 81 & 11.33 & 5 & LINER & (5)\\
NGC~3758 & 0.02985 & 132 & 11.00 & 4.5 & Seyfert 1 & (2)\\
NGC~6090 & 0.02930 & 129 & 11.49 & 4.5 & Seyfert & (2)\\
UGC 4881 & 0.03930 & 175 & 11.60 & 4 & LINER Composite & (1,2)\\
UGC 5101 & 0.03937 & 175 & 12.03 & 5 & Seyfert 1 & (3)\\
VV 283 & 0.03748 & 167 & 11.46 & 5 & Seyfert 2 & (2)\\
VV 705 & 0.04019 & 179 & 11.82 & 4.5 & Composite & (2)\\
\hline
\multicolumn{5}{c}{Reference Sample}\\
\hline
M51A & 0.00155 & 8.58$^e$ & 10.51 & 3 & Seyfert 2 & (6)\\
M51B & 0.00191 & 8.58$^e$ & 9.61 & 3 & LINER & (6)\\
NGC~2976 & 0.00004 & 3.55$^f$ & 8.91 & 2 & None & (8)\\
NGC~3031 & -0.00014 & 3.5$^f$ & 9.62 & 2 & Seyfert 2 & (3)\\
NGC~3077 & 0.00004 & 3.83$^f$ & 8.85 & 2 & None & (6)\\
NGC~3190 & 0.00432 & 19.3$^f$ & 9.82 & 3 & LINER & (8)\\
NGC~3690 & 0.0100 & 43 & 11.76 & 4$^g$ & LINER Composite & (1)\\
NGC~4625 & 0.00212 & 11.75$^e$ & 8.95 & 3 & Seyfert & (9)\\
NGC~5394 & 0.01153 & 50 & 10.69 & 3.5 & Composite & (2)\\
NGC~5395 & 0.01158 & 50 & 10.76 & 3.5 & Seyfert 2 & (9)\\
M101 & 0.00081 & 6.7$^f$ & 10.30 & 3 & None & (7)\\
NGC~5474 & 0.00098 & 6.8$^f$ & 8.60 & 3 & None & (7)\\
\hline
\hline
\end{tabular}
\begin{tablenotes}
\item[a] Luminosity distance $D_L$ for the galaxies at $z > 0.01$ were calculated using the Hubble law with $H_0 = 70$ and scaling by (1 + z). Total IR luminosity is calculated from 5--1000~\micron, following \citet{fri06}.
\item[b] \citet{wei18}
\item[c] `Composite' indicates strong AGN and SF activity coexist.
\item[d] \textbf{References for AGN classification} (1) \citet{var15}; (2) \citet{tob13}; (3) \citet{bau13}; (4) \citet{iva00}; (5) \citet{gar15}; (6) \citet{her16}; (7) \citet{bra15}; (8) \citet{gon15}; (9) \citet{ver10}
\item[e] \citet{mcq17}
\item[f] \citet{dal17}
\item[g] NGC\,3690abc consists of two blended objects (a and b), with a nearby but separate third component c (IC\,694).  The tabulated photometry for NGC\,3690 comes from the ab components only.
\end{tablenotes}
\end{threeparttable}
\end{table*}

\subsection{Photometry}

For most galaxies in the Reference Sample, we assembled photometry for single galaxies rather than the entire merging systems to ensure the most reliable SED fits. This included even the merging systems M51A/B, M101/NGC~5474, NGC~3031/3077, and NGC~5394/5395.  The advanced merger NGC~3690/IC~694 system is an exception; it had to be observed as a single blended entity.   Galaxies in the Late-Stage Merger Sample could only be imaged as a single source.  Table~\ref{tab1} lists the physical parameters and previously known AGN status for each galaxy.  Details about the SED fit for individual objects can be found in Appendix~A.

Photometry for all galaxies in both samples used matched apertures on archival broadband images of up to 33 different bands (following L13). We started with the near- and far-UV bands from \textit{GALEX} \citet{mar05}.  At visible wavelengths, we used \textit{ugriz} imaging from the Sloan Digital Sky Survey (SDSS) Data Release 12 \citep{ala15}.  For the near-IR bands, we used $JHK_s$ imaging from the Two Micron All-Sky Survey (2MASS; \citeauthor{skr06} \citeyear{skr06}). Mid-IR comprised  \textit{Spitzer} Infrared Array Camera (IRAC) 3.6, 4.5, 5.8, and 8.0~\micron\ imaging and Multiband Imaging Photometer for \textit{Spitzer} (MIPS) 24, 70, and 160~\micron\ imaging.  We also used the \textit{IRAS} \citep{neu84} bands at 12, 25, 60, and 100~\micron.  For the late-stage mergers we also incorporated near- and mid-IR imaging from the \textit{Wide-Field Infrared Survey Explorer} (\textit{WISE}, \citeauthor{wri10} \citeyear{wri10}) at 3.4, 4.6, 12, and 22~\micron.  Most of the {\textit WISE} imaging was taken before the W4 filter was recalibrated \citep{bro14}, but because the \textit{WISE} 22~\micron\ data are outweighed in the fitting by the MIPS 24~\micron\ and the IRAS 25~\micron\ data, the difference in the SED fit between the previous and new calibrations of W4 is negligible.  Finally, for the far-IR bands we used archival imaging from the \textit{Herschel} Photoconductor Array Camera and Spectrometer (PACS) 60--90~\micron~, 90--130~\micron~, and 130--210~\micron\ bands as well as the \textit{Herschel} Spectral and Photometric Imaging Receiver (SPIRE) bands at 250, 350, and 500~\micron.  For the \textit{Herschel} imaging we used the \textit{Herschel} Interactive Processing Environment (HIPE), version 14.1 \citep{ott10}.

Some datasets required special handling.  The publicly available archival IRAC mosaics for IRAS~08572+3915 and Mrk~231 were not suitable for photometry.  The Mrk~231 mosaics (specifically, the post-basic calibrated or PBCD mosaics) were saturated in all four IRAC bands.  The 5.8 and 8.0~$\mu$m PBCD mosaics for IRAS~08573+3915 also show conspicuous saturation.  We therefore constructed our own mosaics for these two objects using only the short exposures (0.6~s) from archived IRAC high-dynamic range observations.  After first verifying that the resulting short-exposure mosaics showed no saturation, we used them for our photometry in place of the publicly available IRAC mosaics.  In addition, we adopted the global {\it IRAS} photometry from the {\it IRAS} FSC.

We assembled the non-{\it IRAS} photometry for all 24 galaxies following the procedure described by \citet{wei18}. We started by subtracting the sky background using the Python package \textit{photutils}\footnote[1]{https://github.com/astropy/photutils} and used an elliptical aperture to capture all the flux.  We took care to use the same aperture area to enclose the full galaxy emission regions in all bands and to correct for any background emission.  Our photometric values are consistent with but more precise than results in the open literature typically obtained in pipeline processing of larger datasets.  The photometry can be found in Appendix B.

For the early-stage mergers from L13, we used the photometry stated in the paper and added SDSS \textit{ugriz} photometry, which was processed the same as by \citet{wei18}.  Photometric uncertainties were calculated using the sum in quadrature of Poisson photon noise determined by the photometry and calibration uncertainties, and we adopted the same calibration uncertainties as L13, following the references therein.

\subsubsection{PACS Spectrophotometry}

We supplemented the SEDs for the late-stage merger sample with a previously underutilized resource: spectrophotometric continuum measurements taken from PACS spectral scans.  The PACS spectrophotometric data provide excellent coverage of the peak of the FIR continua.  Multiple observers obtained PACS range or spectral scans of lines of these sources, including some galaxies lacking standard PACS photometry.  Some galaxies were observed many times, and for them the PACS archive has an abundance of spectrophotometry, while others were observed in only a few lines.

The PACS observers used a variety of observing configurations (e.g., chopping throw, integration times, scan lengths, number of repeats), so all the PACS spectrophotometry had to be reduced individually.  We used HIPE 15 and pipeline processing 14.2, which were the most recent versions of each separate program at data collection time.  The sources here are adequately contained within the central $3\times3$ spaxels of the IFU (a practical limit being a diameter of about 15\arcsec), and we used flux density values obtained with the C129 calibration, taking the sum of the central $3\times3$ spaxels.  The task used is extractCentralSpectrum for the chopNod Astronomer's Observing Templates (AOTs).  We obtained the continuum level as the median of flux density values away from the spectral feature.  Because each scan typically has many bad values at the start and end, those were also excluded.  As a consistency check, we performed both automatic and manual flux density measurements, and they were in excellent agreement.

The PACS spectrophotometry required creating custom single-pass filters.  These were 0.5~\micron\ wide, which corresponds to the width of the bins generated when doing off-line spectrophotometry.  These filters consisted of a delta function throughput at the measured continuum wavelength.  In general, we took the continuum data closest to 60, 90, and 150~\micron\ containing the most individual observations, such as the continuum near 63 microns or 88 microns.  This allowed for the most consistent derived photometric values while also remaining near but not on top of the PACS photometry wavelengths.

There are usually hundreds of datapoints in a typical PACS spectral scan, but they are only quasi-independent.  Although the formal flux density uncertainties are small because of the large number of points, we adopted a value of $\pm$10\% as more fairly accounting for the systematic uncertainties, similar to the PACS photometric uncertainties \citep{pal12}.  See Appendix B for the table of PACS spectrophotometry for the 10 galaxies in the late-stage merger sample containing PACS spectra.

We did not obtain PACS spectrophotometric data for the early-stage merging galaxies from the L13 sample because of their low modeled $f_{AGN}$ from CIGALE. The galaxies in the late-stage sample all had large $f_{AGN}$, and the AGN emission models from \citet{fri06} peak at the PACS wavelength range, so we wanted to have the best characterization possible for the emission in the PACS range for the late-stage merging galaxies with high AGN fractions. However, because this adds multiple new data points in a small wavelength range, the risk of over-fitting the SEDs increases.  For the galaxies with large AGN fractions and a high sensitivity to changes in the PACS bands, only three spectrophotometric values were used with similar uncertainties to the PACS photometric data to help characterize the peak without over-fitting. To avoid the issue for the galaxies with low AGN fractions in the early-stage sample, where the emission peak is already well characterized, we omitted PACS spectrophotometric data.

\subsection{Spectroscopy}

Another way to estimate the AGN luminosity fraction is by using spectral lines that separately trace AGN and SF activity.  Specifically, [O~\textsc{iv}] at 25.89~\micron\ and [Ne~\textsc{v}] at 14.32~\micron\ are strong lines that signify the presence of an AGN.  In contrast, the [Ne~\textsc{ii}] line at 12.81~\micron\ is diagnostic of SF activity \citep{dal09, lam12}.  We used spectra taken by \textit{Spitzer}'s InfraRed Spectrograph (IRS; \citeyear{hou04} \citeauthor{hou04}), which provided spectroscopic coverage from 5--36~\micron.  For each galaxy, we used the Short-High (SH) and Long-High (LH) modes, which have resolving power $\lambda/\Delta\lambda = 600$. The SH mode uses a slit of 4.7 x 11.3 arcseconds, while the LH mode uses a slit of 11.1 x 22.3 arcseconds.

We reduced IRS spectra using the Spectroscopic Modeling Analysis and Reduction Tool (SMART; \citet{hig04}), version 8.2.9.  SMART was specifically designed for the IRS and provides an easy-to-use interface to reduce, analyze, and view the spectra. To validate our results from SMART, we compared our final IRS spectra for a few galaxies with those reduced using an earlier version of SMART and placed on the Cornell Atlas of Spitzer/IRS Sources (CASSIS; \citeauthor{leb11} \citeyear{leb11}).  Differences in the two versions of the final spectra were negligible. We used the data stored on CASSIS for spectral line analysis for 14 galaxies in our sample because the spectra reduction was already completed and reliable.  For the remaining 10 galaxies, we used the results from SMART.

For these 10 galaxies, we retrieved the IRS spectra from the \textit{Spitzer} Heritage Archive (SHA).  The observations were taken in both Stare and Map modes.  We extracted Stare observations flux densities and spectra without further corrections, but for the Map images we used the Cube Builder for IRS Spectral Mapping (CUBISM; \citeauthor{smi07} \citeyear{smi07}) to build 3D spectral cubes with 2 spatial and 1 spectral dimension.  The 2 spatial dimensions had pointings in a 3x3 grid with the centre pointing aimed at the nucleus. We confirmed the central spectra contained the galaxy nuclei before extracting the results to files for importing into SMART. Then, we used SMART to fit and extract the three target spectral line features.  To start, we calculated a linear baseline around each spectral line and subtracted it before fitting a Gaussian to the line profile.  For the cases where the signal was not strong or significant contamination caused irreversible blending, SMART would provide a Gaussian fit with limits, so these translated into 3$\sigma$ upper limits on the integrated line flux and the line width.

The [Ne~\textsc{ii}] line is relatively isolated with no other nearby, potentially contaminating spectral lines.  The [Cl~\textsc{ii}] line at 14.37~\micron\ caused no noticeable contamination to the [Ne~\textsc{v}] line.  However, the [O~\textsc{iv}] 25.89~\micron\ line partially overlaps with the [Fe~\textsc{ii}] line at 25.99~\micron, producing a slightly blended line profile.  For lines with significant potential contamination we used a double Gaussian profile to fit the composite (double-line) profiles. The [Fe~\textsc{ii}] contamination did end up forcing the use of an upper limit for some detections of the [O~\textsc{iv}] line as the blending caused both single and double Gaussian fits to fail.

We calculated integrated line fluxes and widths from the Gaussian fits.  We compared the results for [Ne~\textsc{v}]/[Ne~\textsc{ii}] and [O~\textsc{iv}]/[Ne~\textsc{ii}], similar to the analysis of \citet{gen98}. Because [Ne~\textsc{ii}] is a strong tracer of starburst activity, whereas [Ne~\textsc{v}] and [O~\textsc{iv}] are strong indicators of AGN activity, the AGN-to-starburst tracer ratios help determine the dominant source of luminosity for these galaxies (\citeauthor{arm07} \citeyear{arm07}, \citeauthor{sat09} \citeyear{sat09} and references therein).  Higher ratios should indicate larger AGN activity compared to starbursts. Ramos Padilla et al.\ (2018, in preparation) provide a detailed analysis of more spectral line ratios and their correlations with IR colours that indicate the presence of AGN.

\section{Analysis}

\subsection{SED Fitting}
\label{ssec:fitting}

For the SED fitting, we used the Code for Investigating Galaxy Emission (CIGALE; \citeauthor{bur05} \citeyear{bur05}).  Specifically, we used `pcigale' version 0.9.0 in Python, which was released in 2016 April.  In brief, CIGALE operates by constructing a multidimensional grid of model SEDs and identifying the SED model that best fits the data with $\chi^2$ minimization.  The grid dimension is set by the number of user-defined parameters used to define the different galaxy components, e.g., intrinsic AGN and stellar emission spectra, star formation history, dust attenuation, and nebular emission.  After it has tested all user-specified models in its grid, CIGALE then outputs what it identifies as the best-fitting model spectrum and the parameter set that best matches the galaxy data.  CIGALE also outputs parameter uncertainties based on the range of models that are consistent within each galaxy's flux density uncertainties.

In this work, for simplicity we used a `delayed' star formation history model (delayed with respect to the SF timescale), assuming a single starburst with an exponential decay, following 
\begin{equation}
SFR(t) \propto \frac{t e^{-t/\tau}}{\tau^2},
\label{sfh_eq}
\end{equation}
where $\tau$ is the e-folding time of the main stellar population, which dominates the stellar emission \citep{lee10}.  The SF starts at a time `age' before the present day, where `age' is a CIGALE parameter given in the model and defined in Table 2 \citep{cie16}.  We also set the separation between the young and old stellar populations (the stellar separation age) to 10~Myr.  This means that at the time that the galaxy is modeled in CIGALE, every star older than 10~Myr is considered `old' while the rest are considered `young'.  The combination between $\tau$ (defined in CIGALE as $\tau_{main}$), `age', and stellar separation is used as a proxy to model the recent star formation in the period of time defined by `stellar separation'. This parametric SFH model allows for CIGALE to be tuned to the recent SFR and can help determine the stage in some complicated cases.  Tests running other SF history options did not significantly alter our conclusions about the AGN fraction.

For the dust attenuation, we used models jointly described by \citet{cal00} and \citet{lei02} along with the \citet{dal14} models for the dust emission in the far-IR.  The Calzetti law for dust extinction and attenuation is described by the following set of local piece-wise power-laws,
\begin{equation}
\kappa(\lambda) = \frac{A(\lambda)}{E(B - V)_*} = a + \frac{b}{\lambda} + \frac{c}{\lambda^2} + \frac{d}{\lambda^3},
\label{caldust_eq}
\end{equation}
where \textit{a, b, c, d} are constants dependent on the wavelength range.  The dust emission from \citet{dal14} follows a modified blackbody SED with a power-law distribution of dust mass at each temperature,
\begin{equation}
dM_d \propto U^{-\alpha} dU,
\label{daldust_eq}
\end{equation}
where $M_d$ is the dust mass heated by a radiation field at intensity $U$.  The power-law index $\alpha$ was allowed to vary from 1 to 3.  We used the stellar emission models from \citet{bru03} and the standard default nebular emission model included in CIGALE.

For the AGN emission we used the \citet{fri06} AGN emission models, which assume isotropic emission from a central source and emission from a surrounding toroidal dust structure.  The assumed central point-like luminous source was defined with a composite power-law in $\lambda L(\lambda)$.  In particular, from 0.001 to 0.03~\micron, $\lambda L(\lambda) \propto \lambda^{1.2}$; from 0.03 to 0.125~\micron, it is independent of wavelength; and from 0.125~\micron\ to 20~\micron, $\lambda L(\lambda) \propto \lambda^{-0.5}$ (\citeauthor{gra94} \citeyear{gra94}, \citeauthor{nen02} \citeyear{nen02}).  The rest of the IR emission comes from the blackbody emission due to AGN heating of the torus.  The AGN emission was calculated for an intermediate-type AGN with an axis angle of 30\fdg1 (where 0\degr\ corresponds to a Seyfert~1 galaxy viewed pole-on and 90\degr\ corresponds to a Seyfert~2 galaxy viewed edge-on).  The default 30\fdg1 axis angle was provided to CIGALE at the outset, and there was no noticeable distinction between different intermediate angles when tested against the simulated SEDs.  Viewing the AGN from 30\fdg1 provides enough dust extinction from the torus surrounding the accretion disk to completely attenuate the UV emission from the AGN.

We held many CIGALE parameters constant but varied in particular the parameters defining the AGN, dust attenuation, and star formation history to model these galaxies as accurately as possible.  Table \ref{params} lists the values and/or ranges for the parameters used.  With these settings, every CIGALE run calculated the reduced $\chi^2$ for each of 6.3 million model SEDs.  The model SED with the lowest $\chi^2$ was saved along with probability density functions (PDFs) for each parameter and a text file containing the best models along with the estimates and uncertainties for each parameter.  These uncertainties were derived from the 1$\sigma$ standard deviations of the PDFs created by CIGALE for each parameter. Thus the `best-fit' SED was not always from the `most probable' individual parameters as found in the PDFs, but they were generally within the uncertainties, in particular for the AGN fraction.

\begin{figure}
\centering
\includegraphics[width=\linewidth]{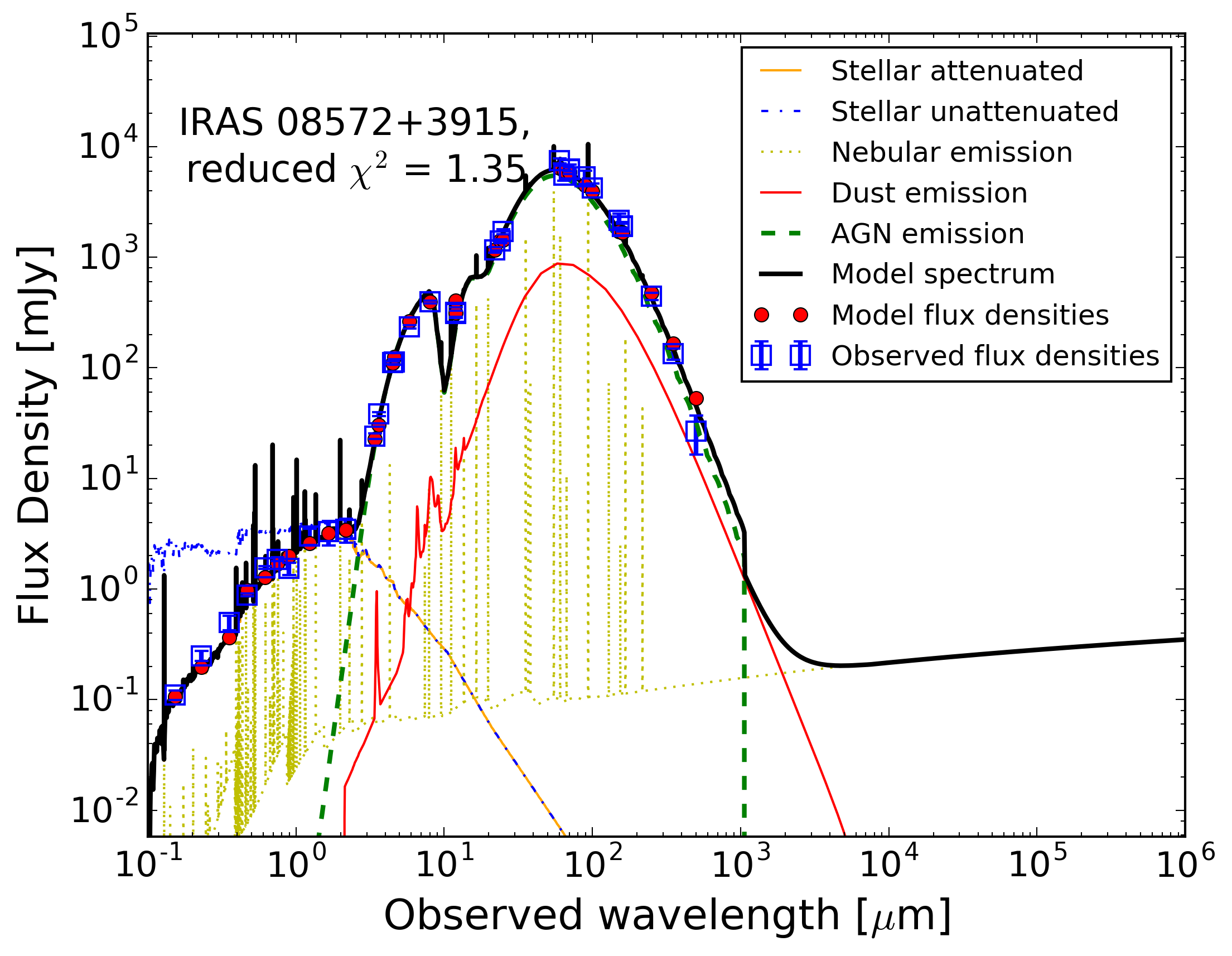}
\caption{An illustration of the data quality and CIGALE SED modeling.  The SED shown (blue symbols in upper panel) is for IRAS~08572+3915.  The best-fit CIGALE model is shown in black.  Red dots indicate CIGALE-derived photometry in the modeled passbands.  The best-fit CIGALE model is the sum of contributions from an AGN (green dashed line), dust-attenuated stellar emission (orange; the intrinsic stellar emission is indicated in blue), nebular  emission (yellow), and dust emission (red).  The bottom panel shows the fractional discrepancies between the model and photometry. The best-fit CIGALE SEDs for all 24 galaxies analyzed in this work are in Figure~\ref{SEDs}. \label{SED}}
\end{figure}

\begin{figure}
\centering
\includegraphics[width=\linewidth]{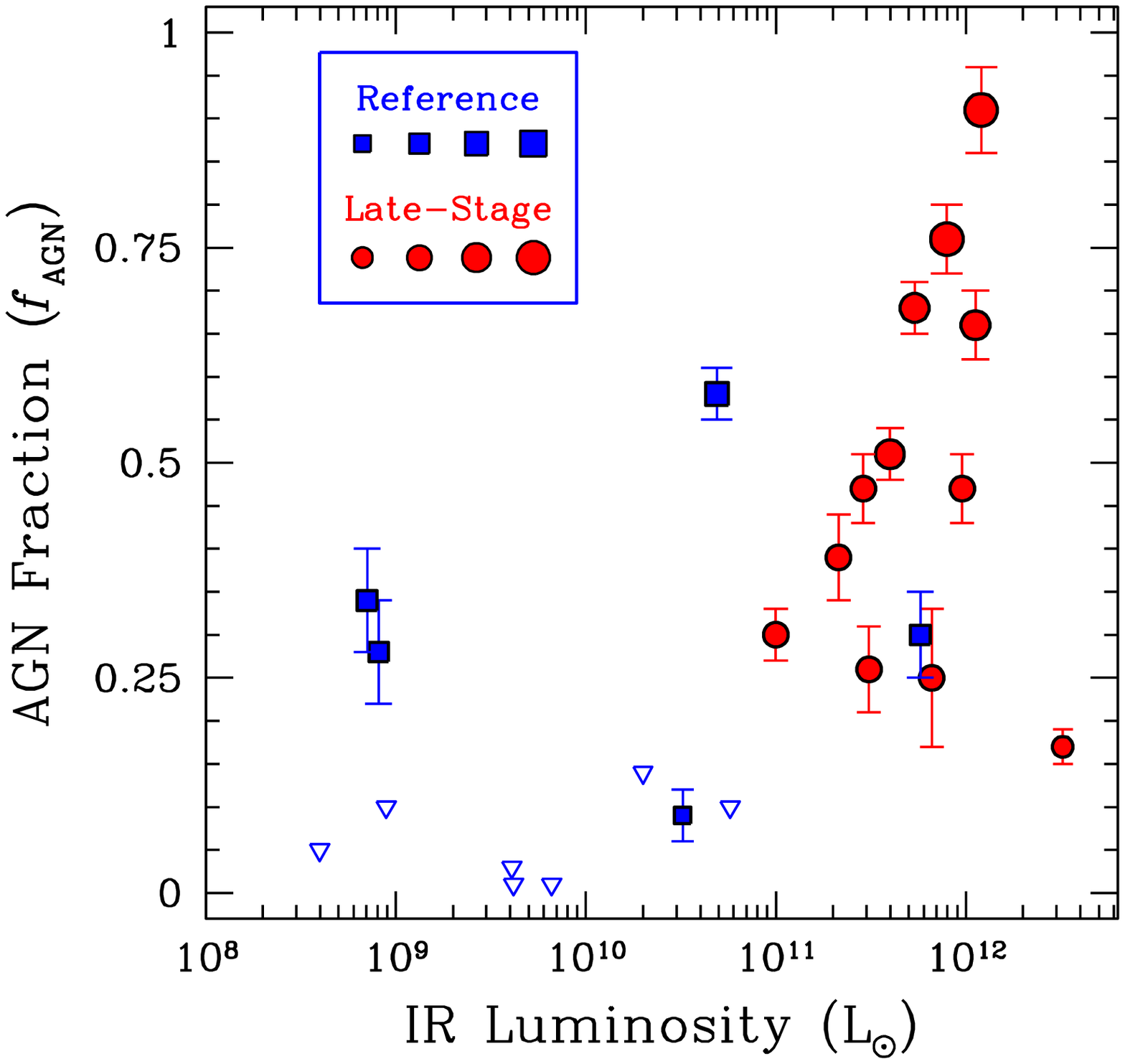}
\caption{Estimated AGN luminosity fractions versus IR luminosity for both the Late-Stage Merger galaxies (red symbols) and the Reference Sample (blue symbols).  The luminosity fractions were measured as a function of IR luminosity between 5 and 1000\,$\mu$m.  Symbol size indicates the percentage AGN luminosity fraction: the smallest symbols are for $f_{AGN}<0.25$, the next largest indicate $0.25\le f_{AGN}<0.50$, and so on.  Triangles indicate 3-$\sigma$ upper limits.
\label{lumAGN}}
\end{figure}

For each best-fit SED identified by CIGALE, we also calculated the AGN luminosity fraction (denoted `fracAGN' by \citeauthor{fri06} \citeyear {fri06} but hereafter referred to as $f_{AGN}$).  $f_{AGN}$ is defined as the AGN contribution to the total IR luminosity from $\sim$5--1000~\micron.  We tested values of $f_{AGN}$ that ranged from 0 to 0.9 (90\% of the total IR luminosity) in increments of 0.1 on all merging galaxies.  $f_{AGN} = 0\%$ accounts for the possibility that an AGN might not contribute to the IR luminosity.  Once the full grid from 0 to 0.9 was tested, CIGALE was run again on each galaxy with a finer but narrower grid for $f_{AGN}$ centered on the best-fit value from the previous run.  Although it is theoretically possible to obtain an AGN fraction close to 1, the probability is extremely low even for the strongest AGN-dominated galaxies in our sample; empirically, we found only one case for which the AGN fraction was significantly larger than 85\%.  For IRAS~08572+3915, $f_{AGN}$ was allowed to exceed 90\% in the model runs with ranges from 0.7 to 0.95 by 0.05 along with 0.99.  CIGALE found a best-fit $f_{AGN}$ value of 91\%, which was obtained through interpolation of the parameter grid points to find the best-fit solution as measured by the reduced $\chi^2$ value.  Table \ref{AGNchi} contains the best-fit $f_{AGN}$ values and reduced $\chi^2$ for each galaxy along with the line ratios described in Section 3.1.  Figure~\ref{SED} shows an example best-fit model, Figure~\ref{$f_{AGN}$pdf} shows the corresponding PDF for $f_{AGN}$, and Figure~\ref{lumAGN} shows $f_{AGN}$ as a function of  100~\micron\ luminosity for both samples. The Late-Stage Sample has larger luminosities than most of the Reference Sample by construction, but nothing selected for or against AGN fraction in either sample.  If anything, AGN of a given luminosity should be easier to detect in low-luminosity galaxies, i.e., in the Reference Sample.

\begin{table*}
\centering
\footnotesize
\caption{CIGALE Parameter Settings Used in This Work}
\label{params}
\begin{tabular}{c c c}
\hline
\hline
Parameter & Definition & Values Tested (range)\\
\hline
\multicolumn{3}{c}{Star Formation History---Delayed Module}\\\\
$\tau_{main}$ & the e-folding time of the main population (Myr) & 50, 500, 1000, 2500, 5000, 7500\\
age & the age of the oldest stars (Gyr) & 0.5, 1, 2, 3, 4, 5, 6\\
sfr\_A & multiplicative factor controlling SFR amplitude & 1.0\\
separation age & separation between young and old stellar populations (Myr) & 10\\
\hline
\multicolumn{3}{c}{\cite{bru03} Stellar Emission Module}\\\\
imf & initial mass function (0 for Salpeter, 1 for Chabrier) & 0\\
metallicity & initial metallicity for the stars & 0.02\\
separation\_age & age of separation between `young' and `old' stellar populations in Myr & 10\\
\hline
\multicolumn{3}{c}{Nebular Emission Module}\\\\
logU & ionization parameter & $-2.0$\\
f\_esc & escape fraction of Lyman continuum photons & 0.0\\
f\_dust & absorption fraction of Lyman continuum photons & 0.0\\
lines\_width & line width in km/s & 300\\
emission & whether or not to include nebular emission & True\\
\hline
\multicolumn{3}{c}{\cite{cal00} and \cite{lei02} Dust Attenuation Module}\\\\
E\_BVs\_young & $E(B-V)_*$, the colour excess of the young stellar continuum light & 0.1, 0.25, 0.4, 0.55, 0.7\\
E\_BVs\_old\_factor & reduction factor for the $E(B-V)_*$ of old vs.  young population & 0.22, 0.44, 0.66, 0.88\\
uv\_bump\_amplitude & amplitude of the 220 nm bump & 0.0\\
powerlaw\_slope & slope delta of the power law attentuation curve & 0.0\\
filters & filters in which attenuation will be calculated & FUV\\
\hline
\multicolumn{3}{c}{\cite{dal14} Dust Module}\\\\
$\alpha$ & slope of the dust temperature distribution in \ref{daldust_eq} & 1, 1.5, 2, 2.5, 3\\
\hline
\multicolumn{3}{c}{\cite{fri06} AGN Module}\\\\
r\_ratio & the ratio between outer and inner radius of AGN torus & 10, 30, 60, 100, 150\\
$\tau$ & the optical depth at 9.7~\micron\ & 0.6, 1, 6, 10\\
$\beta$ & the density radial exponent & $-1, -0.75$\\
$\gamma$ & the density exponential factor & 0, 2\\
opening\_angle & the opening angle of the torus & 60, 100, 140\\
$\psi$ & the angle between equator and line of sight & 30.1\\
& (0 is Type 2 and 89.9 is Type 1) & \\
$f_{AGN}$ & the AGN fraction to the IR luminosity & 0, 0.1, 0.2, 0.3, 0.4, 0.5, 0.6, 0.7, 0.8,\\
 & & 0.9 (0.95, 0.99)\\
\hline
\end{tabular}
\end{table*}

\begin{table*}
\centering
\small
\caption{Derived fractional AGN contribution to the total IR luminosity, SFR, and reduced $\chi^2$ from the CIGALE models, and the measured [Ne \textsc{v}]/[Ne \textsc{ii}] \& [O \textsc{iv}]/[Ne \textsc{ii}] ratios from \textit{Spitzer} IRS spectra.}

\label{AGNchi}
\begin{threeparttable}
\begin{tabular}{c c c c c c}
\hline
\hline
Galaxy Name & $f_{AGN}$ & SFR ($M_{\odot}$ yr$^{-1}$) & Reduced $\chi^2$ & [Ne \textsc{v}]/[Ne \textsc{ii}] & [O \textsc{iv}]/[Ne \textsc{ii}]\\
\hline
\multicolumn{5}{c}{Late-Stage Merger Sample}\\
\hline
IRAS~08572+3915 & $0.91\pm0.05$ & $21.4\pm4.9$ & 1.35 & $< 0.107$ & $< 0.574$\\
IRAS~15250+3609 & $0.47\pm0.04$ & $89.8\pm8.1$ & 2.31 & $< 0.101$ & $< 0.202$\\
Mrk~231 & $0.17\pm0.02$ & $444\pm23$ & 3.18 & $< 0.003$ & $< 0.362$\\
Mrk~273 & $0.66\pm0.04$ & $57.9\pm7.5$ & 1.66 & $0.227\pm0.014$ & $1.206\pm0.091$\\
Mrk~463 & $0.68\pm0.03$ & $30.0\pm2.7$ & 2.95 & $2.038\pm0.136$ & $6.866\pm0.733$\\
NGC~2623 & $0.39\pm0.05$ & $20.0\pm1.4$ & 3.88 & $0.062\pm0.007$ & $0.173\pm0.015$\\
NGC~3758 & $0.30\pm0.03$ & $7.02\pm1.12$ & 1.36 & ... & ...\\
NGC~6090 & $0.26\pm0.05$ & $35.0\pm2.3$ & 2.00 & $< 0.009$ & $0.045\pm0.021$\\
UGC 4881 & $0.51\pm0.03$ & $23.2\pm1.9$ & 1.39 & $< 0.005$ & $0.037\pm0.008$\\
UGC 5101 & $0.76\pm0.04$ & $20.6\pm3.8$ & 5.56 & 0.152$\pm$0.018 & $0.158\pm0.018$\\
VV 283 & $0.47\pm0.04$ & $20.7\pm1.9$ & 1.79 & 0.008$\pm$0.001 & $0.015\pm0.004$\\
VV 705 & $0.25\pm0.08$ & $77.1\pm8.8$ & 0.97 & $< 0.006$ & $< 0.019$\\
\hline
\multicolumn{5}{c}{Reference Sample}\\
\hline
M51A & $0.09\pm0.03$ & $2.73\pm0.14$ & 0.96 & $0.032\pm0.002$ & $0.238\pm0.009$\\
M51B & $< 0.03$ & $< 0.11$ & 0.88 & $< 0.057$ & $< 0.108$\\
NGC~2976 & $0.28\pm0.06$ & $0.078\pm0.004$ & 1.34 & $< 0.133$ & $< 0.041$\\
NGC~3031 & $< 0.01$ & $0.382\pm0.019$ & 1.57 & $< 0.024$ & $0.149\pm0.029$\\
NGC~3077 & $0.34\pm0.06$ & $0.069\pm0.014$ & 0.78 & $< 0.003$ & $< 0.048$\\
NGC~3190 & $< 0.01$ & $0.040\pm0.029$ & 1.22 & $< 0.081$ & $0.113\pm0.029$\\
NGC~3690 & $0.30\pm0.05$ & $59.3\pm3.2$ & 2.01 & $< 0.007$ & $< 0.027$\\
NGC~4625 & $< 0.10$ & $0.105\pm0.005$ & 0.68 & $< 0.141$ & $< 0.044$\\
NGC~5394 & $0.58\pm0.03$ & $3.09\pm0.15$ & 2.19 & $0.005\pm0.002$ & $0.016\pm0.001$\\
NGC~5395 & $< 0.10$ & $6.27\pm0.31$ & 0.48 & $< 0.021$ & $< 0.028$\\
M101 & $< 0.14$ & $4.01\pm0.20$ & 0.68 & $0.111\pm0.041$ & ...\\
NGC~5474 & $< 0.05$ & $0.006\pm0.0003$ & 4.08 & $< 1.02$ & $< 0.38$\\
\hline
\hline
\end{tabular}
\begin{tablenotes}
\item[a] Upper limits are defined at 3$\sigma$
\item[b] `...' indicates the galaxy was not observed in the required wavelengths by IRS
\end{tablenotes}
\end{threeparttable}
\end{table*}

\begin{figure}
\centering
\includegraphics[width=\linewidth]{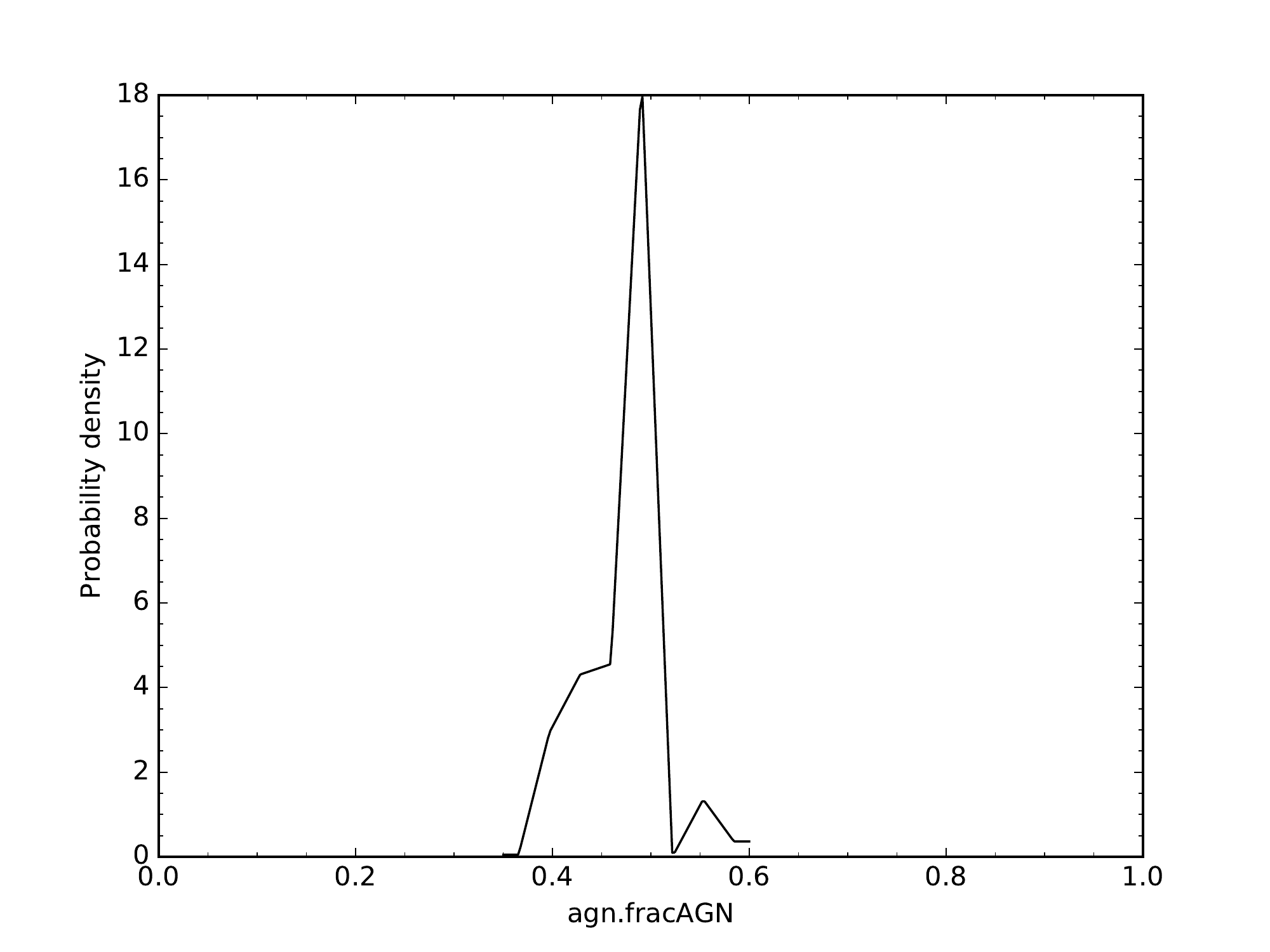}
\caption{A measure of the $f_{AGN}$ probability for IRAS~15250+3609. CIGALE found 0 probability for values of $f_{AGN}$ less than 0.3 and greater than 0.6 \label{$f_{AGN}$pdf}}
\end{figure}

In some cases, CIGALE produced a best-fit model with $f_{AGN}$ = 0 having no estimated uncertainty, even for galaxies known to host an AGN from previous studies.  In general when $f_{AGN} < 0.1$, the uncertainties tend to be large fractions of the best-fit value.  As a result, the AGN model is relatively poorly constrained in low-$f_{AGN}$ cases.  This is because with just photometric data, a weak AGN cannot be distinguished from a slightly increased SFR. CIGALE uses a slightly different, more luminous dust model in the IR to account for the small influx to the SED that the AGN contributes \citep{cie15}.  However, in most of these cases, the CIGALE best-fit model underestimates the far-IR \textit{Herschel}/SPIRE bands, up to a factor of 1.5 or 2.  CIGALE models including non-thermal radio emission were considered in an attempt to better fit the \textit{Herschel}/SPIRE data points, but the radio emission was ultimately discarded as negligible because the added emission in the far-IR was 3 orders of magnitude too small to make up the difference.

\begin{figure}
\centering
\includegraphics[width = \columnwidth]{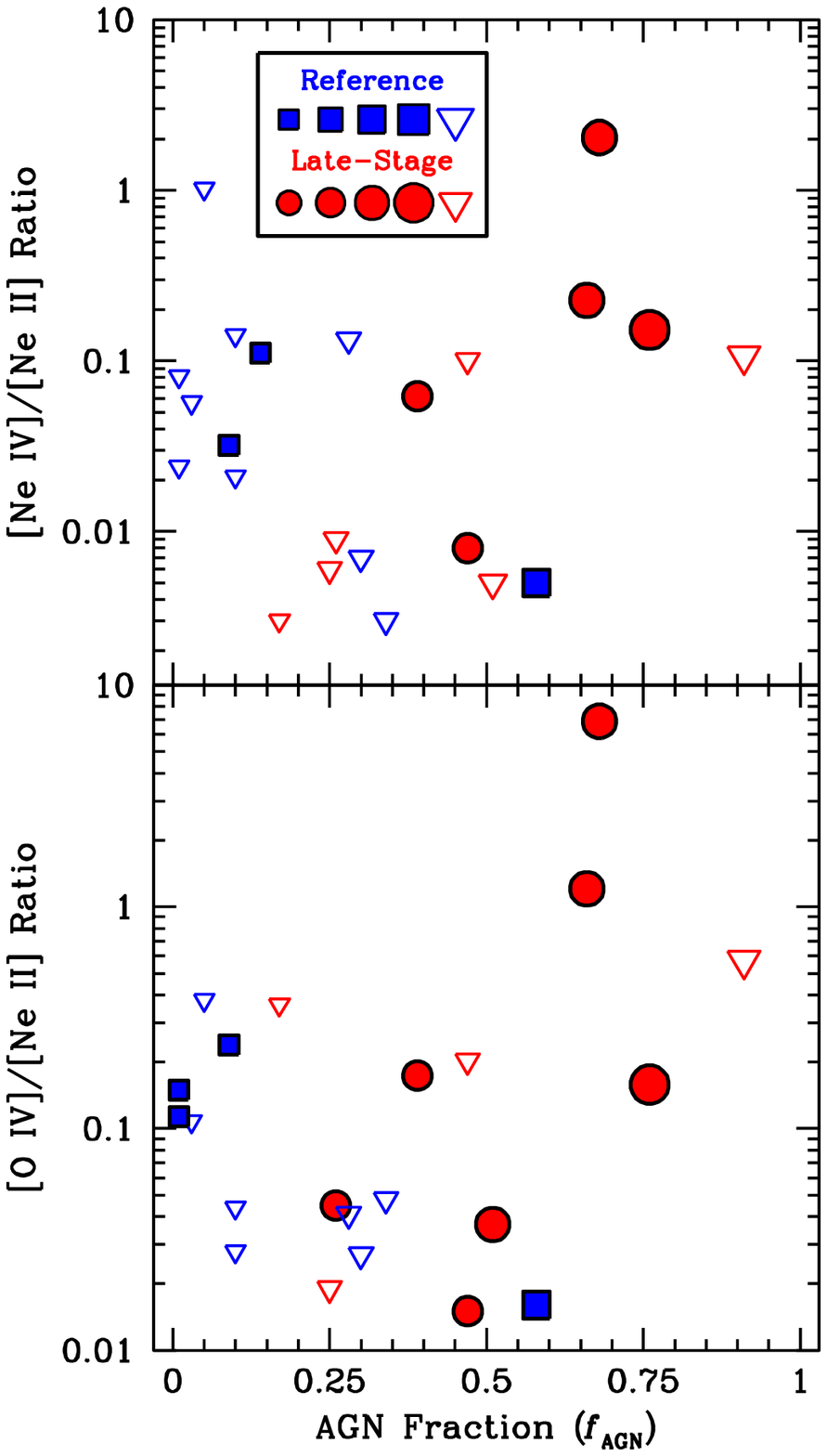}
\caption{Integrated emission-line flux ratios as a function of AGN luminosity fraction.  {\sl Upper panel:} [Ne \textsc{v}]/[Ne \textsc{ii}].  Symbol size indicates the percentage AGN luminosity fraction: the smallest symbols are for $f_{AGN}<0.25$, the next largest indicate $0.25\le f_{AGN}<0.50$, and so on.  Triangles indicate 3-$\sigma$ upper limits.  {\sl Lower panel:} [O \textsc{iv}]/[Ne \textsc{ii}].
\label{line_ratios}}
\end{figure}

For the spectral line analysis, with the results shown in Figure~\ref{line_ratios}, we ran linear regression tests on the combination of both samples for both [Ne~\textsc{v}]/[Ne~\textsc{ii}] vs.  $f_{AGN}$ and [O~\textsc{iv}]/[Ne~\textsc{ii}] vs.  $f_{AGN}$, and the results were not significant at the 3$\sigma$ level. However, when running the linear regression test on the late-stage merger sample only, there was evidence for a linear trend at the 2$\sigma$ level. Further analysis of spectral line ratios, including correlations with flux density ratios, is discussed by Ramos Padilla et al.\ (2018, in preparation).

\subsection{AGN Observables}

Numerous studies (e.g., \citeauthor{ste05}, \citeyear{ste05}, \citeauthor{ste12} \citeyear{ste12}, \citeauthor{don08} \citeyear{don08}, \citeauthor{ros12} \citeyear{ros12}) have demonstrated that galaxy colours can reveal AGN.  Flux ratios such as $f_{24~\micron} / f_{100~\micron}$ and $f_{12~\micron} / f_{24~\micron}$ as well as \textit{K--L} and \textit{L--M} colours have been used to help determine the presence of an AGN.  We performed a linear regression test of $f_{AGN}$ versus every flux ratio in our photometric data to determine whether any ratio showed a significant correlation.  The significant results, with Pearson correlation ratios of magnitude greater than 0.8 and with \textit{p}-values of 0.027 or less (corresponding to a significance level of 3$\sigma$), are summarized in Table \ref{flux_ratios}.

\begin{figure}
\centering
\includegraphics[width=\columnwidth]{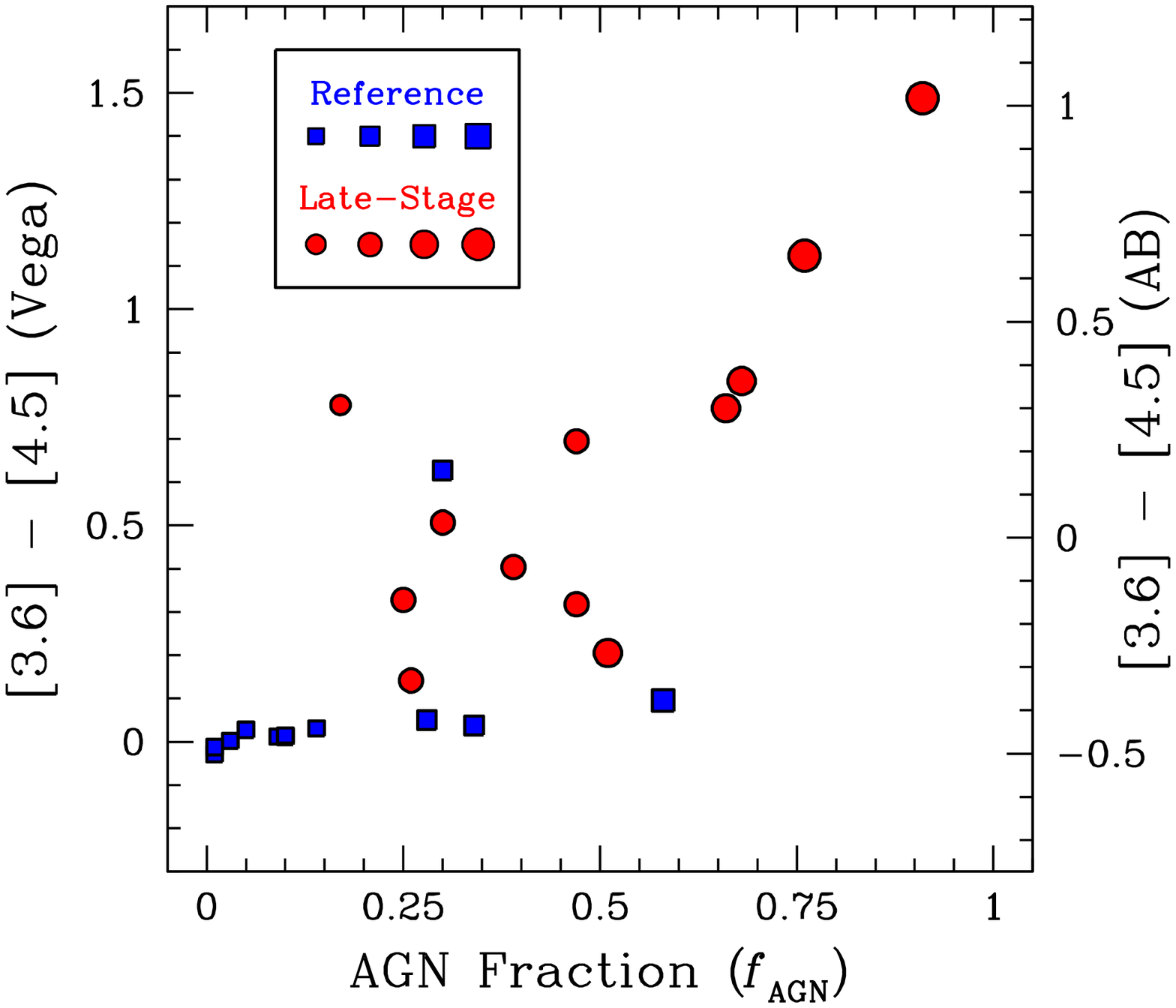}
\caption{IRAC $[3.6]-[4.5]$ color versus AGN luminosity fraction.  Symbol size indicates the percentage AGN luminosity fraction: the smallest symbols are for $f_{AGN}<0.25$, the next largest indicate $0.25\le f_{AGN}<0.50$, and so on.
\label{obsfrac}}
\end{figure}

\begin{figure}
\centering
\includegraphics[width=\columnwidth]{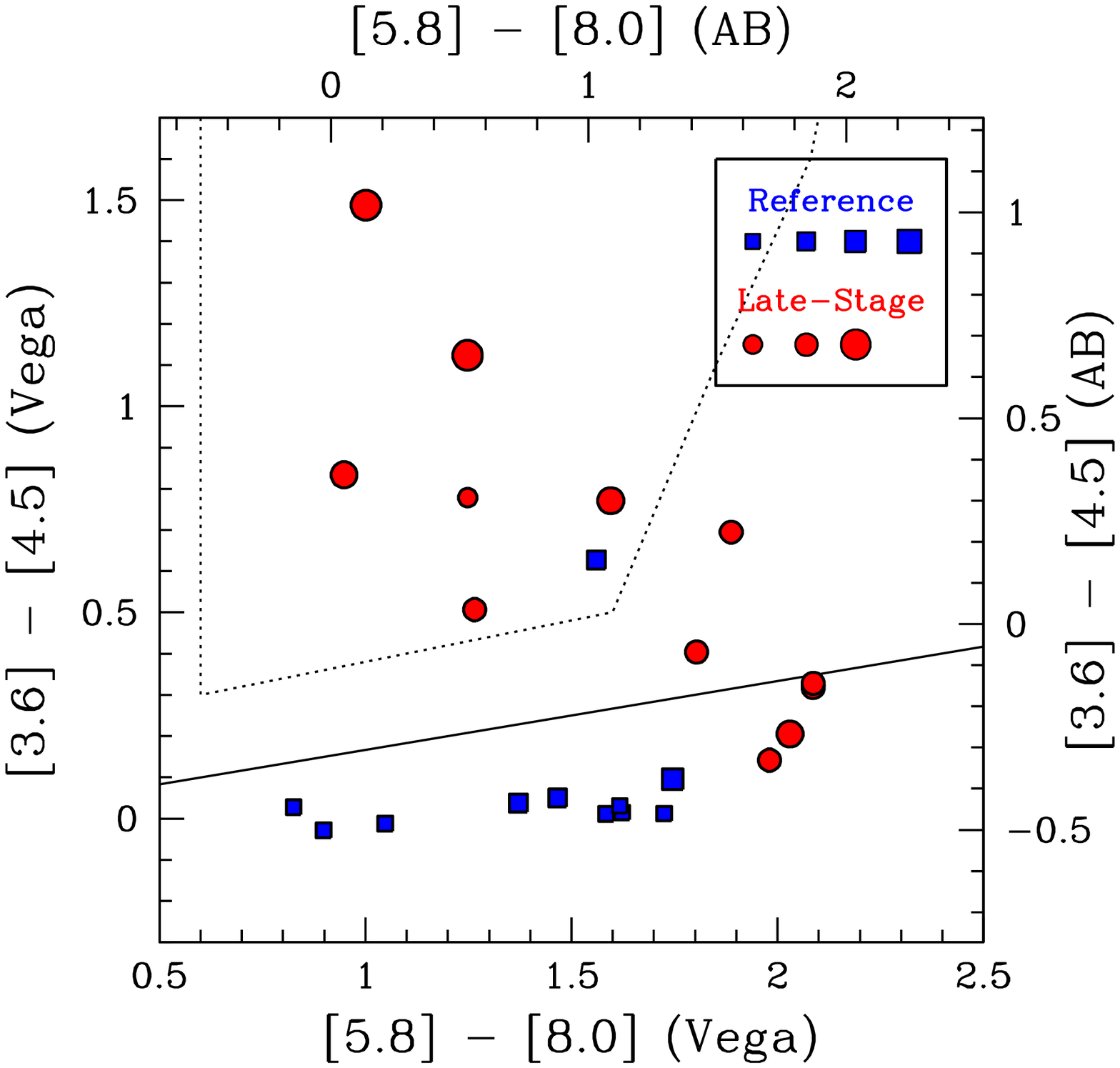}
\caption{IRAC colour-colour plot following Stern et al.\ (2005) for all galaxies in both subsamples analyzed in this work.  Symbol size indicates the percentage AGN luminosity fraction: the smallest symbols are for $f_{AGN}<0.25$, the next largest indicate $0.25\le f_{AGN}<0.50$, and so on.  The dotted line encloses the region in which low-redshift galaxies reside when their IRAC colours are dominated by luminous AGN (the AGN wedge).  The straight line is an empirical boundary below which nearby star-formation-dominated galaxies typically lie.}
\label{stern_wedge}
\end{figure}

$L-M$ and similar colours such as IRAC $[3.6]-[4.5]$ and \textit{WISE} $W1-W2$ are the basis for the \citet{ste05} and \citet{don08} plots showing a significant difference for the AGN-dominated systems. The $[3.6]-[4.5]$~\micron\ colour is significant here at ${\sim}6\sigma$, as seen in Figure~\ref{obsfrac}. However, the corresponding \textit{WISE} colours and mixing of IRAC and \textit{WISE} colours are not significant due to the low sample size of galaxies with \textit{WISE} photometry.  Consistent with the SED analysis described in Sec.~\ref{ssec:fitting}, the early-stage mergers consistently populate the starburst region of the \citet{ste05} IRAC colour--colour plot.  By contrast, even though our analysis shows that all the late-stage merger galaxies contain luminous AGN and moreover that many of them are AGN-dominated, only about half of them have IRAC colours indicating that these objects host luminous AGN. The apparent discrepancy is not surprising because the SED analysis is based on much more information than the simple colour--colour plot; in particular, it attempts to estimate and account for dust obscuration.  The \citet{ste05} plot could miss AGN when stars overwhelm the AGN at IRAC wavelengths or when the AGN is heavily obscured by dust in the near-mid IR. However, the four largest AGN fractions modeled by CIGALE correspond to the four galaxies with the reddest [3.6]--[4.5] colours are Figure~\ref{stern_wedge} contains the Stern colour--colour plot for the 24 galaxies in our sample.  The galaxies from the Reference Sample nearly all fall in this region.  By contrast, the late-stage merger subsample populates both the AGN wedge \emph{and} the star-formation region of the plot, albeit only the extreme red end of the latter.

\citet{san88} showed that the $[25]-[60]$ colour is an AGN tracer, but this colour is not correlated with our measurements of $f_{AGN}$ as given by CIGALE. Two of the galaxies modeled with strong AGN are among the bluest in $[25]-[60]$ colour, and the reddest $[25]-[60]$ measurement belongs to a galaxy modeled with an AGN fraction of $\sim$5\%. Ramos Padilla et al.\ (2018, in preparation) analyzed the correlation this colour has with spectral line ratios in addition to AGN fraction. The UV, optical, and 2MASS near-IR connected with IRAC, IRAS, MIPS, and PACS mid and far-IR have strong correlations with $f_{AGN}$. In particular, the flux ratios of \textit{GALEX} FUV and NUV, Sloan \textit{u, g, r, i, z}, and 2MASS $J$, $H$, and $K_s$ with IRAC 4.5 and 5.8~\micron\ and MIPS 24~\micron\ and 70~\micron\ bands are significantly correlated with AGN fraction. The extremely red colors at the high end of the correlation show that a steep increase in the SED in the near and mid-IR is indicative of an AGN. Photometric data at similar wavelengths show similar correlations; UV--70~\micron\ flux ratios are significant for both MIPS and PACS 70~\micron\ photometry, for example. Also, negative correlations are found between the AGN fraction and the MIPS 70~\micron\ and PACS/SPIRE colours, showing that the presence of an AGN makes the far-IR SED steeper than the expected cool dust power-law.

\section{Simulations}

We tested the reliability of CIGALE by analyzing the SEDs of simulated merging galaxies and comparing the CIGALE-derived galaxy parameters with the known galaxy parameters from the simulations.  Performing such `numerical experiments' using simulations is a very useful means to validate methods of observational inference, as the ground truth is known \textit{a priori} and various uncertainties can be controlled. For previous examples and discussions of this validation process, see \citet{mic14}, \citet{hay15}, \citet{smi15}.  As \citet{lan14} have described, the simulations provide realistic SEDs (see also \citeyear{wei18} \citeauthor{wei18}).  The aim was to determine how well CIGALE recovers $f_{AGN}$ (i.e., the AGN's contribution to the bolometric luminosity) of the simulated galaxies.  The simulated merger SEDs were created using a two-step process.  First, mergers were simulated using a hydrodynamical code \citep{spr05, hay11, lan14}, and then a radiative transfer code was used to generate the emergent light from the simulated mergers and simulate an observation \citep{jon06, jon10}. The hydrodynamic simulations and radiative transfer code used are described in detail by \citet{lan14} and \citet{wei18}. Here we summarize the key aspects of our analysis.

\subsection{Models}

The merger simulations used the TreeSPH \citep{her89} code GADGET-3 \citep{spr05}, which employs a hierarchical tree N-body method to compute gravitational interactions in an N-body cosmological simulation that includes gravity, gas dynamics (via smoothed-particle hydrodynamics), stellar evolution, and other physical mechanisms.  GADGET-3 implements the thermodynamic transport of energy through gas dynamics and radiative heating and cooling and conserves both energy and entropy.  The ISM is modeled with two phases of matter in which cold, dense clouds interact with a hot, diffuse gas medium \citep{spr03}.  The hydrodynamical code models star formation according to the Kennicutt-Schmidt (`K-S') relation \citep{ken98}, an empirical relation between SFR and the volume density of gas.  When the density of gas particles in the simulation surpasses a minimum threshold, gas particles are converted into star particles according to the K-S relation.  GADGET-3 uses sub-resolution models to describe starforming regions because its grid is too coarse to resolve individual cold gas clouds; this limitation directly affects how radiative transfer is modeled.

The AGN contributions to the SEDs were computed from the black hole accretion rate, and the
corresponding AGN feedback was included using the sub-resolution model of \citet{spl05}.  The AGN were represented in the simulations by black hole particles which grow and radiate by accreting surrounding gas \citep{spl05}.  Black hole particles accreted according to the Bondi-Hoyle-Lyttleton model, at the rate
\begin{equation}
\dot{M}_{BH} = \frac{4\pi\alpha G^2M_{BH}^2\rho}{(c_s^2 + v^2)^{3/2}},
\end{equation}
where $\rho$ is the gas density, $c_s$ is the speed of sound in the gas, $v$ is the black hole speed relative to the gas, and $\alpha$ is a system-dependent, dimensionless parameter, usually estimated as between 1 and 2 \citep{bon52}; we took $\alpha = 1.5$.  Because the accretion occurs on spatial scales far below GADGET-3's resolution, the code uses a sub-resolution model to interface black hole particles to surrounding gas particles.  GADGET-3 models accretion of gas particles as a stochastic process.  Each particle near a black hole is assigned a probability of accretion weighted by the estimated gas density near the black hole, the location of the particle relative to the BH, the Bondi accretion rate, and the time step.  GADGET-3 also imposes an upper limit on $\dot{M}_{BH}$ at the Eddington rate, at which the radiation pressure from an AGN overcomes the gravitational attraction of the gas.

As an AGN accretes gas, its accretion disk heats up and radiates energy into the host galaxy.  GADGET-3 treats the thermal energy delivered to the black hole as thermal energy radiated into the surroundings with power
\begin{equation}
L_r = \epsilon_r \dot{M}_{BH} c^2,
\label{LMdot}
\end{equation}
where $\epsilon_r$ is the radiative efficiency, which is set to 10\% in these simulations, the consensus value for efficient black hole accretion. As can be seen in Equation~\ref{LMdot}, the AGN luminosity is directly proportional to the accretion rate, so when the AGN is turned off, as described below, $\epsilon_r = 0$.  In this way, the accreting AGN directly influence surrounding regions of star formation.

We used the 3D polychromatic Monte Carlo dust radiative transfer code SUNRISE \citep{jon06, jon10} to calculate spatially resolved UV--mm SEDs for the simulated galaxies. SUNRISE performs a radiative transfer calculation for the attenuation and re-emission from the dust heated by star formation and AGN activity, as well as the stellar components, to generate `observed' SEDs for the merger.  Merger steps for SED calculation were at regular intervals at 10~Myr near coalescence and at 100~Myr otherwise \citep{lan14}.  SEDs were computed for seven different viewing angles at each step to account for the impact of dust attenuation along different lines of sight.

\begin{figure}
\centering
\includegraphics[width=\columnwidth]{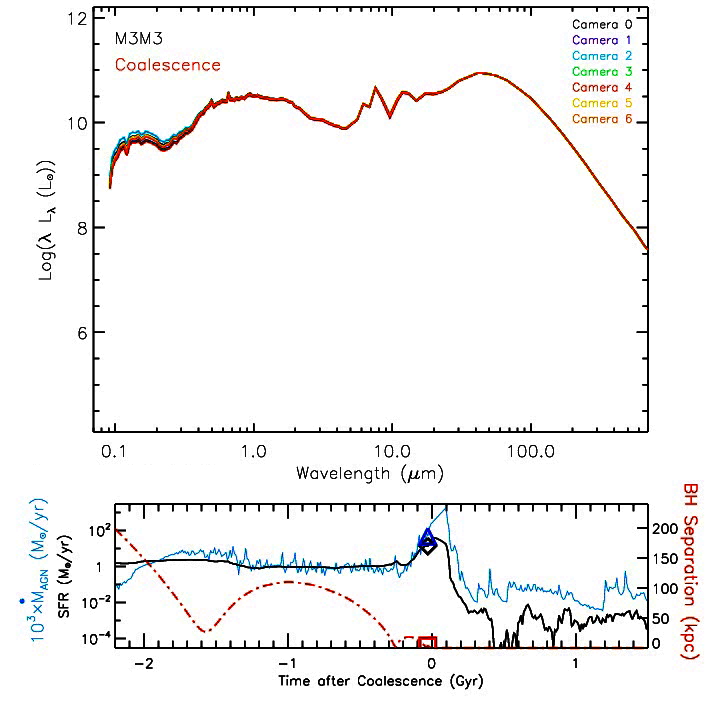}
\caption{The SUNRISE SED output for the M3--M3 major merger.  (Top) The full
SED just prior to coalescence of the two supermassive black holes.  The seven
color-coded viewing angles in this case give nearly identical SEDs.  (Bottom)
The SFR (black), AGN accretion rate (blue) and black hole separation versus merger
time in Gyr with respect to coalescence. The diamond and square markers indicate
the moment corresponding to the SED in the upper panel.}
\label{SED_M3M3}
\end{figure}

Five galaxy models called M4, M3, M2, M1, and M0 with stellar masses respectively of 5, 4.22, 1.18, 0.38, and 0.061 $\times 10^{10}$~\Msol\ were used (see Table 2 of \citeauthor{lan14} \citeyear{lan14}; \citeauthor{rol15} \citeyear{rol15}; \citeauthor{hay11} \citeyear{hay11}).  One further model named c6e was a massive gas-rich galaxy with a halo mass of $9 \times 10^{12}$~\Msol\ and a gas fraction of 60\%, meant to mimic some submillimeter galaxy (SMG) properties.  Figure~\ref{SED_M3M3} shows the simulated SED for the M3--M3 merger case.  We created output files at the specified intervals during the mergers of all combinations of the six galaxy models and then ran SUNRISE to compute the SED for each step and seven viewing angles of every merger.

\subsection{Simulated SEDs Results}

We did not see any evidence that CIGALE's output reliability depended on the particulars of the merger scenario, and the M3--M3 or M4--M4 major merger simulations represent the results.  As \citet{lan14} have already described, those simulations give realistic SEDs \citep{wei18}.  Figure~\ref{SED_M3M3} (lower) illustrates the black hole accretion varying during the M3-M3 merger, peaking (for this example) at about 0.7~\Msol~yr$^{-1}$ shortly after the moment of coalescence.  The accretion rate hovers at a few times $10^{-3}$\Msol~yr$^{-1}$ for most of the early stages of the interaction, even during the first close pass of the two black hole nuclei, but starts to climb to its peak about 12~Myr before coalescence, when the separation is about 150~kpc.  The accretion activity remains above the earlier baseline level for about 30~Myr, during which time the increasingly dense gas in the simulation also produces a burst of star formation, and afterwards the AGN accretion drops to a new baseline nearly 20 times smaller than the pre-merger level.

\begin{figure}
\centering
\includegraphics[width=\columnwidth]{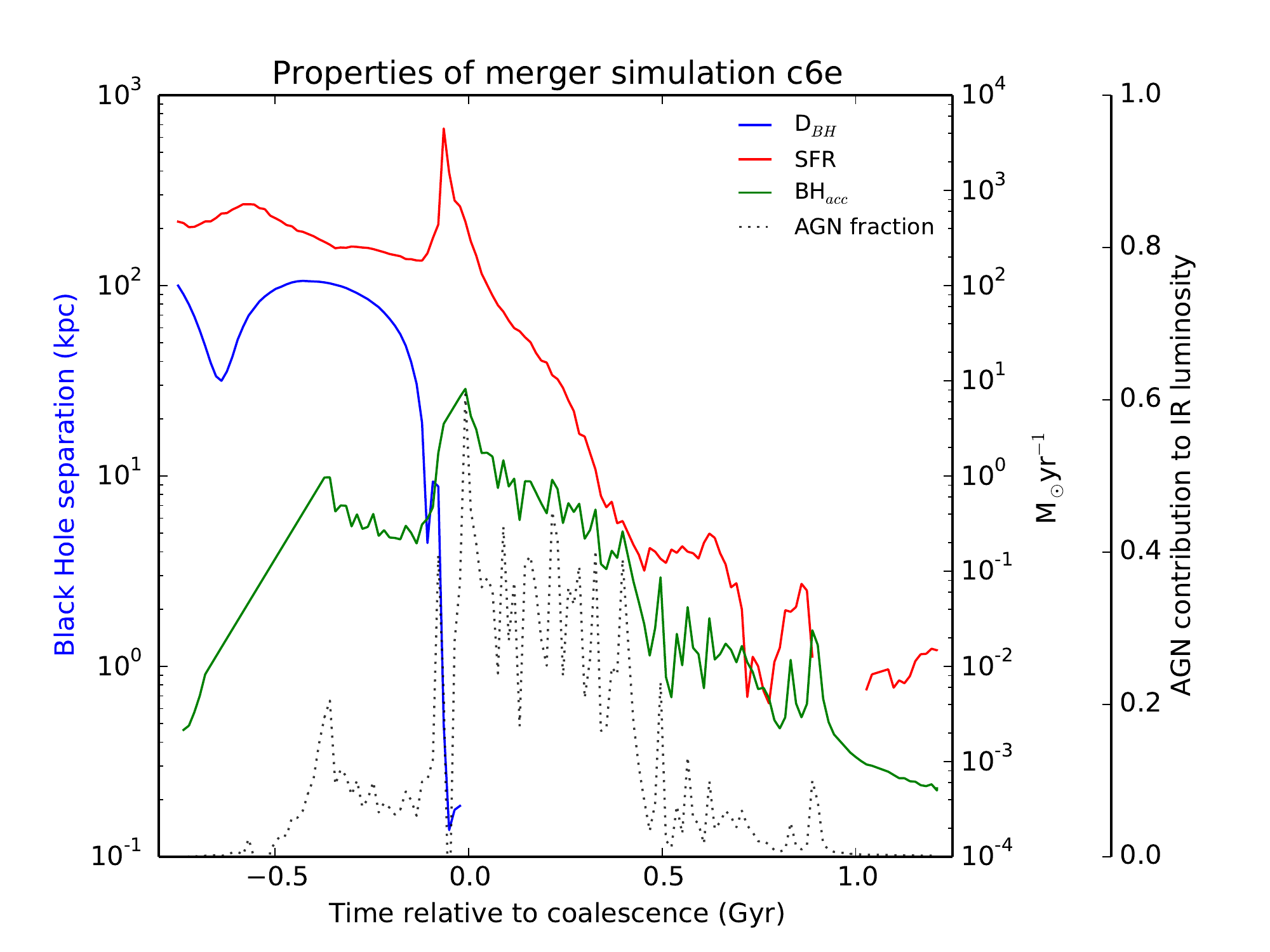}
\caption{The SUNRISE SED output of the c6e--c6e merger.  In this gas-rich example, the black hole separation is shown as the blue line, the star formation rate is indicated in red, and the black hole accretion rate is in green.  The AGN fractional contribution to the lumninosity is shown with the dotted black line; the $L_{AGN}$ was calculated from the accretion rate, and $L_{SFR}$ was calculated from the model's SFR.}
\label{c6e}
\end{figure}

The gas-rich merger simulation `c6e--c6e' has an initial gas fraction of 60\%, in contrast to the other simulated galaxies which, independent of mass, used gas fractions of only 15 and 38\% \citep{cox08, hay13, lan14}.  The c6e--c6e simulation uses a baryonic mass of $4 \times 10^{11}$~\Msol, considerably more than the other simulations, but the same black hole mass of $1.4 \times 10^5$\Msol.  For this merger, Figure~\ref{c6e} plots the AGN fraction along with some other parameters versus time.  The AGN luminosity in this example peaks briefly as high as 55\% at coalescence, and estimates of the SFR based solely on $L_{FIR}$ will be correspondingly too high.

\begin{figure}
\centering
\includegraphics[width=\columnwidth]{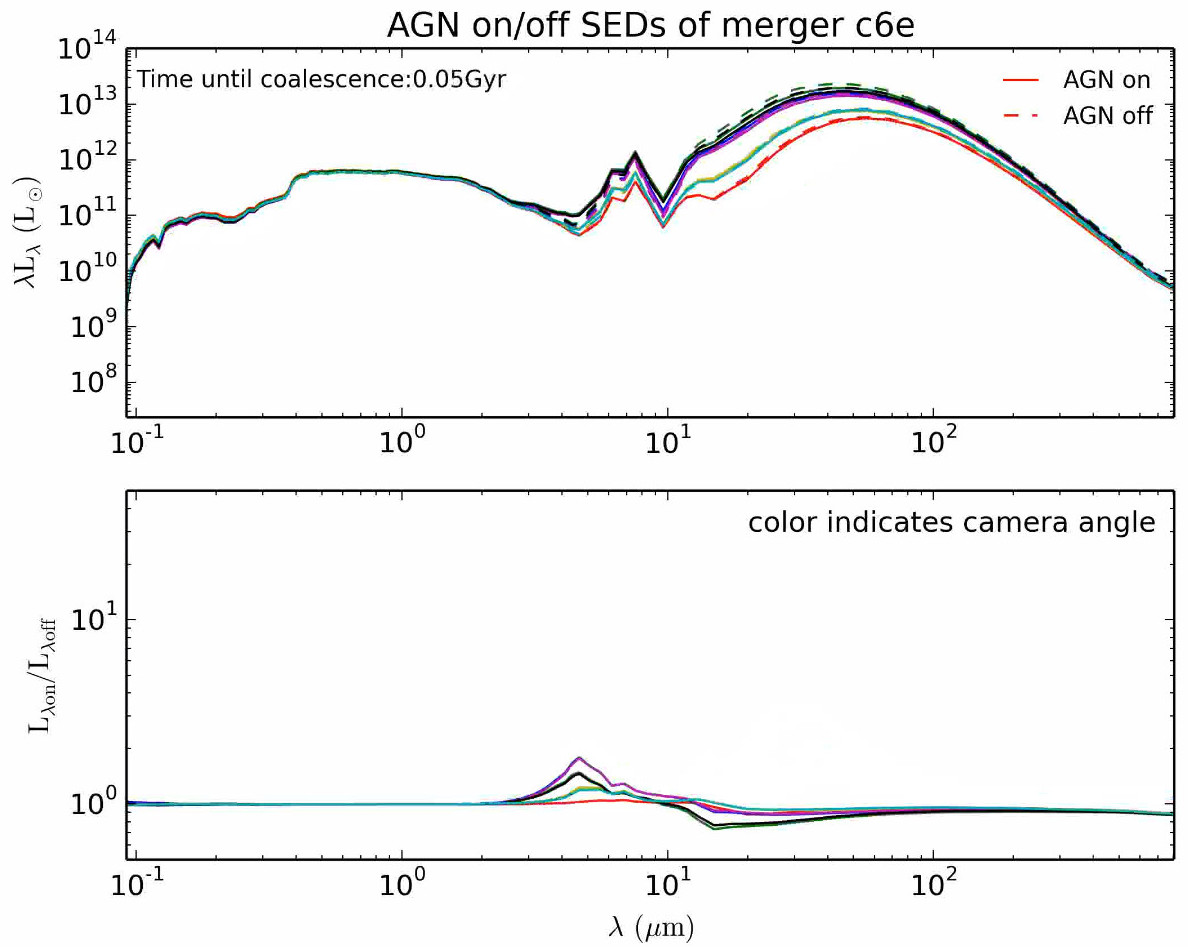}
\caption{AGN on vs. AGN off: the SUNRISE output for the c6e--c6e major merger at 0.7 Gyr. Colors correspond to different viewing angles.  (Top) The SED for the AGN turned on (solid curves) and AGN turned off (dashed curves).  (Bottom) The flux ratios for AGN on / AGN off, showing the spectral differences.}
\label{SED_c6e}
\end{figure}

The SED simulations for c6e--c6e made with the AGN `turned off', that is with $\epsilon_r = 0$, are illustrated in Figure~\ref{SED_c6e}. At the largest viewing angle, (edge-on), the strongest difference between the two cases is a factor of two in the 5--8~\micron\ range. The fact that this part of the spectral range is most sensitive to the AGN fraction confirms what is already well-known from earlier \Sp observations: the IRAC colour-colour diagram as manifest in the so-called Stern wedge is a useful tool to identify AGN \citep{ste05}. However, the reverse is not necessarily true, as Figure~\ref{stern_wedge} shows. Low-luminosity or highly obscured AGN may have blue [3.6]--[4.5] colours or red [5.8]--[8.0] colours, and by using the full SED analysis we can obtain a more reliable accounting of AGN emission and demographics than IRAC colours alone.

\subsection{CIGALE Performance}

\begin{figure}
\centering
\includegraphics[width=0.8\linewidth]{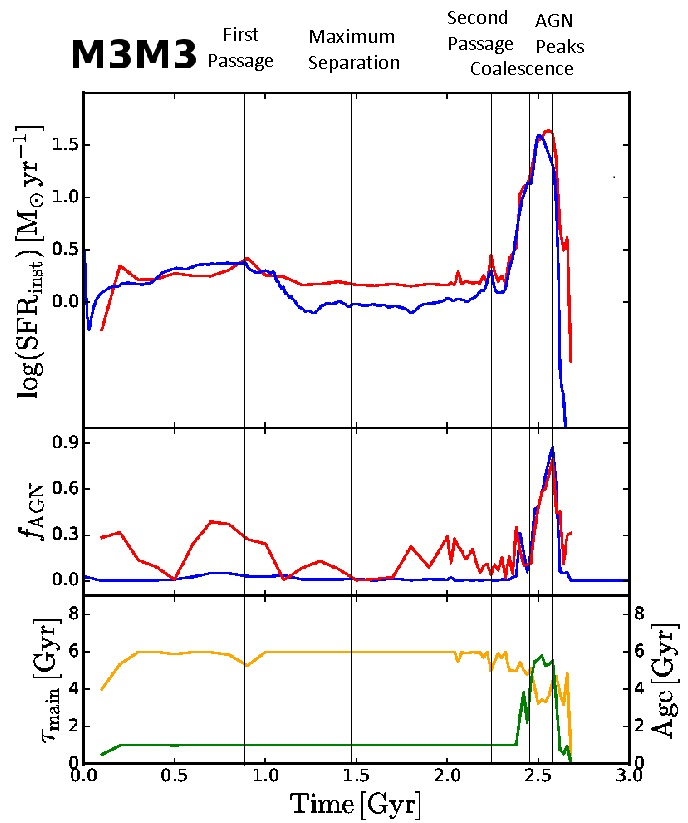}
\caption{CIGALE-derived parameters for the M3--M3 merger; the top panel shows SFR, the middle panel shows $f_{AGN}$, and the bottom panel shows $\tau_{\rm main}$ and age.  The blue curves show the model values output from the GADGET-3 simulations, and the red curves show values derived by CIGALE.  The yellow line in the bottom panel is the derived age, and the green is the derived {\it e}-folding time for the delayed star formation history.}
\label{M3M3_all}
\end{figure}

We compared CIGALE's model results against simulations both with the AGN turned on (a normal $\epsilon_r = 10\%$) and with the AGN turned off ($\epsilon_r = 0$).  The parameters of particular relevance here are: the ratio of the AGN's dust torus radii, the optical depth at 9.7~\micron, the AGN opening angle, the AGN luminosity, and the AGN fraction as estimated both with the \citet{dal14} and the \citet{fri06} methods.  Figure~\ref{M3M3_all} plots the CIGALE-modeled outputs for the SFR and AGN fraction versus the simulated output values as a function of elapsed time with the AGN on.

CIGALE does an excellent job of evaluating $f_{AGN}$ when the AGN is the dominant fraction of luminosity, but it does a poorer job of estimating $f_{AGN}$ when it is below about 15\%.  When CIGALE errs most often it overestimates $f_{AGN}$, but sometimes when $f_{AGN}$ < 0.05 CIGALE returns a value of zero.  The CIGALE $f_{AGN}$ estimate is particularly egregious in the 0.2--1.2~Gyr period, estimating a fraction of as much as 35\% when it is in fact less than 10\%.  This occurs because slight changes to the relative flux densities in the 5--30~\micron\ bands happen in merger phases when the SFR is low and the longer-wavelength emission is therefore also weak.  As a result, these slight changes have an undue impact on CIGALE by causing large shifts in the AGN fraction upward and correspondingly the SFR downward.

In low flux density cases, the full SED spectral output information is used to supplement the calculation of AGN luminosity fraction from the IR bands, which partially corrects for the $f_{AGN}$ estimated by CIGALE.  Similarly, when CIGALE underestimates the observed flux density at wavelengths below 10~\micron\ but overestimates it longward of 10 and 100~\micron\ (as we can see in the SED fit to UGC~5101, for example, in Figure~\ref{SEDs}), additional correction can be applied by shifting upward the allowed range for the $f_{AGN}$ parameter accordingly and/or using a cooler dust model. Not least, spectroscopic information (as per Section 2.3) can also be used to calibrate a CIGALE output when the AGN fraction is low.  In all of the AGN-dominated systems considered here, the AGN fraction is high enough that the CIGALE results are credible.

\section{Discussion}

As shown in Figure~\ref{lumAGN}, the luminosity is strongly correlated with merger stage, which is expected.  The AGN luminosity fraction is also correlated with both, as all of the stage 3.5 or less merging galaxies except for NGC~5394 have AGN luminosity fractions below the values of all of the stage 4 or higher merging galaxies. The lowest $f_{AGN}$ values of 0\% were found in early-stage merging galaxies, and the late-stage merging galaxy with the lowest AGN fraction was Mrk~231, which is classified as a stage 6 at post-coalescence. This galaxy has already completed the merging process, and the AGN luminosity is dropping while the SFR is still relatively high.

Nearly half of the late-stage merging galaxies in our sample of 11 host an AGN that is at least as luminous as the rest of the galaxy in the IR.  These behemoths can even exceed the IR luminosities of their hosts by an order of magnitude, as in IRAS~08572+3915, which at an estimated 91\% surpasses the original limit we had imposed on the AGN fraction at 90\%.  This is consistent with \citet{efs14}, who have also reported an AGN fraction of $\sim$90\%.  More comments on each galaxy in both samples can be found in Appendix A.

Our results are consistent with earlier analysis of merging galaxies.  Timelines have been created of the luminosity of a model merging galaxy, highlighting the period of time when the AGN turns on in the galaxy's later stages, near coalescence (e.g., \citeauthor{lan14} \citeyear{lan14}, \citeauthor{hay14a} \citeyear{hay14a}).  Our results are consistent with the timeline, as the late-stage mergers are more likely to contain an AGN than the early-stage mergers in our sample, and the AGN fractions in the late-stage mergers are usually higher than those in the early-stage mergers.

Figure~\ref{line_ratios} shows that the line ratios in Table \ref{AGNchi} for the early-stage mergers do not correlate with the AGN fraction, but they do show a slight linear trend for the late-stage mergers.  Overall, there was no measurable significant correlation between the [Ne \textsc{v}]/[Ne \textsc{ii}] or [O \textsc{iv}]/[Ne \textsc{ii}] ratio and the AGN fraction in the combined sample.  This could be due to limitations in the measurements, as there were many upper limits for the line ratios in the early-stage merging galaxies in the Reference Sample at low AGN fraction.  Some line ratios in our sample are similar to the line ratios for galaxies shown by \citet{dal09}, but their sample is low-luminosity and unlikely to contain strong AGN.

\section{Conclusions}

The AGN fraction in our Late-Stage Sample is systematically and significantly higher than that measured in our Reference Sample.  For the Late-Stage Sample, $f_{AGN}$ ranges from 17 to 91\%.  The 91\% estimate is for IRAS\,08572+3915, a late-stage ULIRG.  The 17\% estimate is for Mrk\,231, which is classified as a stage 6 post-coalescence merger having a high SFR.  In contrast, in the Reference Sample, $f_{AGN}$ is below 15\% for all but three galaxies.  The difference is probably because the Late-Stage Sample galaxies are advanced in their interaction level, with material flowing to their centers and feeding the AGN that reside there, with the exception of MRK\,231 which is consistent with being in the post-merger stage.

CIGALE SED modeling of late-stage snapshots of a set of SPH merger simulations yields AGN luminosity fractions that are in good agreement in general with the simulation values and also are consistent with values measured in the Late-Stage Merger Sample.  However, CIGALE incorrectly inferred AGN fractions up to 30\% in earlier stages of the simulated mergers when the true value was close to 0.  Galaxies in the Reference Sample with little to no empirical evidence in the SED for AGN activity were not modeled with large $f_{AGN}$, so the SED fits for these galaxies are reliable.

We also measured spectral line ratios for [Ne \textsc{v}]/[Ne \textsc{ii}] and [O \textsc{iv}]/[Ne \textsc{ii}] to provide another tool to estimate the strength of the AGN.  We found no overall correlation above the 2$\sigma$ level in our samples; some strong AGN have comparatively weak line ratios, similar to those of the weaker AGN. The effects of extinction in these mid-IR lines likely plays a significant role. We do, however, find that the late-stage merging galaxies alone do show a possible linear trend between AGN fraction and line ratios.

\section*{Acknowledgements}

We thank the National Science Foundation, the Smithsonian Astrophysical Observatory, and Jonathan McDowell for providing JD the ability to complete this research through the NSF Research Experience for Undergraduates Program held at the SAO.  We would also like to thank Aliza Beverage for her assistance and feedback with the research and writing process, and D.~Burgarella and the CIGALE team for their advice.  The SAO REU program is funded in part by the National Science Foundation REU and Department of Defense ASSURE programs under NSF Grant no.\ 1262851, and by the Smithsonian Institution.  The Flatiron Institute is supported by the Simons Foundation.  HAS, ASW, and JRM-G acknowledge partial support from NASA Grants NNX14AJ61G and NNX15AE56G.  This research has made use of the SIMBAD database, operated at CDS, Strasbourg, France.  This research has made use of the NASA/IPAC Extragalactic Database (NED), operated by the Jet Propulsion Laboratory, California Institute of Technology, under contract with the National Aeronautics and Space Administration.





\begin{thebibliography}{199}
\frenchspacing
\bibitem[Alam et al.(2015)]{ala15} Alam, S.; Albareti, F.D.; Allende Prieto, C.; et al. 2015, \apj, 219, 12
\bibitem[Armus et al.(2007)]{arm07} Armus, L.; Charmandaris, V.; Bernard-Salas, J.; et al. 2007, \apj, 656, 148
\bibitem[Baumgartner et al.(2013)]{bau13} Baumgartner, W.H.; Tueller, J.; Markwardt, C.B.; et al. 2013, \apjs, 207, 19
\bibitem[Bondi(1952)]{bon52} Bondi, H. 1952, \mnras, 112, 195
\bibitem[Brassington et al.(2015)]{bra15} Brassington, N.; Zezas, A.; Ashby, M.L.N.; et al. 2015, \apjs, 218, 6
\bibitem[Brown, Jarrett, \& Cluver(2014)]{bro14} Brown, M.J.I.; Jarrett, T.H.; \& Cluver, M.E. 2014, PASA, 31, 49
\bibitem[Bruzual \& Charlot(2003)]{bru03} Bruzual, G. \& Charlot, S. 2003, \mnras, 344, 1000
\bibitem[Burgarella, Buat, \& Iglesias-Paramo(2005)]{bur05} Burgarella, D.; Buat, V.; \& Iglesias-Paramo, J. 2005, \mnras, 360, 1413
\bibitem[Calzetti et al.(2000)]{cal00} Calzetti, D.; Armus, L.; Bohlin, R.C.; et al. 2000, \apj, 533, 682
\bibitem[Ciesla et al.(2015)]{cie15} Ciesla, L.; Charmandaris, V.; Georgakakis, A.; et al. 2015, \aap, 576, 10
\bibitem[Ciesla et al.(2016)]{cie16} Ciesla, L.; Boselli, A.; Elbaz, D.; et al. 2016, \aap, 585, 43
\bibitem[Corwin, Buta, \& de Vaucouleurs(1994)]{cor94} Corwin, H.G, Buta, R.J. \& de Vaucouleurs, G.  1994, \aj, 108, 212
\bibitem[Cowley et al.(2017)]{cow17} Cowley, W.I.; B\'ethermin, M.; Lagos, C.d.P; et al. 2017, \mnras, 467, 1231
\bibitem[Cox et al.(2008)]{cox08} Cox, T.J.; Jonsson, P.; Somerville, R.S.; et al. 2008, \mnras, 384, 386
\bibitem[Dale et al.(2009)]{dal09} Dale, D.A.; Smith, J.D.T.; Schlawin, E.A.; et al. 2009, \apj, 693, 1821
\bibitem[Dale et al.(2014)]{dal14} Dale, D.A.; Helou, G.; Magdis, G.E.; et al. 2014, \apj, 784, 83
\bibitem[Dale et al.(2017)]{dal17} Dale, D.A.; Cook, D.O.; Roussel, H.; et al. 2017, \apj, 837, 90
\bibitem[Donley et al.(2008)]{don08} Donley, J.L.; Rieke, G.H.; P\'erez-Gonz\'alez, P.G.; et al.  2008, \apj, 687, 111
\bibitem[Drouart et al.(2016)]{dro16} Drouart, G.; Rocca-Volmerange, B.; De Brueck, C.; et al. 2016, \aap, 593, 109
\bibitem[Efstathiou et al.(2014)]{efs14} Efstathiou, A.; Pearson, C.; Farrah, D.; et al. 2014, \mnras, 437, 16
\bibitem[Fernandez-Ontiveros et al.(2016)]{fer16} Fernandez-Ontiveros, J.A.; Spinoglio, L.; Pereira-Santaella, M.; et al. 2016, \apjs, 226, 19
\bibitem[Franceschini et al.(2003)]{fra03} Franceschini, A.; Braito, V.; Persic, M.; et al. 2003, \mnras, 343, 1181
\bibitem[Fritz, Franceschini, \& Hatziminaoglou(2006)]{fri06} Fritz, J.; Franceschini, A.; \& Hatziminaoglou, E. 2006, \mnras, 366, 767
\bibitem[Garc\'ia-Lorenzo et al.(2015)]{gar15} Garc\'ia-Lorenzo, B.; M\'arquez, I.; Buerra-Ballesteros, J.K.; et al. 2015, \aap, 573, 59
\bibitem[Gehrels et al.(2004)]{geh04} Gehrels, N.; Chincarini, G.; Giommi, P.; et al. 2004, \apj, 611, 1005
\bibitem[Genzel et al.(1998)]{gen98} Genzel, R.; Lutz, D.; Sturm, E.; et al. 1998, \apj, 498, 579
\bibitem[Gingold \& Monaghan(1977)]{gin77} Gingold, R.A. \& Monaghan, J.J. 1977, \mnras, 181, 375
\bibitem[Gonz\'alez-Mart\'in et al.(2015)]{gon15} Gonz\'alez-Mart\'in, O.; Masegosa, J.; M\'arquez, I.; et al. 2015, \aap, 578, 74
\bibitem[Granato \& Danese(1994)]{gra94} Granato, G.L. \& Danese, L.  1994, \mnras, 268, 253
\bibitem[Groves et al.(2008)]{gro08} Groves, B.; Dopita, M.A.; Sutherland, R.S.; et al. 2008, \apjs, 176, 438
\bibitem[Gruppioni et al.(2008)]{gru08} Gruppioni, C.; Pozzi, F.; Polleta, M.; et al. 2008, \apj, 701, 850
\bibitem[Hayward et al.(2011)]{hay11} Hayward, C.C.; Du\^san, K.; Jonnson, P.; et al. 2011, \apj, 743, 159
\bibitem[Hayward et al.(2013)]{hay13} Hayward, C.C.; Narayanan, D.; Du\^san, K.; et al. 2013, \mnras, 428, 2529
\bibitem[Hayward et al.(2014a)]{hay14a} Hayward, C.C.; Lanz, L.; Ashby, M.L.N.; et al. 2014, \mnras, 445, 1598
\bibitem[Hayward et al.(2014b)]{hay14b} Hayward, C.C.; Torrey, P.; Springel, V.; et al. 2014, \mnras, 442, 1992
\bibitem[Hayward \& Smith(2015)]{hay15} Hayward, C.C.; \& Smith, D.J.B. 2015, \mnras, 446, 1512
\bibitem[Hern\'andez-Garc\'ia et al.(2016)]{her16} Hern\'andez-Garc\'ia, L.; Masegosa, J.; Gonz\'alez-Mart\'in, O.; et al. 2016, \apj, 824, 7
\bibitem[Hernquist \& Katz(1989)]{her89} Hernquist, L \& Katz, N. 1989, ApJS, 70, 419
\bibitem[Higdon et al.(2004)]{hig04} Higdon, S.J.U.; Devost, D.; Higdon, J.L.; et al. 2004, \pasp, 116, 975
\bibitem[Hong et al.(2015)]{hon15} Hong, J.; Im, M.; Kim, M.; et al. 2015, \apj, 804, 34
\bibitem[Hopkins, Richards, \& Hernquist(2007)]{hop07} Hopkins, P.F.; Richards, G.T.; \& Hernquist, L. 2007, \apj, 654, 731
\bibitem[Houck et al.(2004)]{hou04} Houck, J.R.; Roellig, T.L.; van Cleve, J.; et al. 2004, \apjs, 154, 18
\bibitem[Ivanov et al.(2000)]{iva00} Ivanov, V.D.; Rieke, G.H.; Groppi, C.E.; et al. 2000, \apj, 545, 190
\bibitem[Jonsson(2006)]{jon06} Jonsson, P. 2006, 372, 2
\bibitem[Jonsson, Groves, \& Cox(2010)]{jon10} Jonsson, P.; Groves, B.; and Cox, T.J. 2010, \mnras, 403, 17
\bibitem[Katz, Weinberg, \& Hernquist(1996)]{kat96} Katz, N.; Weinberg, D.H.; \& Hernquist, L. 1996, \apjs, 105, 19
\bibitem[Keel et al.(1985)]{kee85} Keel, W.C.; Kennicutt, R.C.; Jr.; Hummel, E.; et al. 1985, \aj, 90, 708
\bibitem[Kennicutt(1998)]{ken98} Kennicutt, R.C., Jr. 1998, \apj, 498, 541
\bibitem[Knapen et al.(2014)]{kna14} Knapen, J.H.; Erroz-Ferrer, S.; Roa, J.; et al. 2014, \aap, 569, 91
\bibitem[Lackner et al.(2014)]{lac14} Lackner, C.N.; Silverman, J.D.; Salvato, M.; et al. 2014, \aj, 148, 137
\bibitem[LaMassa et al.(2012)]{lam12} LaMassa, S.M.; Heckman, T.M.; Ptak, A.; et al. 2012, \apj, 758, 1
\bibitem[Lanz et al.(2013)]{lan13} Lanz, L.; Zezas, A.; Brassington, N.; et al. 2013, \apj, 768, 90 (L13)
\bibitem[Lanz et al.(2014)]{lan14} Lanz, L.; Hayward, C.C.; Zezas, A.; et al. 2014, \apj, 785, 39
\bibitem[Lebouteiller et al.(2011)]{leb11} Lebouteiller, V.; Barry, D.J.; Spoon, H.W.W.; et al. 2011, \apjs, 196, 8
\bibitem[Lee et al.(2010)]{lee10} Lee, S.-K.; Ferguson, H.C.; Somerville, R.S.; et al. 2010, \apj, 725, 1644
\bibitem[Leitherer et al.(1999)]{lei99} Leitherer, C.; Schaerer, D.; Goldader, J.D.; et al. 1999, \apjs, 123, 3
\bibitem[Leitherer, Calzetti, \& Martins(2002)]{lei02} Leitherer, C.; Calzetti, D.; \& Martins, L.P. 2002, \apj, 574, 114
\bibitem[Lucy(1977)]{luc77} Lucy, L.B.  1977, \aj, 82, 1013
\bibitem[Martin et al.(2005)]{mar05} Martin, C.D.; Fanson, J.; Schiminovich, D.; et al. 2005, \apj, 619, L1
\bibitem[McQuinn et al.(2017)]{mcq17} McQuinn, K.B.W.; Skillman, E.D.; Dolphin, A.E.; et al. 2017, \aj, 154, 51
\bibitem[Micha{\l}owski et al.(2014)]{mic14} Micha{\l}owski, M.J.; Hayward, C.C.; Dunlop, J.S.; et al. 2014, \aap, 571, 75
\bibitem[Moshir, Kopman, \& Conrow(1992)]{mos92} Moshir, M.; Kopman, G.; \& Conrow, T.A.O.  1992, Pasadena: IPAC, Caltech
\bibitem[Nenkova, Ivezi\'c, \& Elitzur(2002)]{nen02} Nenkova, M.; Ivezi\'c, \v Z.; \& Elitzur, M.  2002, \apj, 570, 9
\bibitem[Neugebauer et al.(1984)]{neu84} Neugebauer, G.; Habing, H.J.; van Duinen, R.; et al. 1984, \apj, 278, 1
\bibitem[Ott(2010)]{ott10} Ott, S.  2010, \aspc, 434, 139
\bibitem[Paladini et al.(2012)]{pal12} Paladini, R.; Linz, H.; Altieri, B.; \& Ali, B. 2012, Assessment Analysis of the Extended Emission Calibration for the PACS Red Channel, Document: PICC-NHSC-TR-034 (Pasadena, CA: NHSC)
\bibitem[Pilbratt et al.(2010)]{pil10} Pilbratt, G.L.; Riedinger, J.R.; Passvogel, T.; et al. 2010, \aap, 518, 1
\bibitem[Ramos Padilla et al.(2018, in preparation)]{ram18} Ramos Padilla, A.F.,; et al. 2018
\bibitem[Rosario et al.(2012)]{ros12} Rosario, D.J.; Santini, P.; Lutz, D.; et al. 2012, \aap, 545, 45
\bibitem[Rosenberg et al.(2014)]{ros14} Rosenberg, M.J.F.; Meijerink, R.; Israel, F.P.; et al. 2014, \aap, 568, 90
\bibitem[Rosenberg et al.(2015)]{ros15} Rosenberg, M.J.F.; van der Werf, P.P.; Aalto, S.; et al. 2015, \apj, 801, 72
\bibitem[Rosenthal et al.(2015)]{rol15} Rosenthal, L.; Hayward, C.C.; Smith, H.; et al. 2015, AAS, 225, 143.47
\bibitem[Safarzadeh et al.(2016)]{saf16} Safarzadeh, M.; Hayward, C.C.; Ferguson, H.; \& Somerville, R.S. 2016, \apj, 818, 62
\bibitem[Sanders \& Mirabel(1996)]{san96} Sanders, D.B.  and Mirabel, I.F.; ARA\&A, 34, 749
\bibitem[Sanders et al.(1988)]{san88} Sanders, D.B.; Soifer, B.T.; Elias, J.H.; et al. 1988, \apj, 325, 74
\bibitem[Sanders et al.(2003)]{san03} Sanders, D.B.; Mazzarella, J.M.; Kim, D.-C.; et al. 2003, \aj, 126, 1607
\bibitem[Satyapal et al.(2009)]{sat09} Satyapal, S.; B\"{o}ker, T.; McAlpine, W.; et al. 2009, \apj, 704, 439
\bibitem[Skrutskie et al.(2006)]{skr06} Skrutskie, M.F.; Cutri, R.M.; Steining, R.; et al. 2006, \aj, 131, 1163
\bibitem[Smith et al.(2007)]{smi07} Smith, J.D.T.; Armus, L.; Dale, D.A.; et al. 2007, \pasp, 119, 1133
\bibitem[Smith et al.(2010)]{smi10} Smith, H.A.; Li, A.; Li, M.P.; et al. 2010, \apj, 716, 490
\bibitem[Smith \& Hayward(2015)]{smi15} Smith, D.J.B.; \& Hayward, C.C. 2015, \mnras, 453, 1497
\bibitem[Snyder et al.(2013)]{sny13} Snyder, G.F.; Hayward, C.C.; Sajina, A.; et al. 2013, \apj, 768, 168
\bibitem[Springel \& Hernquist(2003)]{spr03} Springel, V.; \& Hernquist, L.  2003, IAUS, 208, 273
\bibitem[Springel(2005)]{spr05} Springel, V.  2005, \mnras, 364, 1105
\bibitem[Springel et al.(2005)]{spl05} Springel, V.; White, S.D.M.; Jenkins, A.; et al. 2005, \nat, 435, 629
\bibitem[Springel(2010)]{spr10} Springel, V.  2010, ARA\&A, 48, 391
\bibitem[Stern et al.(2005)]{ste05} Stern, D.; Eisenhardt, P.; Gorjian, V.; et al. 2005, \apj, 631, 163
\bibitem[Stern et al.(2012)]{ste12} Stern, D.; Assef, R.J.; Benford, D.J.; et al. 2012, \apj, 753, 30
\bibitem[Toba et al.(2013)]{tob13} Toba, Y.; Oyabu, S.; Matsuhara, H.; et al. 2013, PASJ, 65, 113
\bibitem[Toomre \& Toomre(1972)]{too72} Toomre, A.; and Toomre, J.  1972, \apj, 178, 623
\bibitem[de Vaucouleurs et al.(1991)]{vau91} de Vaucouleurs, G.; de Vaucouleurs, A.; Corwin, H.G.; et al. 1991, New York: Springer-Verlag
\bibitem[Vaddi et al.(2016)]{vad16} Vaddi, S.; O'Dea, C.P.; Baum, S.A.; et al. 2016, \apj, 818, 182
\bibitem[Vardoulaki et al.(2015)]{var15} Vardoulaki, E.; Charmandaris, V.; Murphy, E.J.; et al. 2015, \aap, 574, 4
\bibitem[V\'eron-Cetty \& V\'eron(2010)]{ver10} V\'eron-Cetty, M.-P.  and V\'eron, P.  2010, \aap, 518, 10
\bibitem[Veilleux, Kim, \& Sanders(2002)]{vei02} Veilleux, S.; Kim, D.C.; \& Sanders, D.B.  2002, \apj, 143, 315
\bibitem[Villforth et al.(2017)]{vil17} Villforth, C.; Hamilton, T.; Pawlik, M.M.; et al. 2017, \mnras, 466, 812
\bibitem[Wang et al.(2014)]{wan14} Wang, L.; Rowan-Robinson, M.; Norberg, P.; et al. 2014, \mnras, 442, 2739
\bibitem[Weiner et al.(2018, in preparation)]{wei18} Weiner, A.S.; Smith, H.A.; Ashby, M.L.N.; et al. 2018
\bibitem[Wenger et al.(2002)]{wen00} Wenger, M.; Ochsenbein, F.; Egret, B.; et al. 2000, \aaps, 143, 9 
\bibitem[Werner et al.(2004)]{wer04} Werner, M.W.; Roellig, T.L.; Low, F.J.; et al. 2004, \apj, 154, 1
\bibitem[Willett et al.(2013)]{wil13} Willett, K.W.; Lintott, C.J.; Bamford, S.P.; et al. 2013, \mnras, 435, 2835
\bibitem[Williams et al.(2017)]{wil17} Williams, W.L.; Calistro Rivera, G.; Best, P.N.; et al. 2017, arXiv, 1711.10504
\bibitem[Wright et al.(2010)]{wri10} Wright, E.L.; Eisenhardt, P. R. M.; Mainzer, A. K.; et al. 2010, \aj, 140, 1868
\nonfrenchspacing
\end{thebibliography}




\appendix

\section{Notes on individual galaxies}

Group I: The Late-Stage Merger Sample

\textbf{IRAS~08572+3915}: IRAS~08572+3915 is a ULIRG. Its very steep spectrum from 2 to 20~\micron\ implies an extremely powerful AGN, which approaches 91\% of the total IR luminosity coming from the galaxy.  This is consistent with \citet{efs14}, who also found an AGN luminosity fraction around 0.9, and \citet{dal14}, who estimated an AGN contribution of 85\%.

\textbf{IRAS~15250+3609}: IRAS~15250+3609 is a ULIRG. The 2--10~\micron\ slope is steep enough to imply an AGN contribution to the SED, which CIGALE estimates at $\sim$47\% of the IR luminosity. \cite{fra03} defines this galaxy as not AGN-dominated but still containing a LINER-type nucleus.

\textbf{Mrk~231}: Mrk~231 is a ULIRG that is the most IR-luminous system in the sample.  We measure the AGN contribution at $\sim$17\% of the IR luminosity, the lowest value in the late-stage sample, lower than the value found by \citet{ros15} by almost a factor of 5, and lower than that from \citet{fri06} by almost a factor of 2. This low fraction is because the data at wavelengths greater than 60~\micron\ are well fit by the dust model with no AGN contribution needed.

\textbf{Mrk~273}: Mrk~273 is a ULIRG.  The steep 3--24~\micron\ spectrum implies a large AGN contribution that CIGALE estimates at $\sim$66\% of the IR luminosity. This is higher than the value given by \citet{ros15} by around a factor of 2.

\textbf{Mrk~463}: Mrk~463 is a LIRG.  The 3--24~\micron\ SED is well fit by an AGN model, and the derived AGN contribution is $\sim$68\% of the IR luminosity.  The spectral line ratios for [Ne \textsc{v}]/[Ne \textsc{ii}] and [O \textsc{iv}]/[Ne \textsc{ii}] are a factor of 5 larger than others in this sample due to relatively low [Ne \textsc{ii}] flux, indicating very little star formation is occurring.  That the line ratio is so particularly strong is surprising, because an AGN strong enough to ionize neon and oxygen that heavily would also be expected to have powerful UV emission, which is not seen in Mrk~463.  That no enhanced UV emission is seen is presumably due to a high internal extinction.  Mrk~463 is also a luminous X-ray source, another indication of strong AGN activity.

\textbf{NGC~2623}: NGC~2623 is a LIRG with the 3.6--24~\micron\ data well fit by an AGN model.  We measure the AGN contribution to be $\sim$39\%, but the $\chi^2$ value is relatively large.

\textbf{NGC~3758}: NGC~3758 is a LIRG and the least luminous galaxy in the late-stage merger sample. The 3.6--24~\micron\ SED is well fit by an AGN model, but the estimated AGN contribution is $\sim$30\%.

\textbf{NGC~6090}: NGC~6090 is a LIRG.  The steep 4.5--8.0~\micron\ SED implies an AGN is present, but the estimated AGN luminosity fraction is $\sim$26\%, one of the lowest in the late-stage merger sample. This is still higher than the 10\% value measured by \citet{dal14}.

\textbf{UGC 4881}: UGC 4881 is a LIRG. We measure the AGN contribution to be $\sim$51\%, although \citet{dal14} modeled it without an AGN. The red 5.8--100~\micron\ colours suggest an AGN is present

\textbf{UGC 5101}: UGC 5101 is a ULIRG. We measure the AGN contribution to be $\sim$76\%, although the best-fit model is markedly worse than most of the late-stage mergers. \citet{dal14} estimated a value nearly a factor of 5 smaller. The red 4.5--24~\micron\ SED implies an AGN is present.

\textbf{VV 283}: VV 283 is a LIRG.  We measure an AGN contribution of $\sim$47\%, and the red 5.8--24~\micron\ SED implies an AGN is present.

\textbf{VV 705}: VV 705 is a LIRG. We measure an AGN contribution of $\sim$25\%.  This is corroborated by the results of \citet{dal14}, who measured an AGN luminosity fraction of 25\%. The red 5.8--24~\micron\ SED implies an AGN is present.\\
Group II: The Reference Sample

\textbf{M51A}: M51A, also known as the Whirlpool Galaxy, is a well-known spiral galaxy with an elliptical companion.  CIGALE modeled M51A with an AGN at 9\% of the IR luminosity.  \citet{her16} modeled it as an obscured AGN, while L13 calculated a best-fit model that did not have an AGN.  Nothing in the SED demands the presence of an AGN.

\textbf{M51B}: M51B is the companion to M51A, and it was modeled with an AGN at $<3$\% of the IR luminosity.  \citet{her16} classified it as a LINER galaxy, but L13 did not calculate an AGN luminosity fraction for it, and nothing in the SED requires an AGN to be present.

\textbf{NGC~2976}: NGC~2976 is a spiral galaxy in the M81 group. CIGALE modeled NGC~2976 with $f_{AGN}$ = 28\%, but previous results from \citet{gon15} and L13 produced models without AGN.  Nothing in the SED requires an AGN.

\textbf{NGC~3031}: NGC~3031, also known as M81, is a nearby spiral galaxy.  CIGALE modeled the galaxy with $f_{AGN} \leq 0.01$, although L13 reported a total IR AGN luminosity fraction of 4\% and a maximum of 16\% in the 8--35~\micron\ range. The galaxy nucleus has a unique dust spectrum \citep{smi10}, and modeling based on standard templates is unreliable. However, nothing in the total-galaxy SED requires an AGN.

\textbf{NGC~3077}: NGC~3077 is a low-luminosity irregular galaxy.  We measure the AGN contribution at $\sim$34\%, which is not consistent with results from \citet{her16} and L13.  NGC~3077 is a fairly isolated companion of NGC~3031 = M81, 46\arcmin~ away \citep{kna14}, indicating it is in the earliest stages of merging.  It does show signs of previous galaxy interaction, but nothing in the SED requires an AGN.

\textbf{NGC~3190}: NGC~3190 is an edge-on spiral galaxy with prominent dust lanes.  CIGALE places a 3$\sigma$ upper limit for the AGN contribution at 1\%, although it has been shown to have a LINER-type nucleus \citep{gon15}.  L13 also described a best-fit model with no AGN contribution, and nothing in the SED suggests an AGN.

\textbf{NGC~3690}: NGC~3690 is the most IR-luminous galaxy in the Reference Sample and its only LIRG \citep{ros14}.  This galaxy is nearing final pass; although not at coalescence, it is still classified as a late-stage merger.  We measure the AGN contribution at $\sim$30\%.  \citet{var15} have shown it to be a composite galaxy, containing a LINER-type nucleus while also undergoing SF.  L13 described the SF but did not mention an AGN.  \citet{dal14} characterized the AGN contribution to the SED at 50\%. The red 3.6--24~\micron\ SED suggests an AGN is present.

\textbf{NGC~4625}: NGC~4625 is a peculiar spiral with a blue SED.  CIGALE fits $f_{AGN} \leq 3\%$.  \citet{ver10} classified it as a Seyfert galaxy (type unknown), but L13 did not define an AGN contribution for NGC~4625. Nothing in the SED requires an AGN.

\textbf{NGC~5394}: NGC~5394 is a companion of NGC~5395 in the middle stages of merging at a projected separation of 28 kpc.  We measure the AGN contribution at $\sim$58\%.  \citet{tob13} stated NGC~5394 as a composite, although L13 did not place an AGN in NGC~5394. The red 4.5--24~\micron\ SED suggests an AGN is present.

\textbf{NGC~5395}: NGC~5395 is the larger spiral companion of NGC~5394.  CIGALE fits $f_{AGN}$ < 10\% to NGC~5395, consistent with both \citet{ver10} and L13 calling it a LINER and attributing to it 3--12\% of the bolometric and mid-IR luminosities. Nothing in the SED requires an AGN.

\textbf{M101}: M101, also known as the Pinwheel Galaxy, is a nearby spiral galaxy showing some tidal disruptions in its outer arms with multiple small companions, including NGC~5474.  CIGALE posits $f_{AGN}$ < 14\%, although \citet{bra15} defined it as not having an AGN, and L13 also gave a best-fit model with no AGN. The red 5.8--8.0 colour is consistent with either an AGN or high SFR.

\textbf{NGC~5474}: NGC~5474 is a smaller companion to M101 at a projected separation of 87 kpc.  CIGALE models NGC~5474 with $f_{AGN} \leq 5\%$.  \citet{bra15} and L13 did not fit a model containing an AGN to the data, and nothing in the SED requires an AGN.

\setcounter{figure}{2}
\begin{figure*}
\centering
\subfloat{\includegraphics[width=0.5\linewidth]{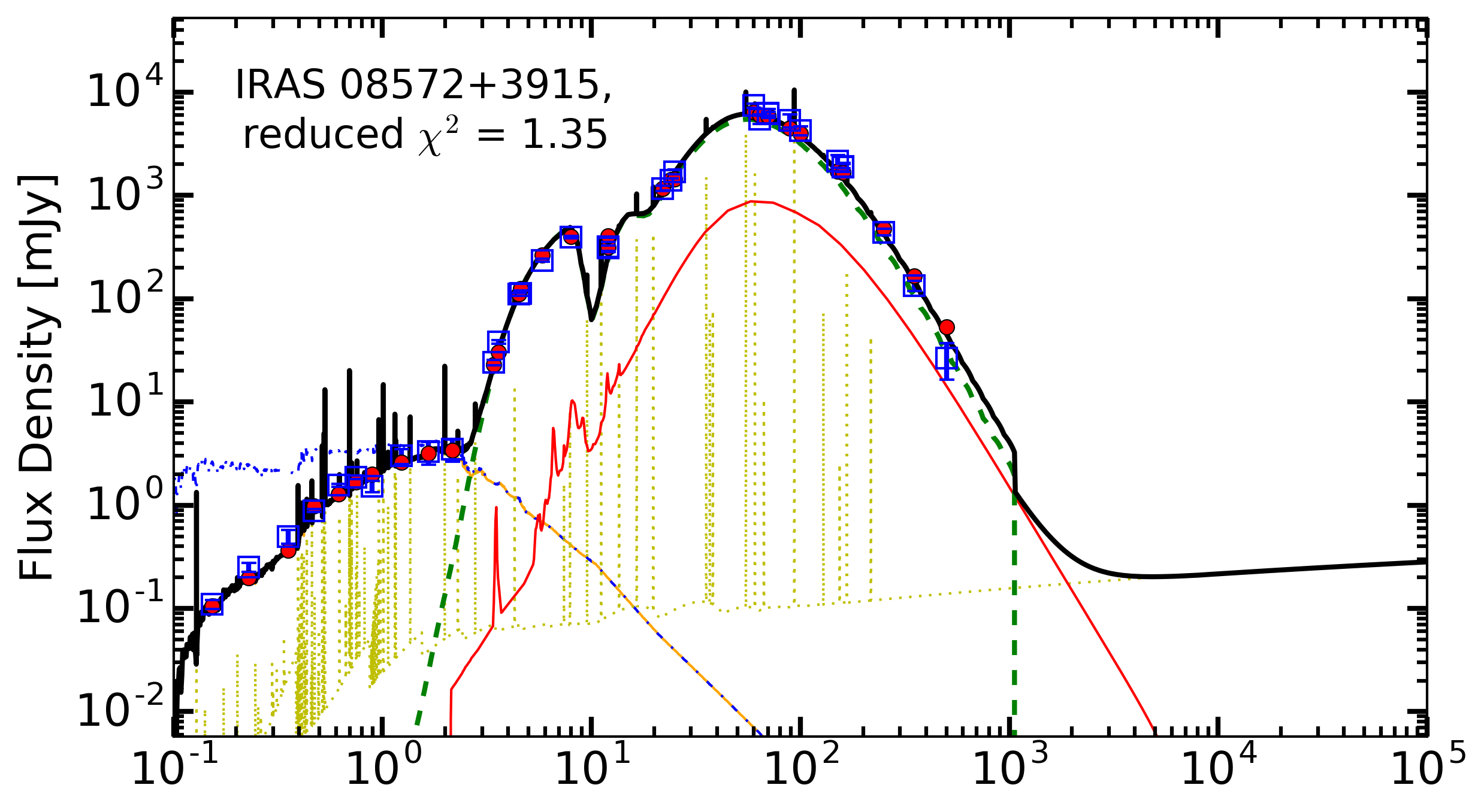}}
\subfloat{\includegraphics[width=0.5\linewidth]{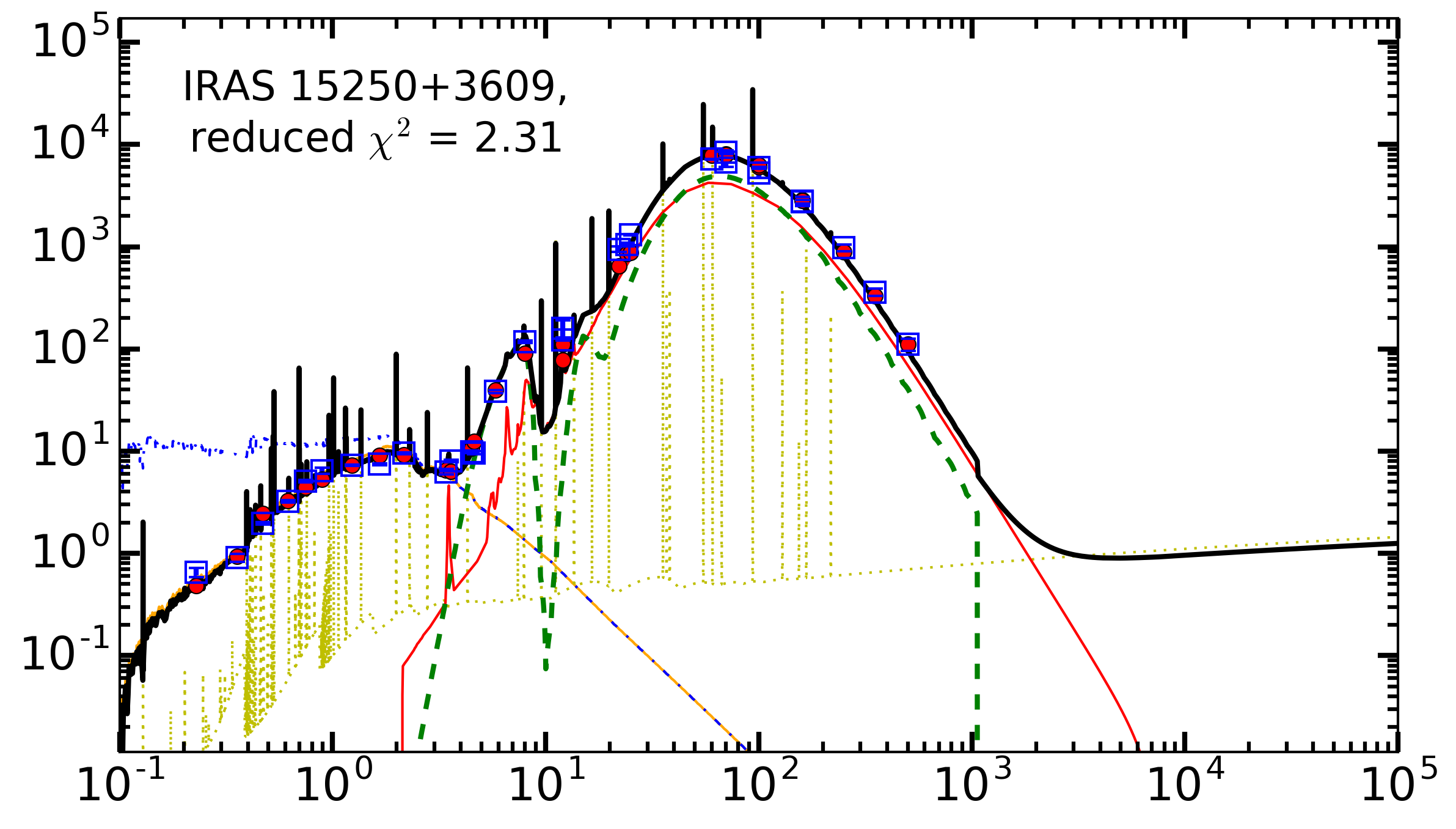}}\\
\subfloat{\includegraphics[width = 0.5\linewidth]{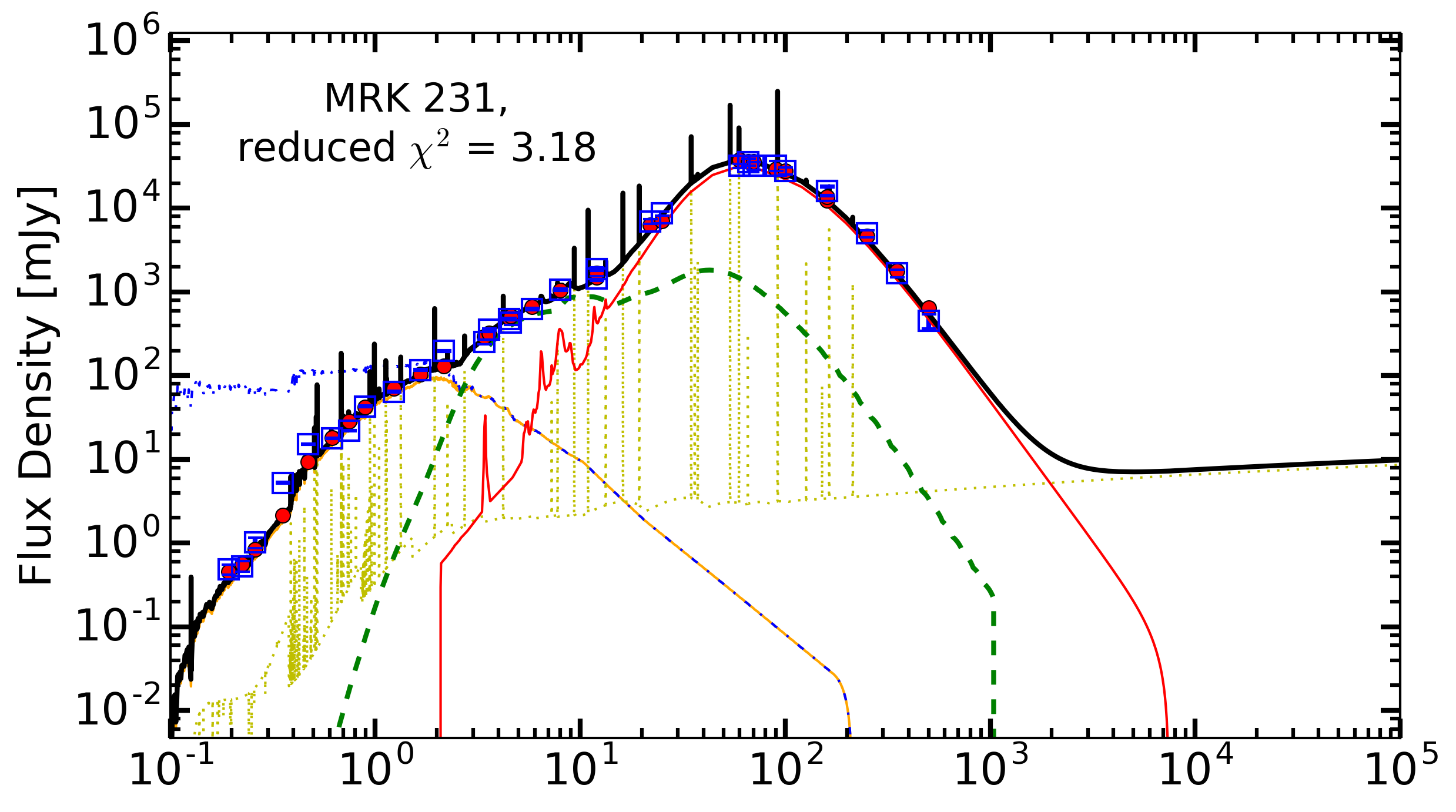}}
\subfloat{\includegraphics[width = 0.5\linewidth]{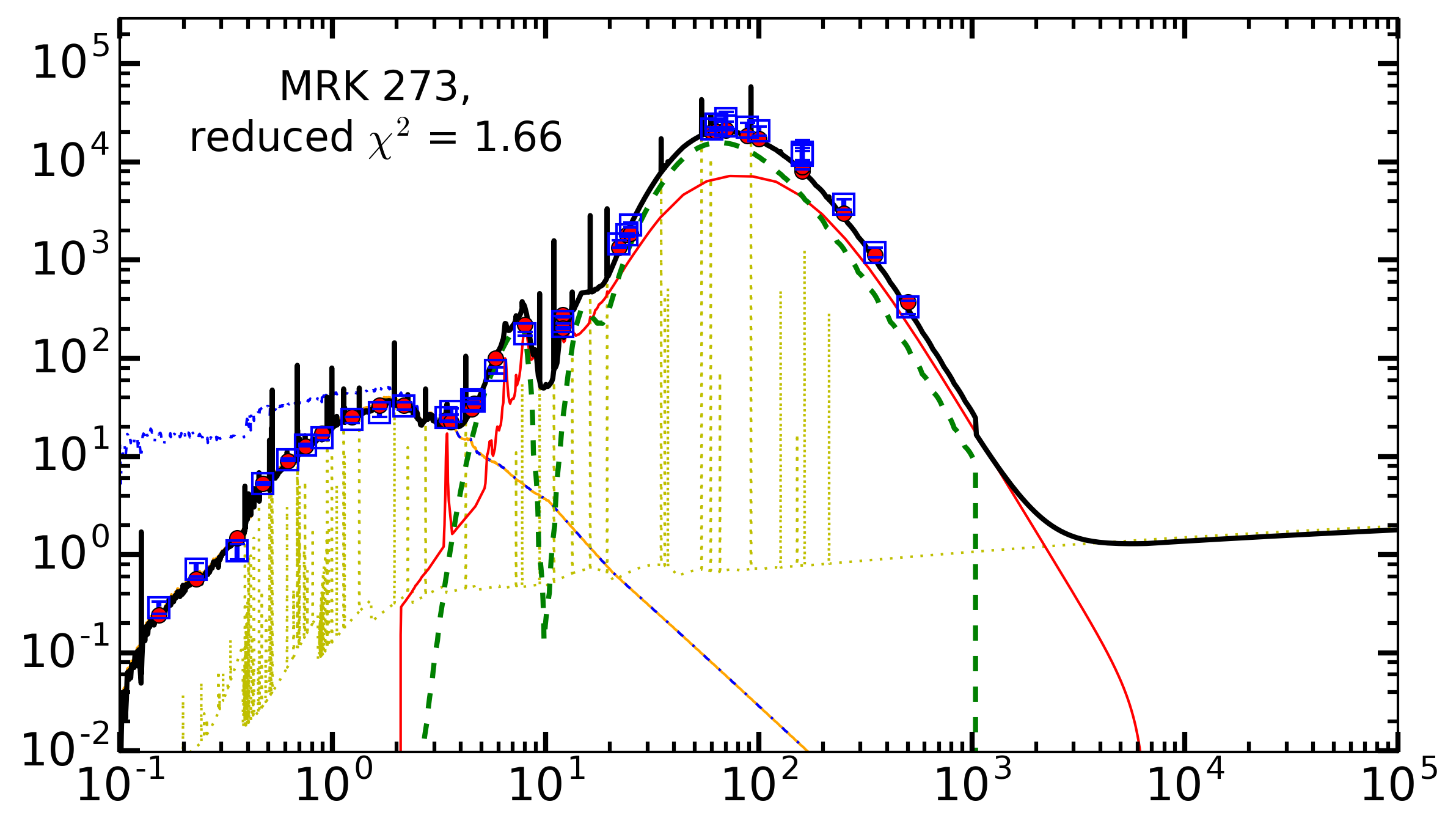}}\\
\subfloat{\includegraphics[width = 0.5\linewidth]{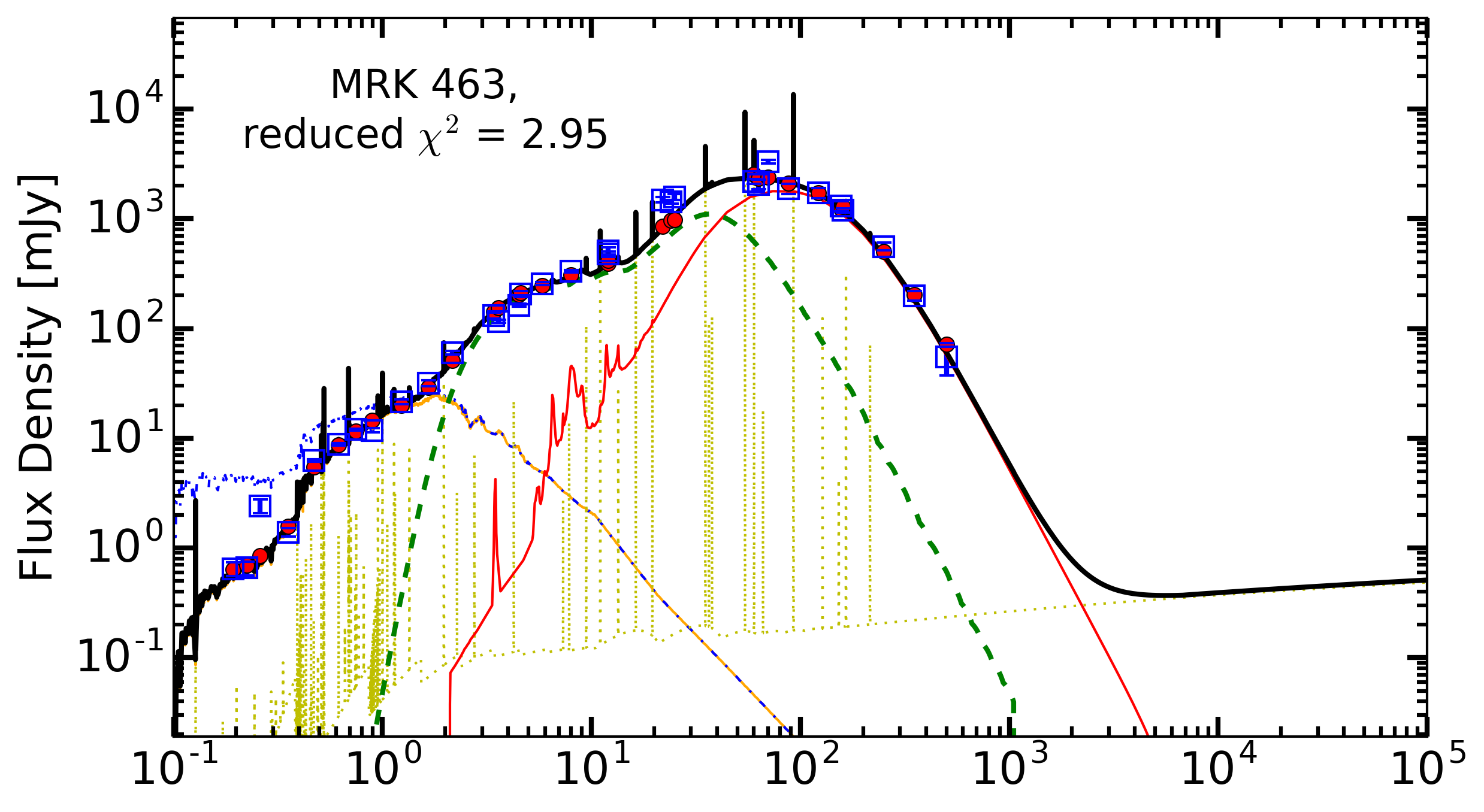}}
\subfloat{\includegraphics[width = 0.5\linewidth]{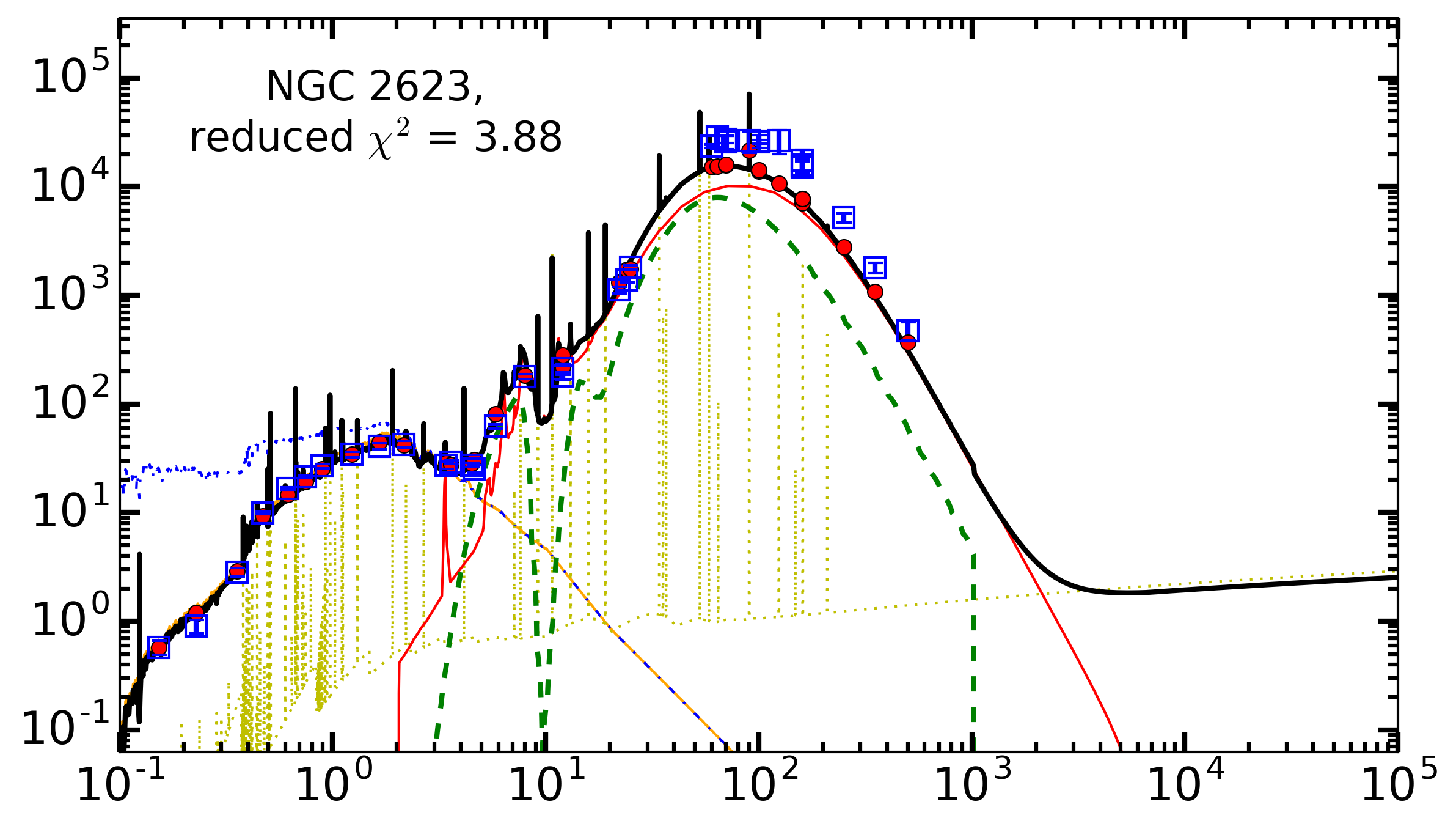}}\\
\subfloat{\includegraphics[width = 0.5\linewidth]{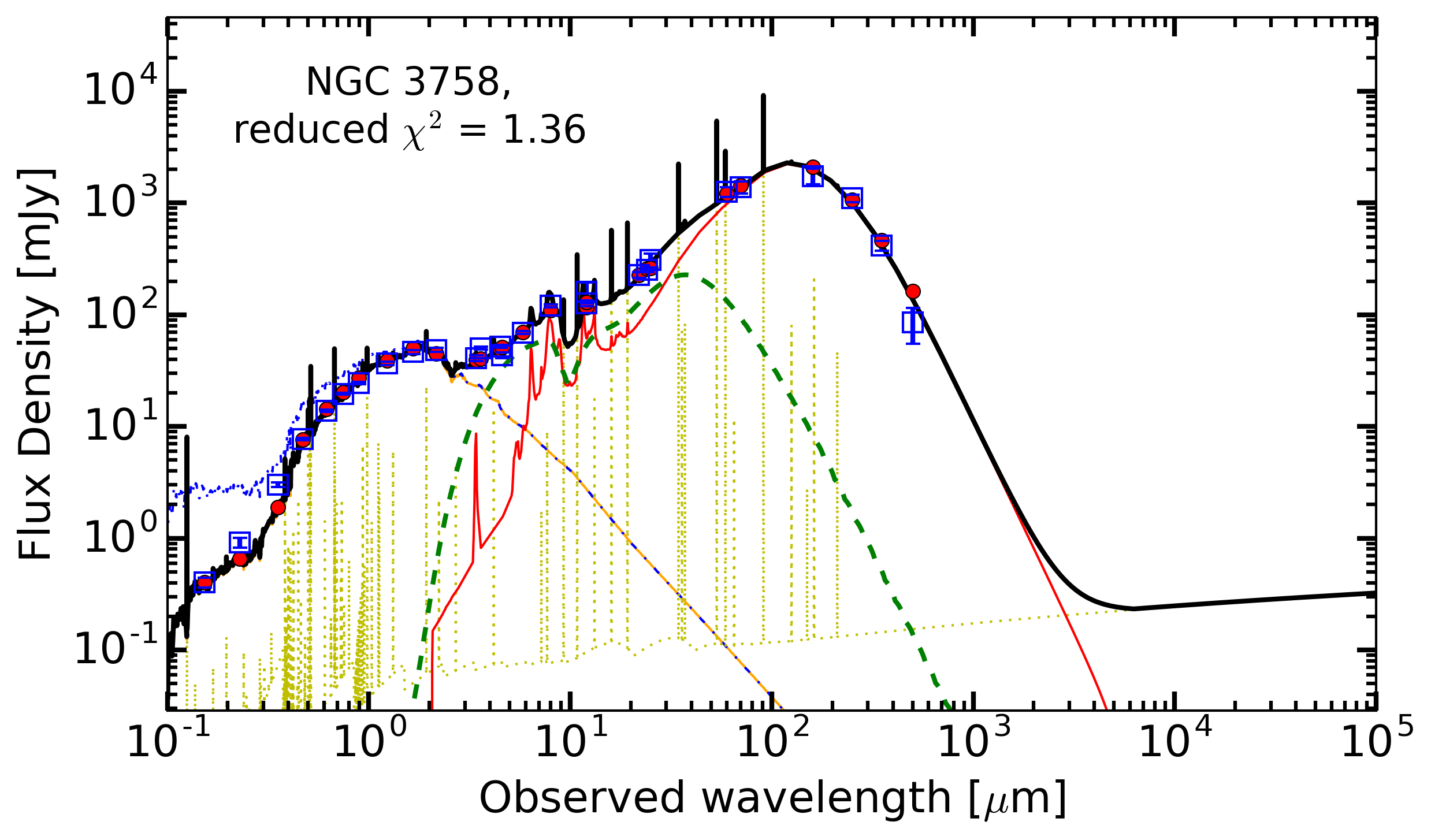}}
\subfloat{\includegraphics[width = 0.5\linewidth]{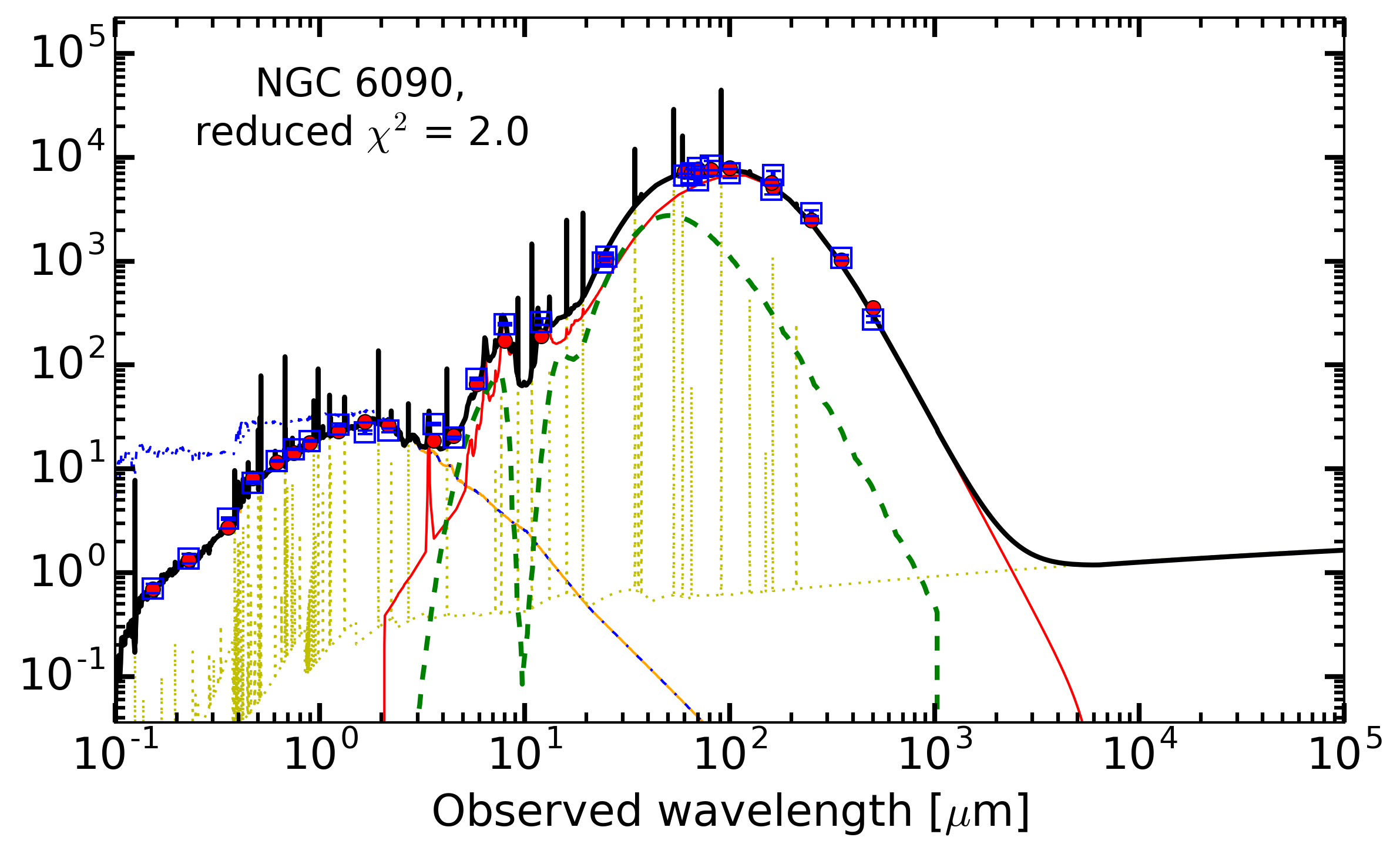}}\\
\end{figure*}
\begin{figure*}
\ContinuedFloat
\centering
\subfloat{\includegraphics[width = 0.5\linewidth]{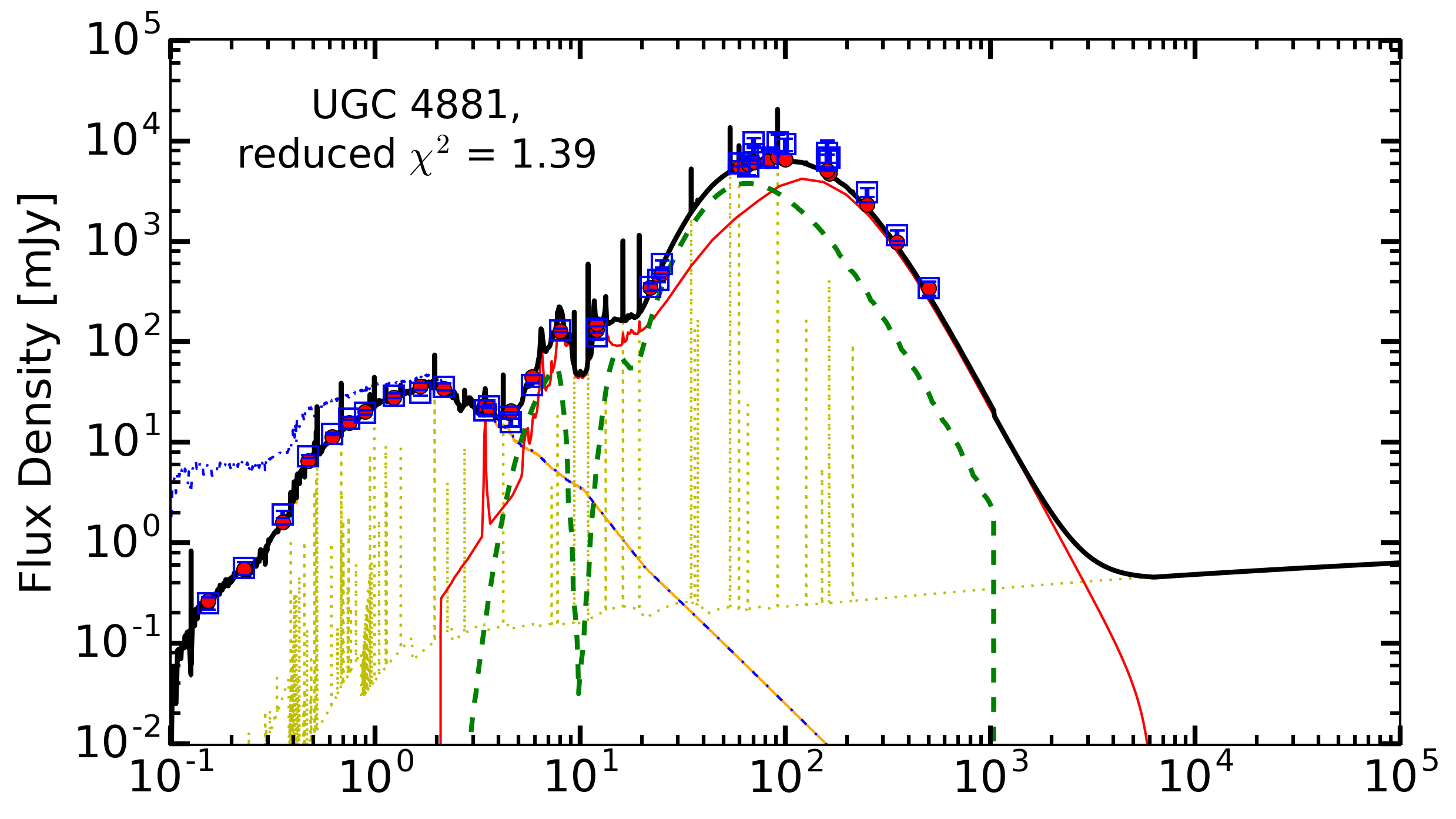}}
\subfloat{\includegraphics[width = 0.5\linewidth]{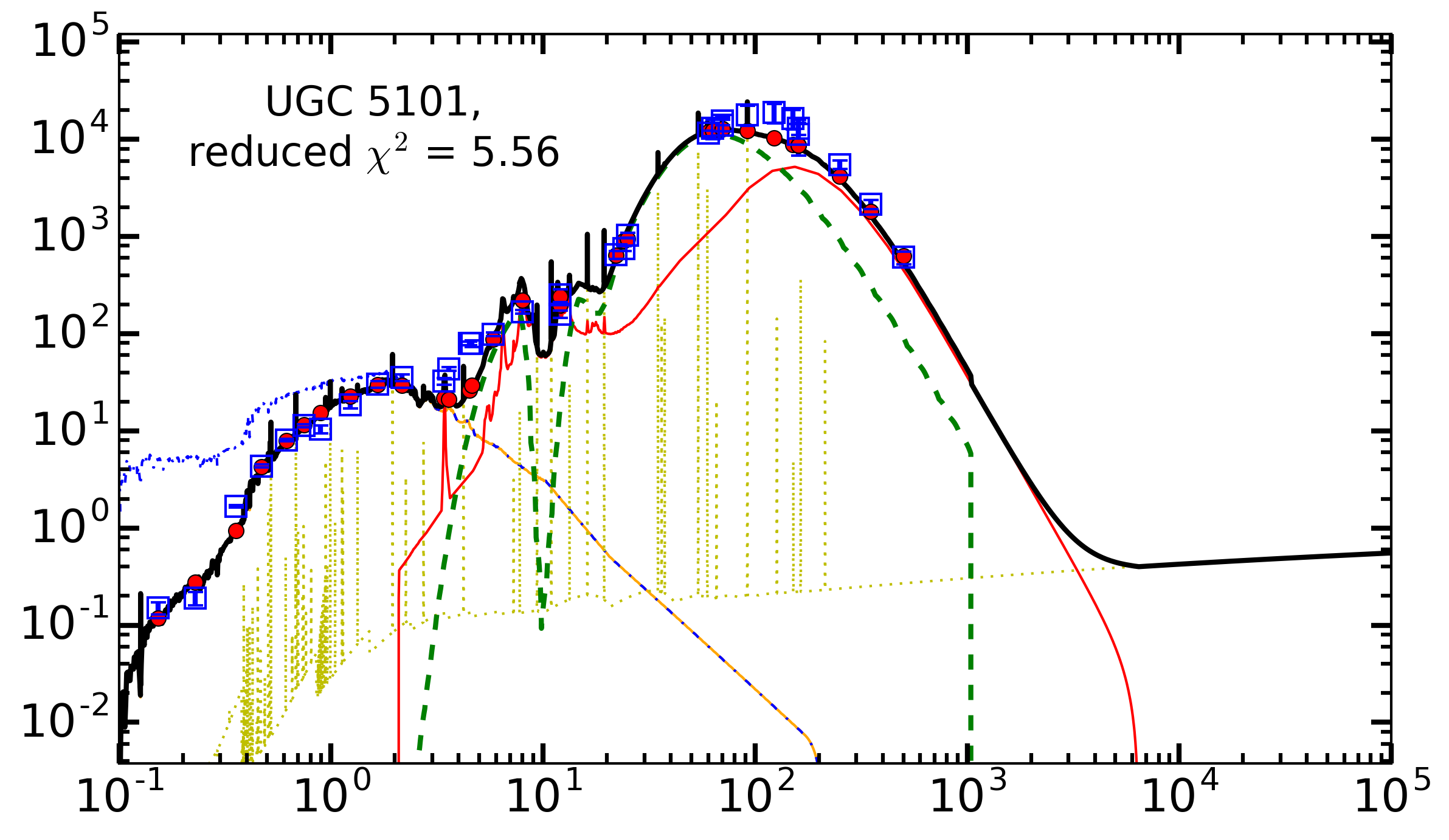}}\\
\subfloat{\includegraphics[width = 0.5\linewidth]{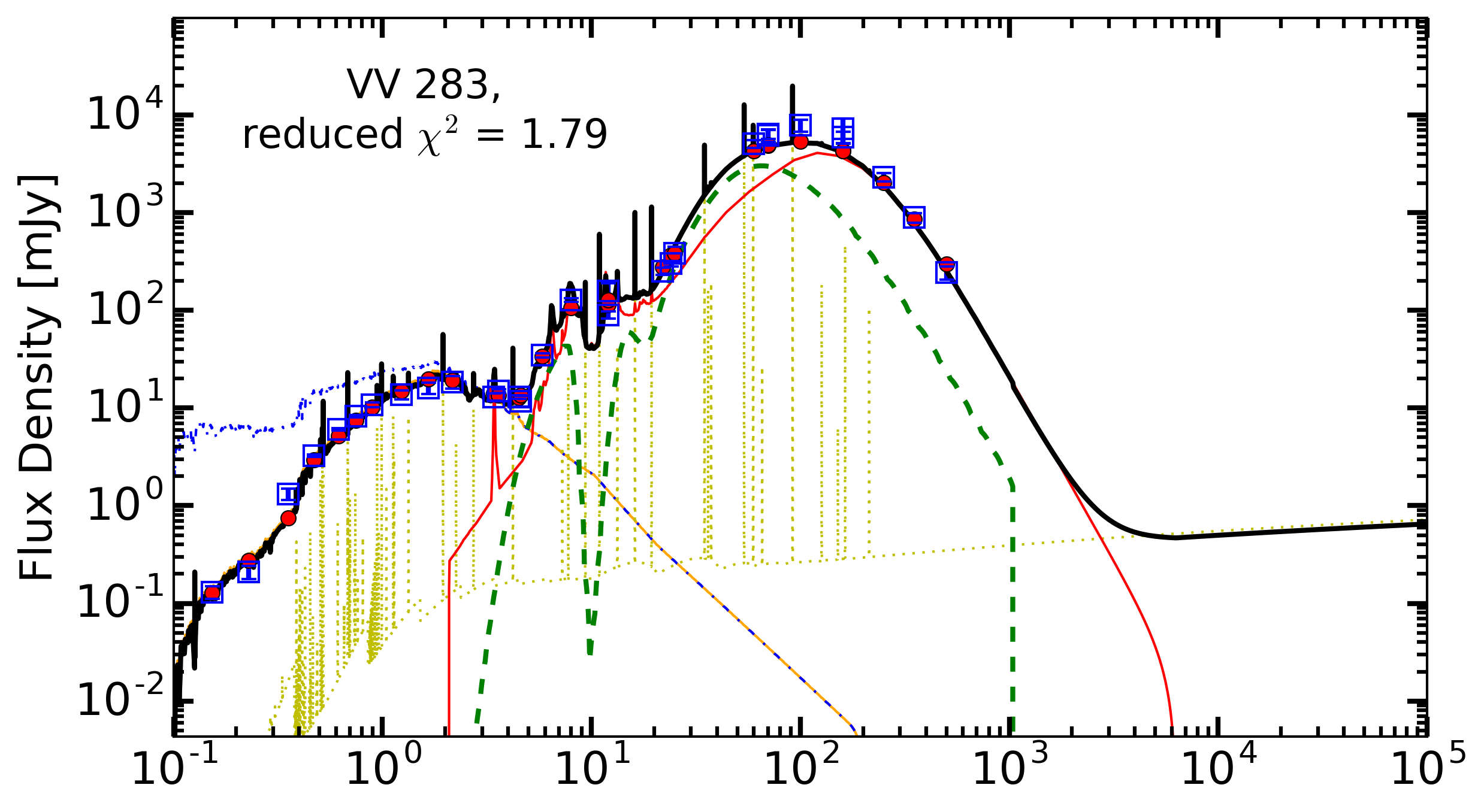}}
\subfloat{\includegraphics[width = 0.5\linewidth]{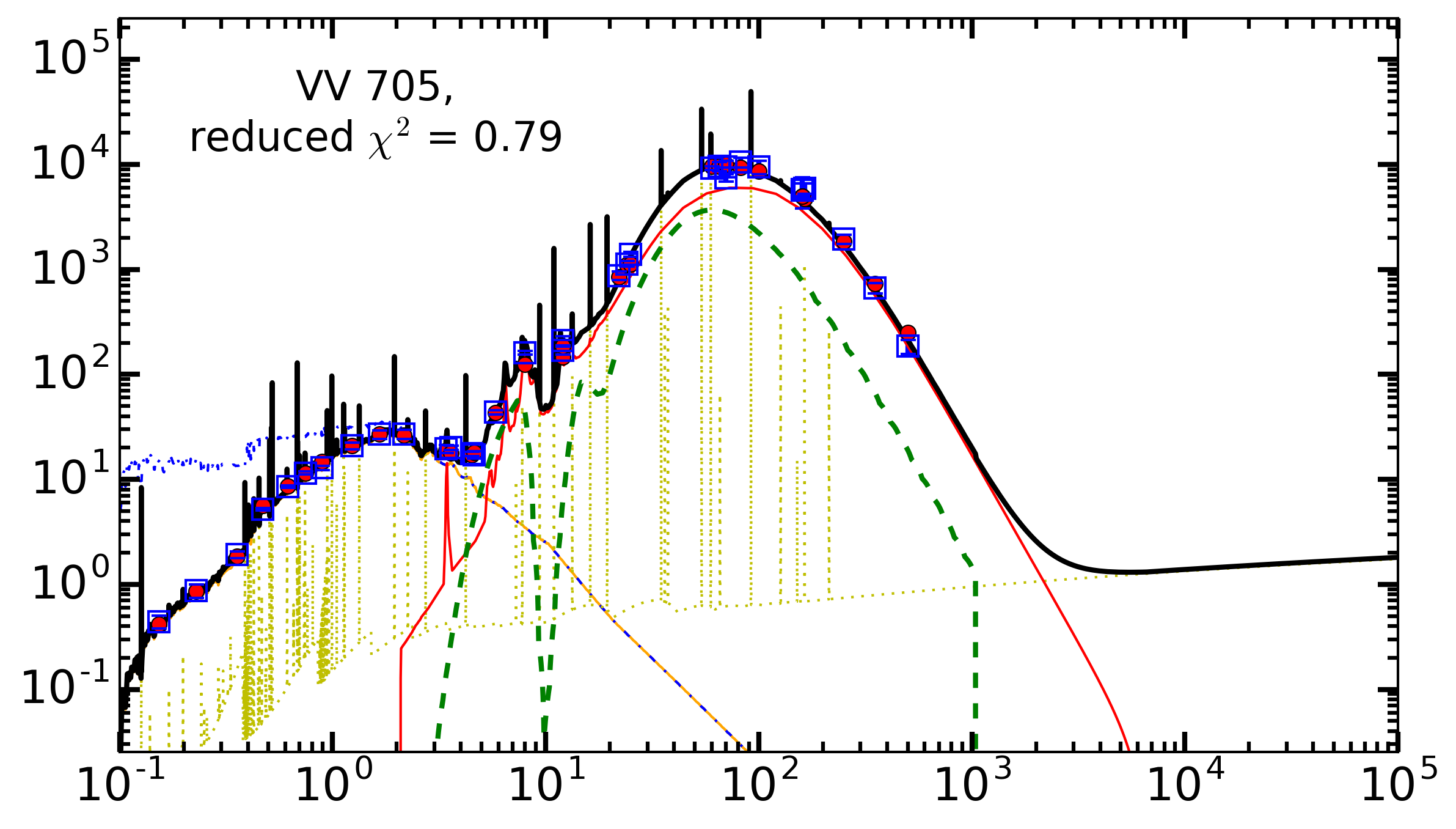}}\\
\subfloat{\includegraphics[width = 0.5\linewidth]{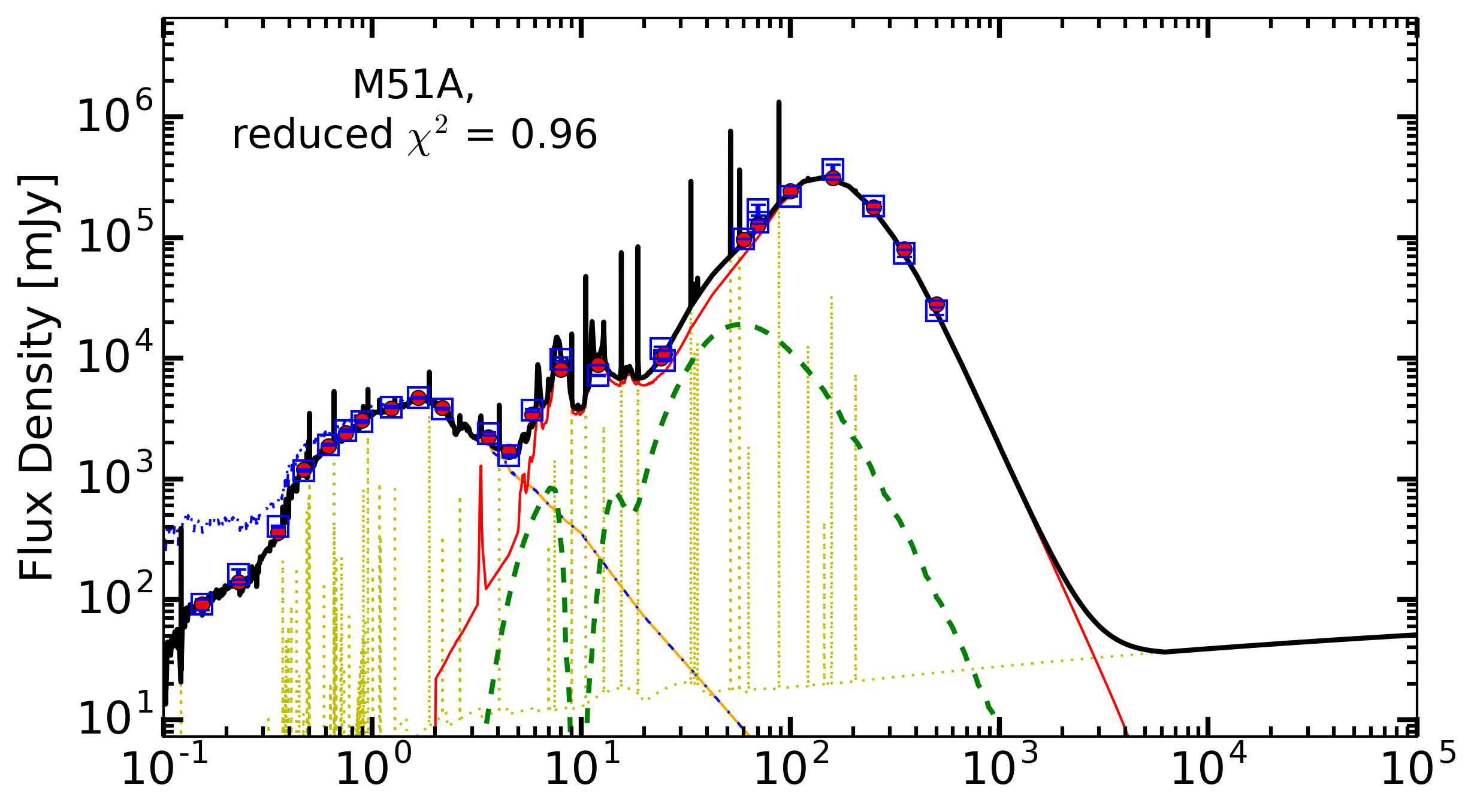}}
\subfloat{\includegraphics[width = 0.5\linewidth]{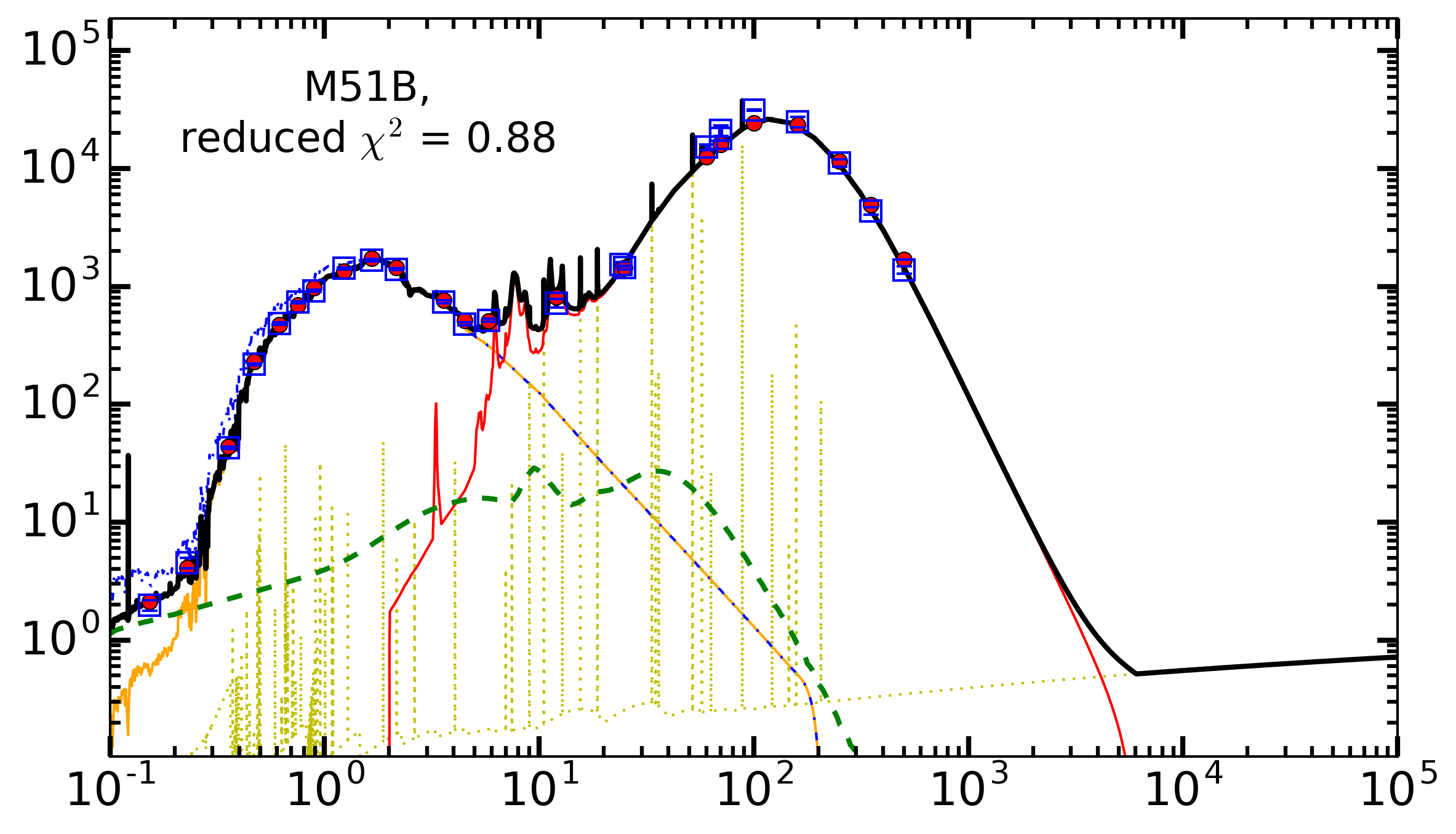}}\\
\subfloat{\includegraphics[width = 0.5\linewidth]{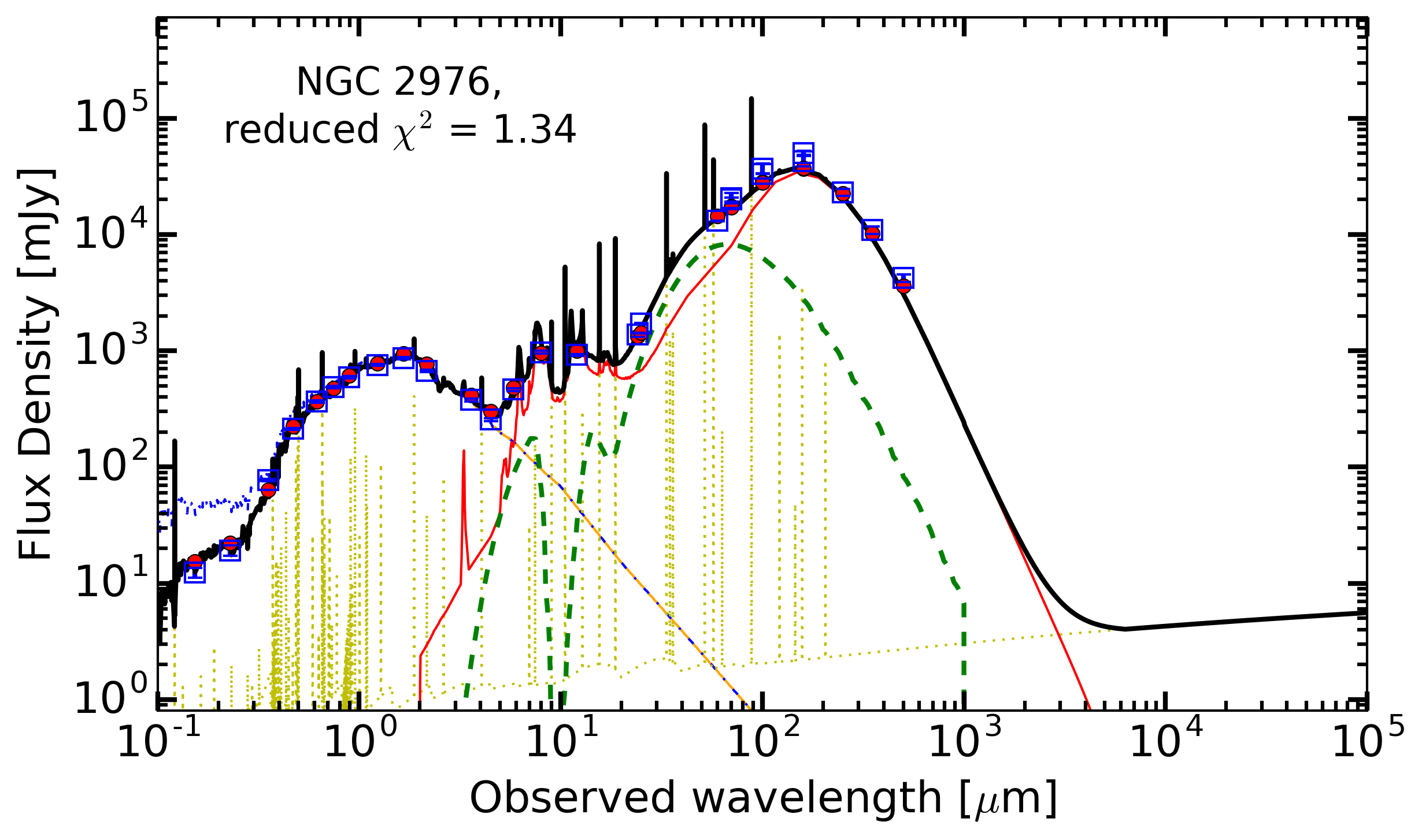}}
\subfloat{\includegraphics[width = 0.5\linewidth]{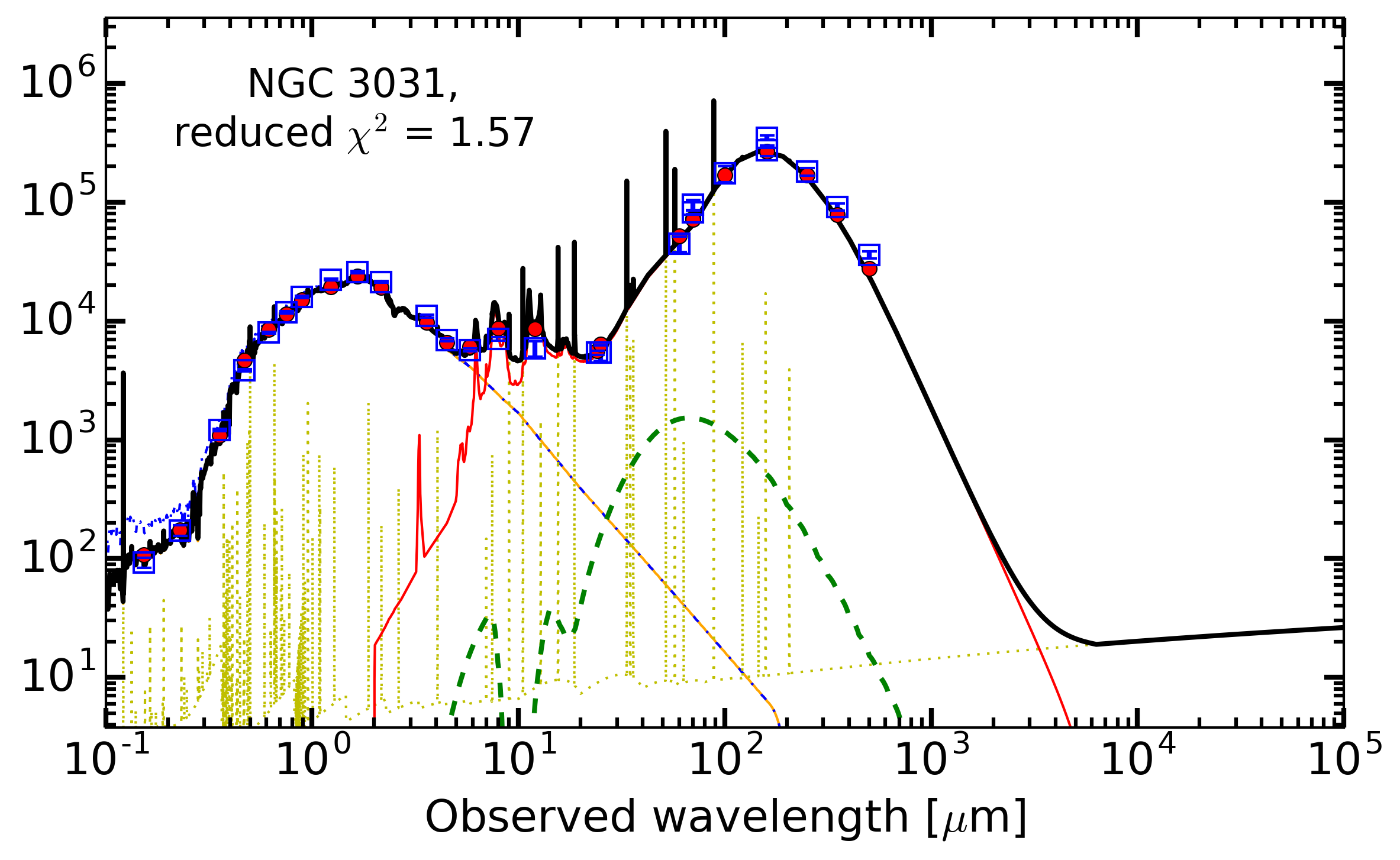}}\\
\end{figure*}
\begin{figure*}
\ContinuedFloat
\centering
\subfloat{\includegraphics[width = 0.5\linewidth]{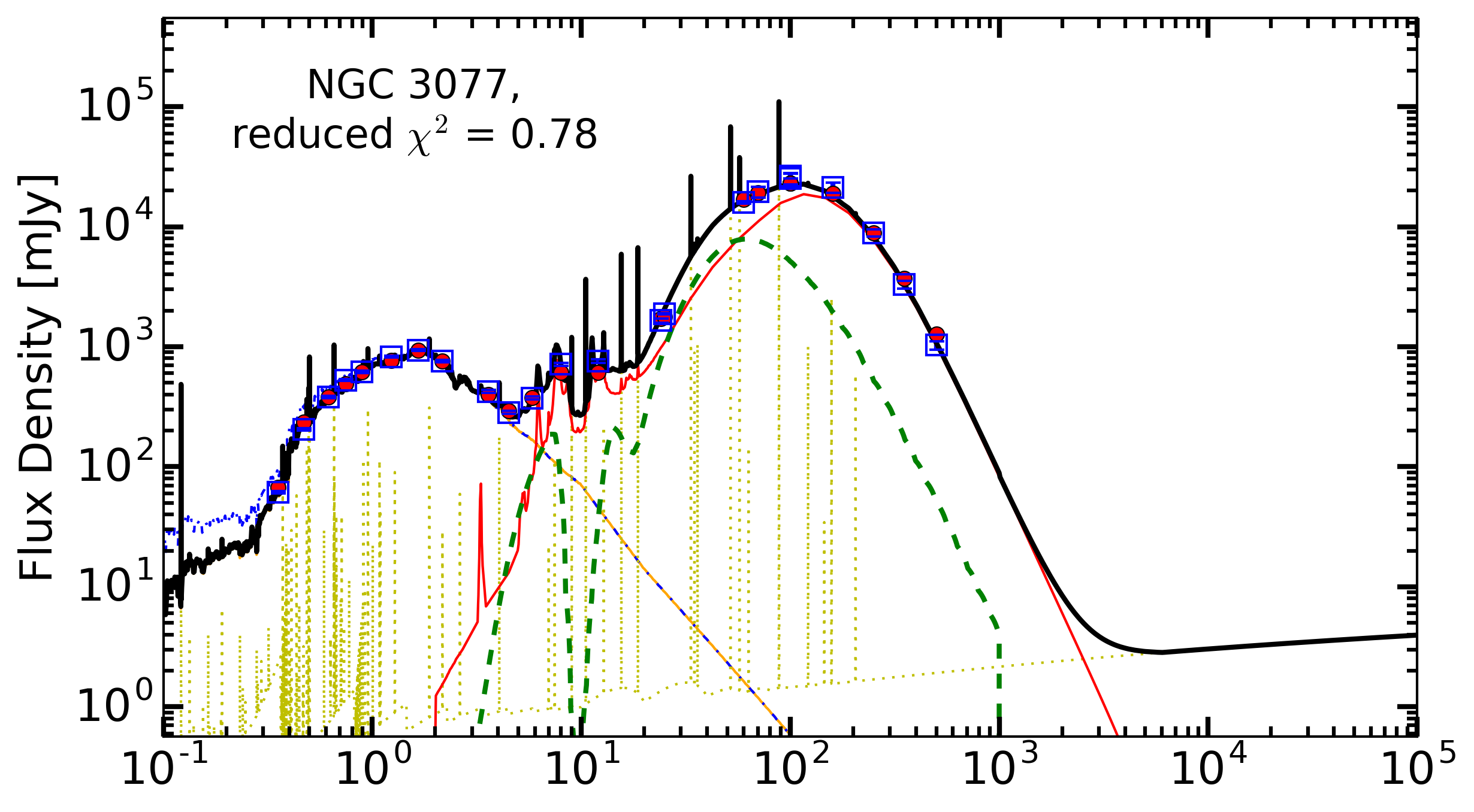}}
\subfloat{\includegraphics[width = 0.5\linewidth]{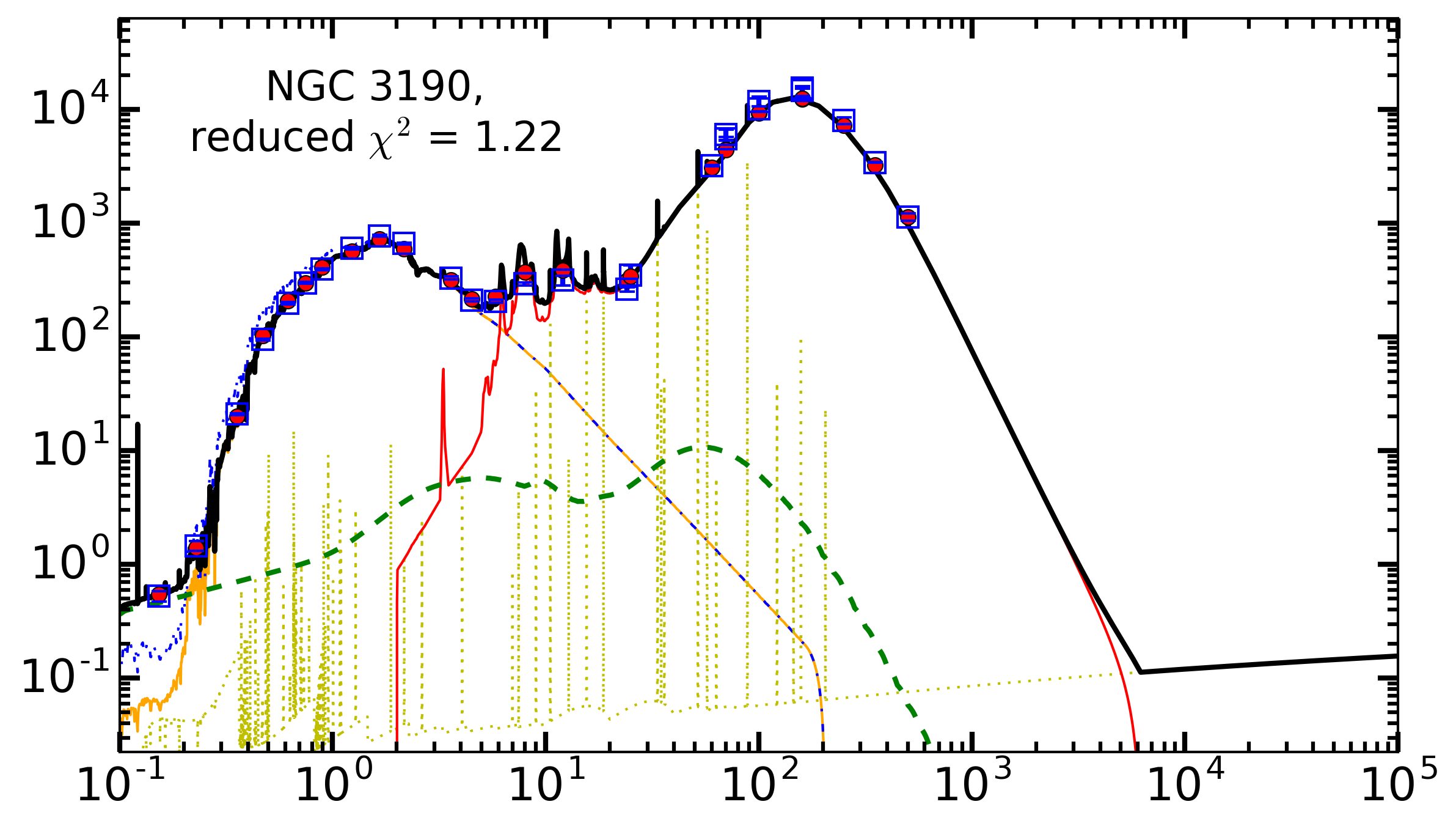}}\\
\subfloat{\includegraphics[width = 0.5\linewidth]{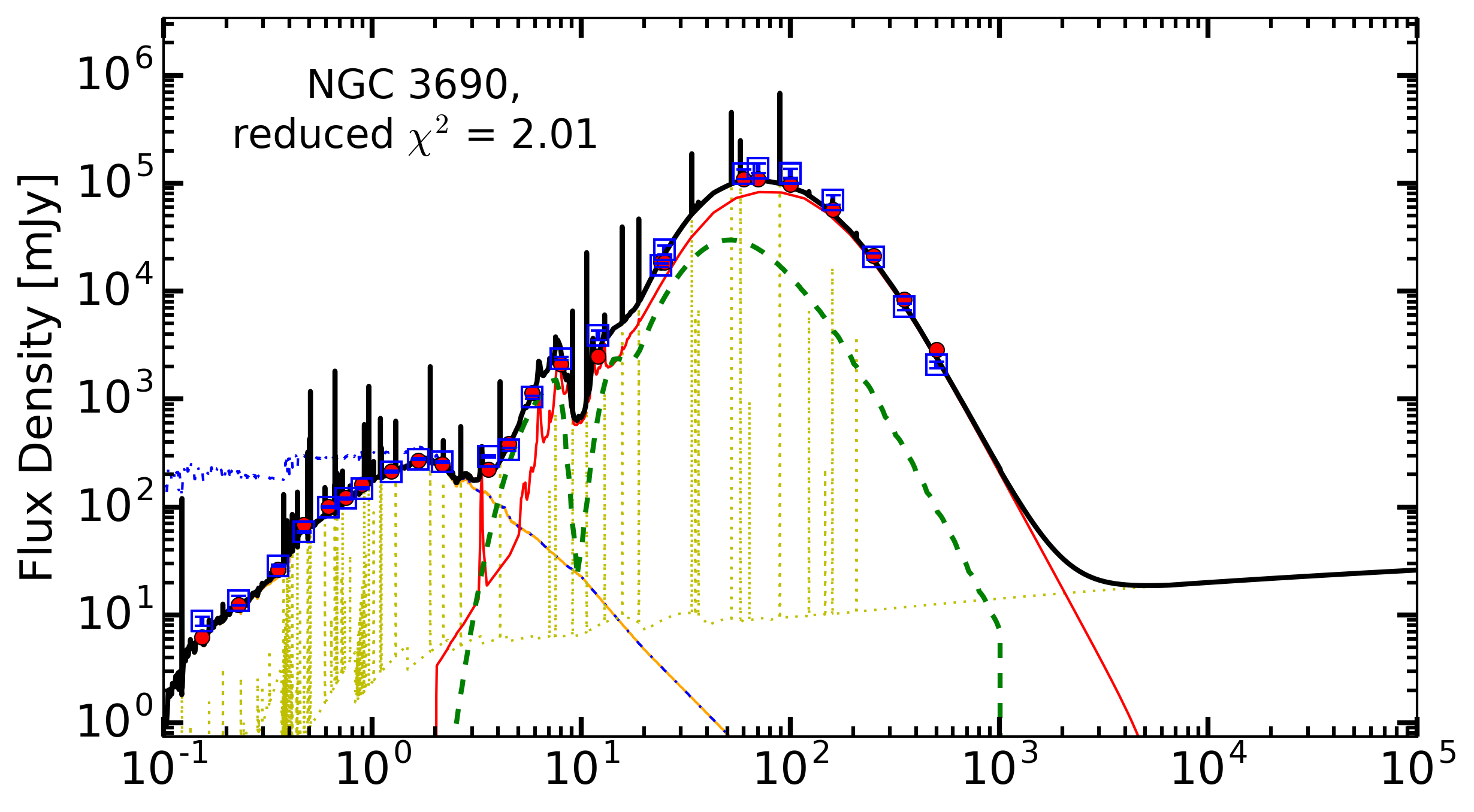}}
\subfloat{\includegraphics[width = 0.5\linewidth]{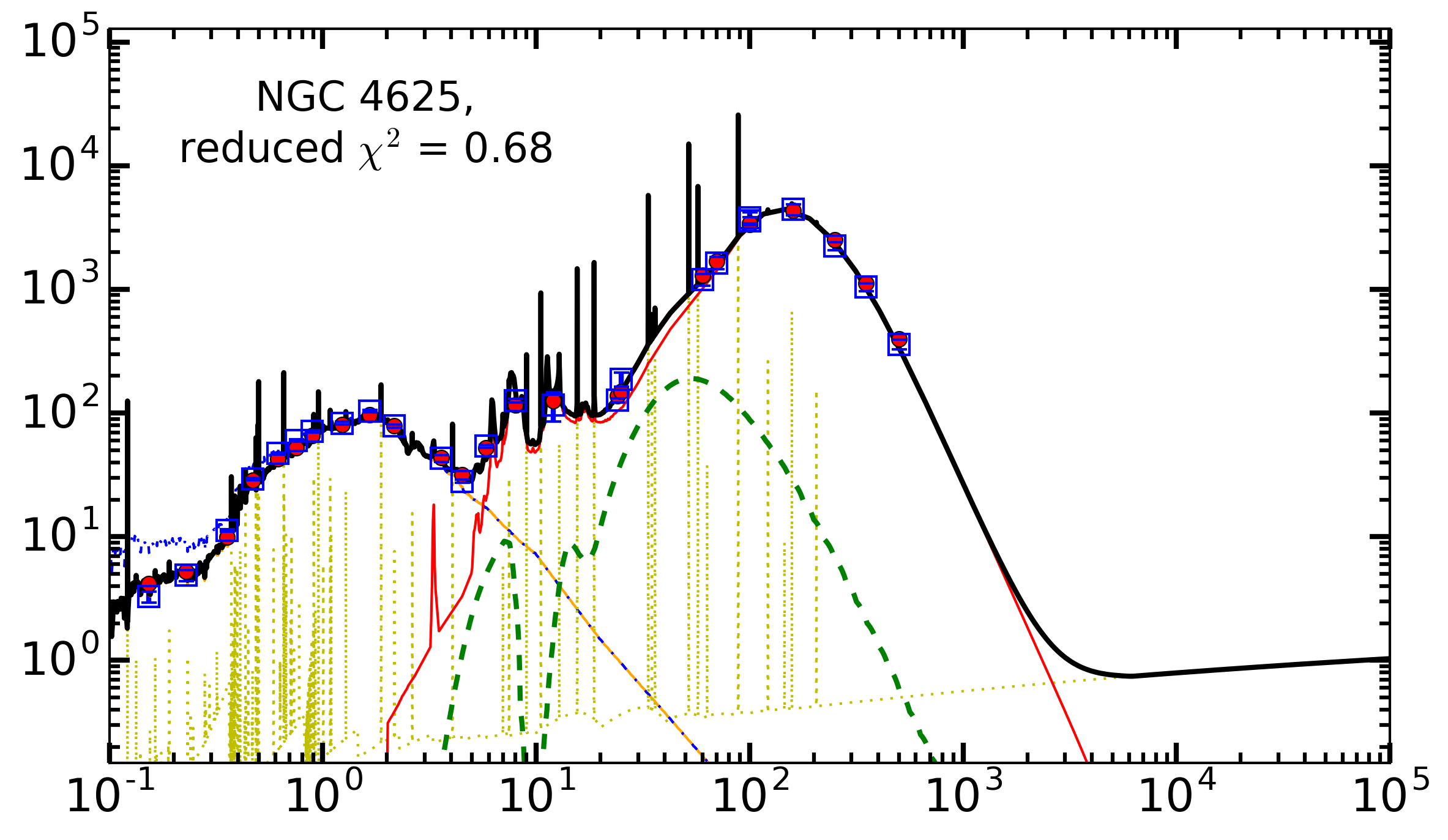}}\\
\subfloat{\includegraphics[width = 0.5\linewidth]{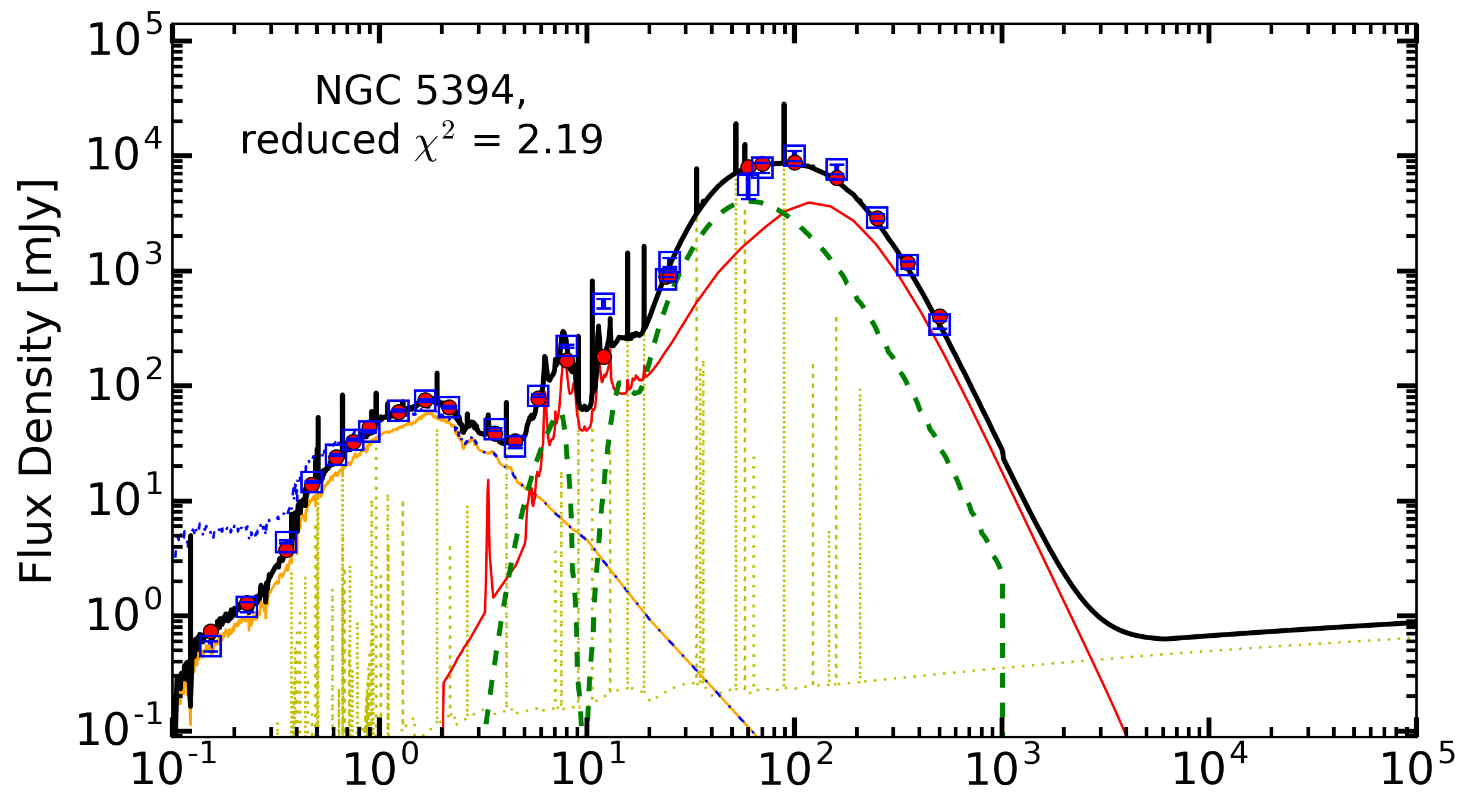}}
\subfloat{\includegraphics[width = 0.5\linewidth]{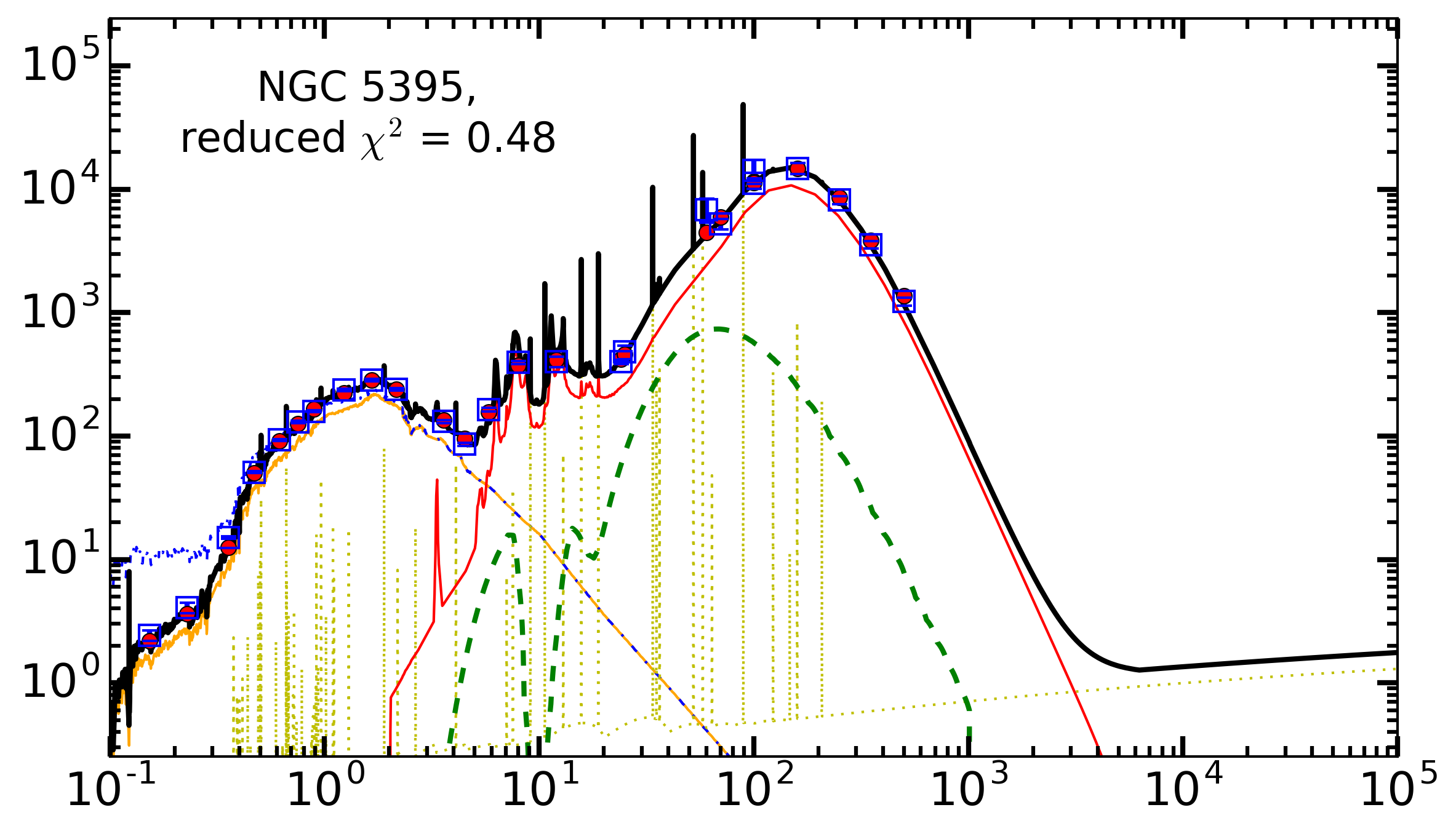}}\\
\subfloat{\includegraphics[width = 0.5\linewidth]{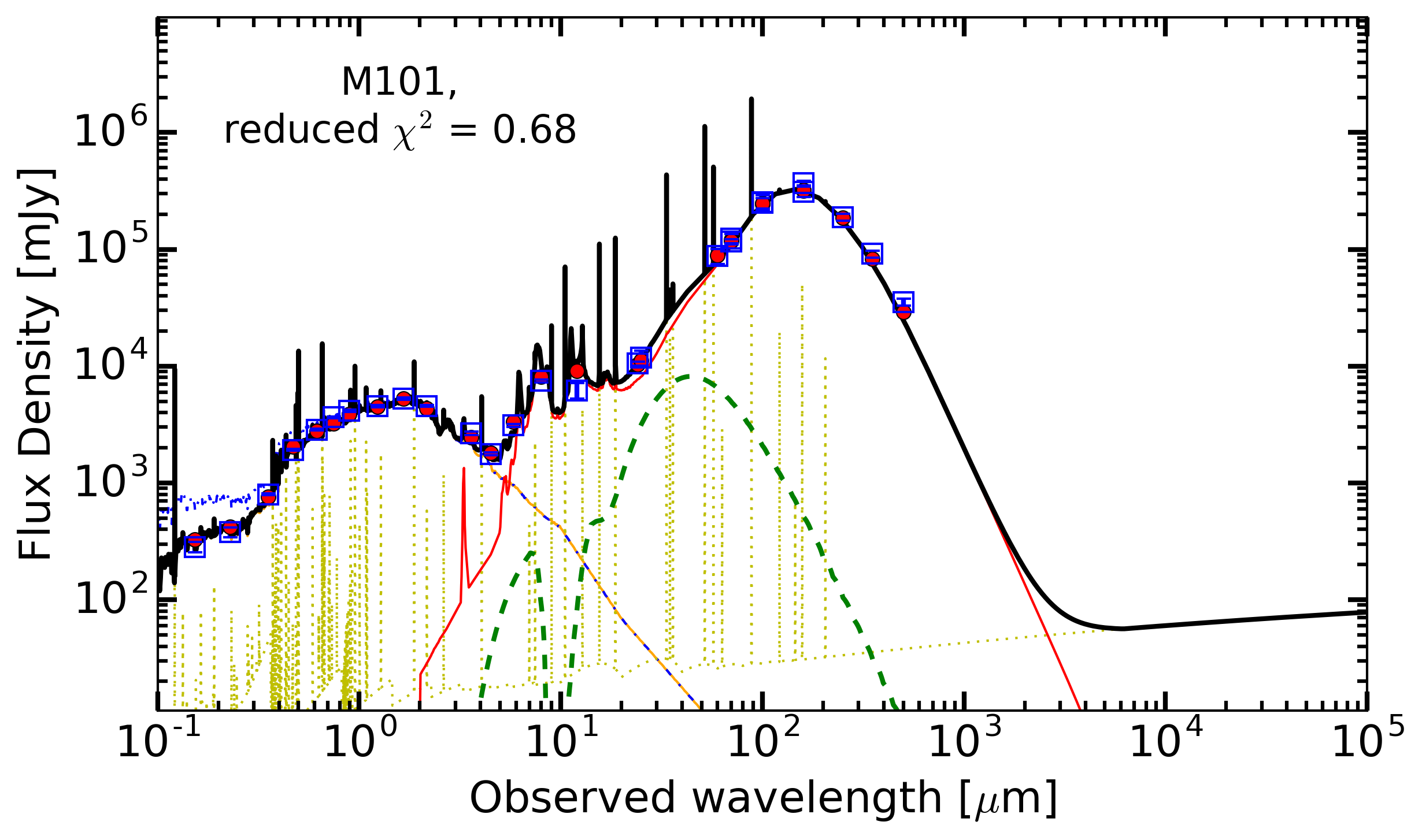}}
\subfloat{\includegraphics[width = 0.5\linewidth]{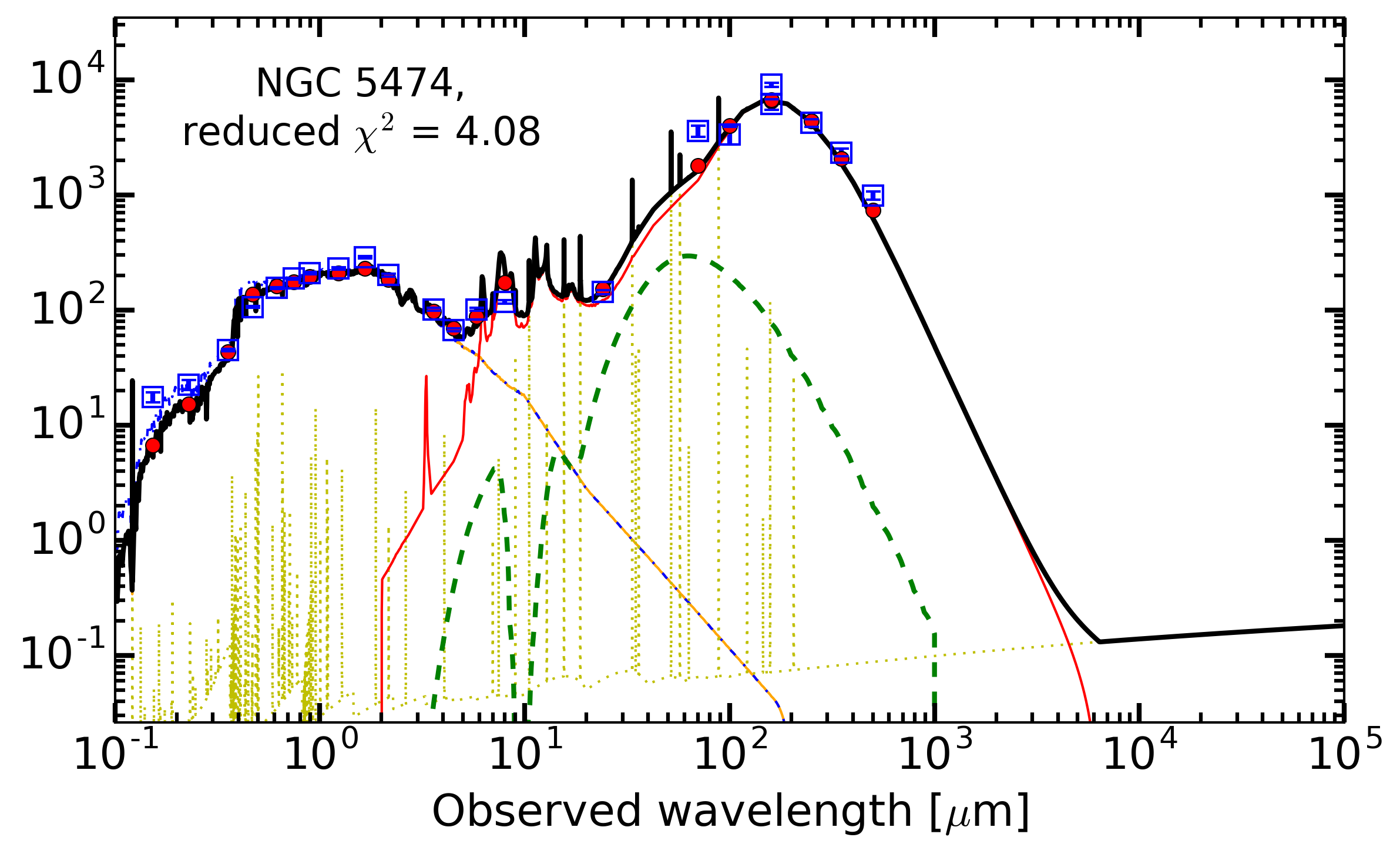}}\\
\caption{Best-fit SED models for the 24 galaxies in the sample containing the nebular emission (gold dotted lines), both attenuated stellar emission (orange) and non-attenuated stellar emission (blue dot-dashed), dust emission (red solid), and AGN emission (green dashed).  The red dots are the best model flux densities and the blue squares mark the observed flux densities with 1$\sigma$ error bars. \label{SEDs}}
\end{figure*}

\section{Photometry and PACS Spectrophotometric Data}

\begin{table*}
\footnotesize
\centering
\caption{\textit{GALEX} and \textit{Swift} UV Photometry}
\label{tab2}
\begin{tabular}{c c c c c c}
\hline
\hline
 & \multicolumn{2}{c}{\textit{GALEX}} & \multicolumn{3}{c}{\textit{Swift}} \\
Galaxy Name & FUV & NUV & UVOT\_UVW2 & UVOT\_UVM2 & UVOT\_UVW1 \\
 & (mJy) & (mJy) & (mJy) & (mJy) & (mJy) \\
\hline
\multicolumn{6}{c}{Late-Stage Merger Sample}\\
IRAS~08572+3915 & $0.110\pm0.012$ & $0.130\pm0.014$ & ... & ... & ...\\
IRAS~15250+3609 & ... & $0.652\pm0.066$ & ... & ... & ...\\
Mrk~231 & ... & ... & $0.487\pm0.050$ & $0.531\pm0.056$ & $1.02\pm0.11$\\
Mrk~273 & $0.290\pm0.041$ & $0.71\pm0.11$ & ... & ... & ...\\
Mrk~463 & ... & ... & $0.645\pm0.066$ & $0.659\pm0.069$ & $2.42\pm0.24$\\
NGC~2623 & $0.58\pm0.06$ & $0.89\pm0.09$ & ... & ... & ...\\
NGC~3758 & $0.403\pm0.040$ & $0.919\pm0.092$ & ... & ... & ...\\
NGC~6090 & $0.705\pm0.071$ & $1.37\pm0.14$ & ... & ... & ...\\
UGC 4881 & $0.250\pm0.036$ & $0.56\pm0.08$ & ... & ... & ...\\
UGC 5101 & $0.150\pm0.021$ & $0.190\pm0.031$ & ... & ... & ...\\
VV 283 & $0.130\pm0.019$ & $0.210\pm0.032$ & ... & ... & ...\\
VV 705 & $0.440\pm0.062$ & $0.87\pm0.13$ & ... & ... & ...\\
\hline
\multicolumn{6}{c}{Reference Sample}\\
M51A & $90.9\pm9.09$ & $162.0\pm16.2$ & ... & $1110.\pm63$ & $1650.\pm95$\\
M51B & $1.98\pm0.20$ & $4.53\pm0.45$ & $89.90\pm6.23$ & $282.0\pm18.7$ & $551.0\pm37.1$\\
NGC~2976 & $12.4\pm1.24$ & $19.2\pm1.92$ & $85.10\pm10.90$ & $200.0\pm25.5$ & $314.0\pm40.1$\\
NGC~3031 & $93.0\pm9.0$ & $173.0\pm17.0$ & ... & $2970.\pm83$ & $6100.\pm180.$\\
NGC~3077 & ... & ... & $90.70\pm11.70$ & $243.0\pm30.9$ & $418.0\pm53.3$\\
NGC~3190 & $0.53\pm0.05$ & $1.46\pm0.15$ & $16.50\pm1.63$ & $60.50\pm5.83$ & $126.0\pm12.2$\\
NGC~3690 & $8.76\pm0.88$ & $13.6\pm1.36$ & ... & ... & ...\\
NGC~4625 & $3.27\pm0.33$ & $4.88\pm0.49$ & $14.50\pm0.88$ & $28.90\pm1.09$ & $41.80\pm1.76$\\
NGC~5394 & $0.55\pm0.05$ & $1.20\pm0.12$ & $5.37\pm1.16$ & $14.10\pm2.85$ & $22.80\pm4.66$\\
NGC~5395 & $2.42\pm0.24$ & $4.05\pm0.41$ & $23.90\pm5.06$ & $61.60\pm12.50$ & $103.0\pm21.1$\\
M101 & $283.0\pm28.0$ & $380.0\pm38.0$ & ... & $2020.\pm175$ & $2610.  \pm248$\\
NGC~5474 & $17.5\pm1.75$ & $22.5\pm2.25$ & ... & $131.0\pm19.4$ & $176.0\pm27.6$\\
\hline
\hline
\end{tabular}
\end{table*}

\begin{table*}
\footnotesize
\centering
\caption{SDSS Photometry}
\label{tab3}
\begin{tabular}{c c c c c c}
\hline
\hline
 & \multicolumn{5}{c}{SDSS}\\
Galaxy Name & u & g & r & i & z \\
 & (mJy) & (mJy) & (mJy) & (mJy) & (mJy) \\
\hline
\multicolumn{6}{c}{Late-Stage Merger Sample}\\
IRAS~08572+3915 & $0.510\pm0.075$ & $0.870\pm0.028$ & $1.58\pm0.047$ & $1.61\pm0.068$ & $1.59\pm0.20$\\
IRAS~15250+3609 & $0.906\pm0.084$ & $1.96\pm0.055$ & $3.25\pm0.088$ & \\
Mrk~231 & $5.34\pm0.23$ & $14.90\pm0.43$ & $17.10\pm0.48$ & $20.90\pm0.59$ & $39.20\pm1.29$\\
Mrk~273 & $1.09\pm0.19$ & $5.29\pm0.17$ & $9.23\pm0.28$ & $13.10\pm0.41$ & $15.50\pm0.83$\\
Mrk~463 & $1.39\pm0.12$ & $6.27\pm0.13$ & $8.81\pm0.19$ & $12.00\pm0.26$ & $11.80\pm0.50$\\
NGC~2623 & $2.84\pm0.15$ & $9.95\pm0.20$ & $16.69\pm0.34$ & $21.3\pm0.44$ & $26.8\pm0.78$\\
NGC~3758 & $3.02\pm0.13$ & $7.72\pm0.16$ & $13.80\pm0.29$ & $19.60\pm0.41$ & $24.50\pm0.61$\\
NGC~6090 & $3.31\pm0.08$ & $7.37\pm0.15$ & $11.90\pm0.24$ & $15.50\pm0.32$ & $18.60\pm0.47$\\
UGC 4881 & $1.92\pm0.16$ & $7.23\pm0.21$ & $12.10\pm0.36$ & $17.30\pm0.52$ & $19.80\pm0.87$\\
UGC 5101 & $1.680\pm0.051$ & $4.35\pm0.15$ & $8.06\pm0.26$ & $11.30\pm0.38$ & $10.4\pm1.0$\\
VV 283 & $1.32\pm0.17$ & $3.25\pm0.11$ & $6.07\pm0.19$ & $8.25\pm0.27$ & $10.80\pm0.54$\\
VV 705 & $1.92\pm0.14$ & $5.18\pm0.16$ & $8.55\pm0.26$ & $11.50\pm0.35$ & $12.90\pm0.60$\\
\hline
\multicolumn{6}{c}{Reference Sample}\\
M51A & $402.0\pm8.1$ & $1170\pm23$ & $1920\pm38$ & $2520\pm50$ & $2980\pm60$\\
M51B & $42.7\pm0.88$ & $218.0\pm4.4$ & $484.0\pm9.7$ & $734.0\pm14.7$ & $920.0\pm18.4$\\
NGC~2976 & $77.1\pm1.57$ & $213.0\pm4.3$ & $366.0\pm7.3$ & $488.0\pm9.8$ & $593.0\pm11.9$\\
NGC~3031 & $1210\pm24$ & $3850\pm77$ & $7980\pm160$ & $11800\pm240$ & $15900\pm320$\\
NGC~3077 & $61.4\pm1.3$ & $205.0\pm4.0$ & $381.0\pm8.0$ & $528.0\pm11.0$ & $617.0\pm12.0$\\
NGC~3190 & $21.0\pm0.44$ & $94.7\pm1.9$ & $198.0\pm4.0$ & $299.0\pm6.0$ & $394.0\pm7.9$\\
NGC~3690 & $28.5\pm0.59$ & $58.7\pm1.2$ & $100.0\pm2.0$ & $120.0\pm2.4$ & $148.0\pm3.0$\\
NGC~4625 & $11.3\pm0.26$ & $29.4\pm0.59$ & $47.1\pm0.94$ & $60.1\pm1.21$ & $71.1\pm1.48$\\
NGC~5394 & $4.38\pm0.13$ & $14.6\pm0.29$ & $25.1\pm0.51$ & $34.0\pm0.69$ & $39.9\pm0.92$\\
NGC~5395 & $15.0\pm0.33$ & $51.0\pm1.02$ & $93.0\pm1.86$ & $130.0\pm2.6$ & $159.0\pm3.2$\\
M101 & $800.0\pm16.0$ & $1920\pm38$ & $2860\pm57$ & $3650\pm73$ & $4160\pm83$\\
NGC~5474 & $44.9\pm0.93$ & $106.0\pm2.0$ & $155.0\pm3.0$ & $187.0\pm4.0$ & $209.0\pm4.0$\\
\hline
\hline
\end{tabular}
\end{table*}

\begin{table*}
\footnotesize
\centering
\caption{2MASS and \textit{IRAS} Photometry}
\label{tab4}
\begin{tabular}{c c c c c c c c}
\hline
\hline
 & \multicolumn{3}{c}{2MASS} & \multicolumn{4}{c}{\textit{IRAS}} \\
Galaxy Name & $J$ & $H$ & $K_s$ & 12~\micron\ & 25~\micron\ & 60~\micron\ & 100~\micron\ \\
 & (mJy) & (mJy) & (mJy) & (mJy) & (mJy) & (mJy) & (mJy) \\
\hline
\multicolumn{8}{c}{Late-Stage Merger Sample}\\
IRAS~08572+3915 & $2.91\pm0.51$ & $3.91\pm0.84$ & $3.87\pm0.82$ & $318\pm35$ & $1700\pm90$ & $7430\pm370$ & $4770\pm150$\\
Mrk~231 & $60.8\pm2.2$ & $111.0\pm3.8$ & $192.0\pm6.1$ & $1870\pm90$ & $8660\pm430$ & $32000\pm1600$ & $29700\pm1000$\\
Mrk~273 & $23.7\pm1.7$ & $27.9\pm2.5$ & $32.6\pm2.8$ & $235\pm27$ & $2280\pm130$ & $21700\pm870$ & $22500\pm900$\\
Mrk~463 & $21.5\pm1.2$ & $31.8\pm2.0$ & $60.2\pm2.1$ & $510\pm40$ & $1580\pm90$ & $2180\pm110$ & $1920\pm210$\\
NGC~2623 & $34.3\pm1.2$ & $40.8\pm1.7$ & $42.6\pm1.8$ & $210\pm20$ & $1810\pm40$ & $23700\pm930$ & $25900\pm1000$\\
NGC~3758 & $36.8\pm1.3$ & $46.4\pm1.8$ & $48.4\pm2.0$ & $160\pm30$ & $309\pm43$ & $1260\pm130$ & $2410\pm190$\\
NGC~6090 & $26.5\pm0.8$ & $22.5\pm1.0$ & $23.5\pm1.2$ & $260\pm20$ & $1110\pm40$ & $6660\pm270$ & $9400\pm1000$\\
UGC 4881 & $29.1\pm1.7$ & $31.2\pm2.2$ & $35.8\pm2.8$ & $135\pm31$ & $599\pm48$ & $5960\pm360$ & $10300\pm1100$\\
UGC 5101 & $18.6\pm1.7$ & $30.1\pm2.5$ & $35.3\pm2.9$ & $250\pm40$ & $1030\pm60$ & $11500\pm810$ & $19900\pm1400$\\
VV 283 & $13.7\pm1.5$ & $16.0\pm2.1$ & $18.5\pm2.8$ & $157\pm33$ & $386\pm66$ & $5070\pm460$ & $7950\pm480$\\
VV 705 & $20.8\pm1.6$ & $26.9\pm2.5$ & $26.9\pm2.7$ & $210\pm20$ & $1390\pm70$ & $9210\pm370$ & $10000\pm900$\\
\hline
\multicolumn{8}{c}{Reference Sample}\\
M51A & $3940\pm79$ & $4690\pm94$ & $3810\pm77$ & $7210\pm75$ & $9560\pm77$ & $97400\pm190$ & $221000\pm300$\\
M51B & $1420\pm28$ & $1670\pm34$ & $1400\pm28$ & $721.0\pm57.7$ & $1450\pm51$ & $15200\pm800$ & $31300\pm370$\\
NGC~2976 & $757.0\pm15.3$ & $862.0\pm17.6$ & $670.0\pm14.1$ & $920.0\pm20.0$ & $1710\pm20$ & $13100\pm30$ & $33400\pm340$\\
NGC~3031 & $22300\pm446$ & $25700\pm515$ & $21300\pm427$ & $5860\pm879$ & $5420\pm813$ & $44700\pm6710$ & $174000\pm26100$\\
NGC~3077 & $826.0\pm17.0$ & $937.0\pm19.0$ & $759.0\pm16.0$ & $760.0\pm23.0$ & $1880\pm25.0$ & $15900\pm390$ & $26500\pm1320$\\
NGC~3190 & $602.0\pm12.1$ & $773.0\pm15.6$ & $661.0\pm13.4$ & $315.0\pm31.5$ & $351.0\pm75.7$ & $3190\pm35$ & $10100\pm510$\\
NGC~3690 & $212.0\pm4.4$ & $278.0\pm5.9$ & $262.0\pm5.5$ & $3900\pm400$ & $24100\pm2400$ & $122000\pm12500$ & $123000\pm12500$\\
NGC~4625 & $83.1\pm1.97$ & $104.0\pm2.7$ & $78.7\pm2.4$ & $117.0\pm31.6$ & $188.0\pm24.6$ & $1200\pm132$ & $3580\pm250$\\
NGC~5394 & $60.5\pm1.36$ & $74.4\pm1.76$ & $65.1\pm1.75$ & $520.0\pm50.0$ & $1190\pm110$ & $5620\pm1410$ & ...\\
NGC~5395 & $237.0\pm4.8$ & $285.0\pm5.9$ & $240.0\pm5.1$ & $400.0\pm40.0$ & $480.0\pm60.0$ & $6860\pm1500$ & $14200\pm3100$\\
M101 & $4540\pm92$ & $5270\pm107$ & $4570\pm94$ & $6200\pm930$ & $11800\pm1770$ & $88000\pm13200$ & $253000\pm37900$\\
NGC~5474 & $229.0\pm5.0$ & $288.0\pm7.0$ & $201.0\pm6.0$ & ... & ... & $1330\pm67$ & ...\\
\hline
\hline
\end{tabular}
\end{table*}

\begin{table*}
\footnotesize
\centering
\caption{\textit{WISE} Photometry for the Late-Stage Merger Sample; it was not used for the Reference Sample (see Section 2)}272
\label{tab45}
\begin{tabular}{c c c c c}
\hline
\hline
Galaxy Name & 3.4~\micron\ & 4.6~\micron\ & 12~\micron\ & 22~\micron\\
& (mJy) & (mJy) & (mJy) & (mJy) \\
\hline
\multicolumn{5}{c}{Late-Stage Merger Sample}\\
IRAS~08572+3915 & $24.0\pm1.44$ & $113.0\pm6.78$ & $308.0\pm18.5$ & $1170\pm70$\\
Mrk~231 & $246.0\pm20.9$ & $420.0\pm35.6$ & $1360.0\pm115.0$ & $6080\pm530$\\
Mrk~273 & $24.8\pm1.58$ & $36.4\pm2.45$ & $212.0\pm14.2$ & $1460\pm130$\\
Mrk~463 & $131.0\pm7.86$ & $206.0\pm12.4$ & $476.0\pm28.6$ & $1490\pm90$\\
NGC~2623 & $27.4\pm1.64$ & $25.4\pm1.52$ & $181.3\pm10.9$ & $1120\pm70$\\
NGC~3758 & $40.9\pm2.46$ & $43.4\pm2.61$ & $127.0\pm7.63$ & $227.0\pm14.2$\\
NGC~6090 & ... & ... & ... & ...\\
UGC 4881 & $20.7\pm1.76$ & $15.8\pm1.35$ & $114.0\pm9.69$ & $352.0\pm30.7$\\
UGC 5101 & $32.4\pm2.75$ & $79.0\pm6.71$ & $158.0\pm13.4$ & $646.0\pm56.8$\\
VV 283 & $13.0\pm1.11$ & $11.8\pm1.51$ & $89.6\pm7.61$ & $252.0\pm22.1$\\
VV 705 & $19.5\pm1.65$ & $17.2\pm1.47$ & $169.0\pm14.4$ & $873.0\pm81.9$\\
\hline
\hline
\end{tabular}
\end{table*}

\begin{table*}
\footnotesize
\centering
\caption{\textit{Spitzer}/IRAC and MIPS Photometry}
\label{tab5}
\begin{tabular}{c c c c c c c c}
\hline
\hline
 & \multicolumn{4}{c}{IRAC} & \multicolumn{3}{c}{MIPS} \\
Galaxy Name & 3.6~\micron\ & 4.5~\micron\ & 5.8~\micron\ & 8.0~\micron\ & 24~\micron\ & 70~\micron\ & 160~\micron\ \\
 & (mJy) & (mJy) & (mJy) & (mJy) & (mJy) & (mJy) & (mJy) \\
\hline
\multicolumn{8}{c}{Late-Stage Merger Sample}\\
IRAS~08572+3915 & $38.6\pm1.2$ & $98.3\pm3.0$ & $257.0\pm7.7$ & $350\pm10$ & $1390\pm56$ & $6160\pm250$ & $1850\pm74$\\
IRAS~15250+3609 & $8.02\pm0.22$ & $9.84\pm0.29$ & $38.3\pm1.2$ & $118.0\pm3.5$ & $1050\pm40$ & $8410\pm330$ & $2750\pm110$\\
Mrk~231 & $357.0\pm10.9$ & $473.0\pm14.4$ & $626.0\pm19.7$ & $1070\pm30$ & ... & ... & ...\\
Mrk~273 & $28.8\pm1.9$ & $37.9\pm1.6$ & $75.2\pm5.1$ & $ 177.0\pm7.5$ & $1800\pm70$ & $27400\pm4700$ & $12500\pm4100$\\
Mrk~463 & $116.0\pm3.5$ & $163.0\pm4.9$ & $256.0\pm7.7$ & $332.0\pm10.0$ & $1430\pm60$ & $3310\pm140$ & $1190\pm50$\\
NGC~2623 & $29.2\pm0.88$ & $27.4\pm0.82$ & $62.4\pm1.9$ & $178.0\pm9.3$ & $1390\pm60$ & $27300\pm2200$ & $15200\pm1500$\\
NGC~3758 & $50.3\pm1.5$ & $51.9\pm1.6$ & $69.1\pm2.1$ & $120.0\pm3.6$ & $250\pm10$ & ... & ...\\
NGC~6090 & $27.0\pm0.9$ & $19.9\pm0.6$ & $73.6\pm2.3$ & $247.0\pm7.4$ & $970\pm39.0$ & $7890\pm320$ & ...\\
UGC 4881 & $22.9\pm1.0$ & $17.90\pm0.76$ & $37.0\pm1.6$ & $130.0\pm5.5$ & $415\pm24$ & $9740\pm960$ & $7600\pm2500$\\
UGC 5101 & $43.5\pm1.9$ & $79.2\pm3.4$ & $98.3\pm4.2$ & $168.0\pm7.1$ & $753\pm43$ & $15400\pm1400$ & $11300\pm4500$\\
VV 283 & $15.00\pm0.64$ & $13.00\pm0.55$ & $34.6\pm1.5$ & $128.0\pm5.4$ & $300\pm17$ & $6390\pm730$ & $7190\pm2030$\\
VV 705 & $20.00\pm0.85$ & $17.50\pm0.75$ & $43.5\pm1.9$ & $161.0\pm6.8$ & $1130\pm64$ & $7480\pm650$ & $5690\pm1880$\\
\hline
\multicolumn{8}{c}{Reference Sample}\\
M51A & $2370\pm71$ & $1550\pm47$ & $3690\pm111$ & $9810\pm294$ & $12000\pm480$ & $135000\pm5400$ & ...\\
M51B & $739.0\pm22.2$ & $479.0\pm14.4$ & $515.0\pm15.4$ & ... & $1540\pm62$ & $18000\pm720$ & ...\\
NGC~2976 & $378.0\pm11.3$ & $256.0\pm7.7$ & $465.0\pm14.0$ & $972.0\pm29.2$ & $1380\pm55$ & $20000\pm800$ & $50300\pm2000$\\
NGC~3031 & $11000\pm330$ & $6930\pm208$ & $5700\pm171$ & $7060\pm212$ & $5410\pm216$ & $82400\pm3300$ & $348000\pm13900$\\
NGC~3077 & $424.0\pm13.0$ & $284.0\pm9.0$ & $374.0\pm11.0$ & $716.0\pm21.0$ & $1650\pm66$ & ... & ...\\
NGC~3190 & $330.0\pm9.9$ & $211.0\pm6.3$ & $206.0\pm6.2$ & $293.0\pm8.8$ & $262.0\pm10.6$ & $5530\pm224$ & $15400\pm224$\\
NGC~3690 & $295.0\pm8.8$ & $340.0\pm10.2$ & $1040\pm31$ & $2370\pm71$ & $17400\pm700$ & ... & ...\\
NGC~4625 & $43.0\pm2.6$ & $28.1\pm0.85$ & $54.1\pm1.6$ & $126.0\pm3.8$ & $127.0\pm5.2$ & ... & ...\\
NGC~5394 & $41.9\pm1.26$ & $29.6\pm0.89$ & $82.1\pm2.47$ & $222.0\pm6.7$ & $846.0\pm33.9$ & ... & ...\\
NGC~5395 & $131.0\pm3.9$ & $85.9\pm2.58$ & $164.0\pm4.9$ & $396.0\pm11.9$ & $400.0\pm16.1$ & ... & ...\\
M101 & $2660\pm80$ & $1770\pm53$ & $3110\pm93$ & $7470\pm224$ & $10500\pm420$ & $117000\pm4700$ & $369000\pm14800$\\
NGC~5474 & $101.0\pm3.0$ & $67.0\pm2.0$ & $101.0\pm3.0$ & $117.0\pm4.0$ & $143.0\pm6.0$ & ... & $9050\pm379$\\
\hline
\hline
\end{tabular}
\end{table*}

\begin{table*}
\footnotesize
\centering
\caption{\textit{Herschel}/PACS and SPIRE Photometry}
\label{tab6}
\begin{tabular}{c c c c c c c}
\hline
\hline
 & \multicolumn{3}{c}{PACS} & \multicolumn{3}{c}{SPIRE} \\
Galaxy Name & 75~\micron\ & 110~\micron\ & 170~\micron\ & 250~\micron\ & 350~\micron\ & 500~\micron\ \\
 & (mJy) & (mJy) & (mJy) & (mJy) & (mJy) & (mJy) \\
\hline
\multicolumn{7}{c}{Late-Stage Merger Sample}\\
IRAS~08572+3915 & $6190\pm620$ & $4120\pm410$ & $1830\pm180$ & $446\pm33$ & $131\pm16$ & $2.67\pm10.3$\\
Mrk~231 & $31400\pm4500$ & $26800\pm3800$ & $14900\pm2100$ & $4990\pm500$ & $1670\pm180$ & $456\pm87$\\
Mrk~273 & $23300\pm2300$ & $20600\pm2200$ & $11600\pm1300$ & $3710\pm460$ & $1200\pm130$ & $334\pm52$\\
Mrk~463 & ... & ... & ... & $559\pm45$ & $199\pm19$ & $55.4\pm18.1$\\
NGC~2623 & $25900\pm2600$ & $26300\pm2600$ & $15800\pm1600$ & $5180\pm0520$ & $1790\pm200$ & $473\pm90$\\
NGC~3758 & $1390\pm170$ & ... & $1740\pm280$ & $1100\pm90$ & $416\pm42$ & $85.1\pm30.4$\\
NGC~6090 & $6020\pm600$ & $7020\pm700$ & $4860\pm490$ & $2910\pm200$ & $1080\pm80$ & $280\pm20$\\
UGC 4881 & $7410\pm1050$ & $9270\pm1310$ & $6350\pm930$ & $3090\pm320$ & $1150\pm140$ & $344\pm55$\\
UGC 5101 & $14000\pm2000$ & ... & $13000\pm1900$ & $5490\pm550$ & $2140\pm220$ & $610\pm92$\\
VV 283 & $6170\pm880$ & $7840\pm1110$ & $5900\pm840$ & $2320\pm240$ & $891\pm97$ & $244\pm38$\\
VV 705 & $9440\pm1340$ & $9420\pm1330$ & $5770\pm820$ & $1950\pm200$ & $667\pm78$ & $185\pm29$\\
\hline
\multicolumn{7}{c}{Reference Sample}\\
M51A & $170000\pm17000$ & ... & $367000\pm36700$ & $183000\pm12800$ & $74300\pm5200$ & $24700\pm1730$\\
M51B & $21100\pm2110$ & ... & $24800\pm2480$ & $11200\pm790$ & $4360\pm310$ & $1380\pm102$\\
NGC~2976 & $20600\pm2060$ & $37000\pm3700$ & $42900\pm4300$ & $23000\pm1610$ & $10900\pm800$ & $4220\pm300$\\
NGC~3031 & $94700\pm9500$ & ... & $272000\pm27300$ & $180000\pm12600$ & $91000\pm6380$ & $35900\pm2520$\\
NGC~3077 & $19500\pm2000$ & $25100\pm2510$ & $21200\pm2130$ & $8870\pm626$ & $3290\pm241$ & $1030\pm84$\\
NGC~3190 & $6020\pm614$ & $11800\pm1180$ & $14600\pm1470$ & $7950\pm558$ & $3420\pm243$ & $1130\pm84$\\
NGC~3690 & $138000\pm13800$ & $124000\pm12400$ & $69900\pm6990$ & $20800\pm1460$ & $7170\pm503$ & $2070\pm146$\\
NGC~4625 & $1640\pm187$ & $3810\pm389$ & $4450\pm455$ & $2240\pm160$ & $1050\pm79$ & $360.0\pm31.8$\\
NGC~5394 & $7860\pm787$ & $10000\pm1010$ & $7650\pm766$ & $2920\pm205$ & $1120\pm81$ & $339.0\pm26.6$\\
NGC~5395 & $5250\pm528$ & $11200\pm1120$ & $14800\pm1480$ & $8190\pm574$ & $3570\pm251$ & $1230\pm88$\\
M101 & $123000\pm12300$ & $249000\pm24900$ & $312000\pm31200$ & $189000\pm13200$ & $91400\pm6400$ & $35300\pm2480$\\
NGC~5474 & $3580\pm403$ & $3320\pm582$ & $6120\pm648$ & $4210\pm304$ & $2330\pm176$ & $985.0\pm83.0$\\
\hline
\hline
\end{tabular}
\end{table*}

\begin{table*}
\centering
\footnotesize
\caption{The PACS Spectrophotometric (SP) Data Points, with wavelength, number of observations, and average flux; ... signifies unknown number of observations were used.}
\label{PACS_SP}
\begin{tabular}{c c c c c c c c c c}
\hline
\hline
& \multicolumn{3}{c}{PACS `Blue' SP} & \multicolumn{3}{c}{PACS `Green' SP} & \multicolumn{3}{c}{PACS `Red' SP}\\
\hline
Galaxy Name & $\lambda$ (\micron) & \# & Flux Density (mJy) & $\lambda$ (\micron) & \# & Flux Density (mJy) & $\lambda$ (\micron) & \# & Flux Density (mJy)\\
\hline
IRAS~08572+3915 & 64 & 386 & $5510\pm550$ & 89 & 497 & $5320\pm530$ & 151 & 60 & $2160\pm220$\\
Mrk~231 & 66 & 701 & $35700\pm3570$ & 90 & 714 & $32200\pm3220$ & 160 & 490 & $16200\pm1620$\\
Mrk~273 & 63 & 419 & $24400\pm2440$ & 88 & 307 & $22500\pm2250$ & 160 & 154 & $12100\pm1210$\\
Mrk~463 & 63 & ... & $2060\pm210$ & 88 & ... & $1870\pm200$ & 157 & ... & $1300\pm140$\\
NGC~2623 & 64 & 71 & $28600\pm2860$ & 90 & 61 & $26300\pm2630$ & 160 & 49 & $17200\pm1720$\\
NGC~6090 & 65 & 209 & $7000\pm700$ & 81 & 234 & $8370\pm840$ & 163 & 57 & $6740\pm670$\\
UGC 4881 & 66 & 154 & $5150\pm520$ & 92 & 153 & $9640\pm960$ & 164 & 69 & $6680\pm670$\\
UGC 5101 & 63 & 284 & $12400\pm1240$ & 92 & 48 & $17800\pm1780$ & 151 & 122 & $16400\pm1640$\\
VV 705 & 65 & 206 & $9580\pm960$ & 82 & 128 & $10500\pm1050$ & 164 & 65 & $5910\pm590$\\
\hline
\end{tabular}
\end{table*}

\section{AGN Observables Linear Fit Analysis}

\begin{table*}
\centering
\caption{The linear regression test results of $f_{AGN}$ vs. colours showing the strongest correlations (Pearson \textit{r} $>$ 0.8) and their significance across both subsamples (requiring at least 3$\sigma$ significance).}
\begin{tabular}{ccccc}
\hline
\hline
Flux Ratio & Number of systems & Slope & \textit{r} & \textit{p}wo\\
\hline
\textit{GALEX} FUV--MIPS 24~\micron & 20 & $6.61\pm1.08$ & $0.82$ & $8.47\times10^{-6}$\\
\textit{GALEX} FUV--\textit{IRAS} 25~\micron & 18 & $6.25\pm1.03$ & $0.83$ & $1.7\times10^{-5}$\\
\textit{GALEX} FUV--\textit{IRAS} 60~\micron & 19 & $6.13\pm0.99$ & $0.83$ & $1.03\times10^{-5}$\\
\textit{GALEX} FUV--PACS 70~\micron & 20 & $6.11\pm1.05$ & $0.81$ & $1.55\times10^{-5}$\\
\textit{GALEX} FUV--MIPS 70~\micron & 13 & $5.36\pm1.14$ & $0.82$ & $0.06\times10^{-2}$\\
\textit{GALEX} NUV--\textit{IRAS} 12~\micron & 20 & $3.99\pm0.65$ & $0.82$ & $7.8\times10^{-6}$\\
\textit{GALEX} NUV--MIPS 24~\micron & 21 & $6.37\pm0.88$ & $0.86$ & $7.06\times10^{-7}$\\
\textit{GALEX} NUV--\textit{IRAS} 25~\micron & 19 & $6.07\pm0.84$ & $0.87$ & $1.41\times10^{-6}$\\
\textit{GALEX} NUV--\textit{IRAS} 60~\micron & 20 & $5.92\pm0.85$ & $0.85$ & $1.6\times10^{-6}$\\
\textit{GALEX} NUV--PACS 70~\micron & 21 & $5.79\pm0.88$ & $0.83$ & $2.67\times10^{-6}$\\
\textit{GALEX} NUV--MIPS 70~\micron & 14 & $5.23\pm0.95$ & $0.85$ & $1.31\times10^{-4}$\\
Sloan \textit{u}--MIPS 24~\micron & 23 & $6.34\pm0.96$ & $0.82$ & $1.55\times10^{-6}$\\
SloaNGC n \textit{u}--MIPS 70~\micron & 15 & $4.98\pm0.94$ & $0.83$ & $148.28\times10^{-6}$\\
Sloan \textit{g}--MIPS 24~\micron & 23 & $6.63\pm1.01$ & $0.82$ & $1.77\times10^{-6}$\\
Sloan \textit{g}--PACS 70~\micron & 23 & $6.04\pm0.97$ & $0.81$ & $3.45\times10^{-6}$\\
Sloan \textit{g}--MIPS 70~\micron & 15 & $5.4\pm1.02$ & $0.83$ & $0.01\times10^{-2}$\\
Sloan \textit{r}--MIPS 24~\micron & 23 & $6.69\pm1.03$ & $0.82$ & $2.02\times10^{-6}$\\
Sloan \textit{r}--MIPS 70~\micron & 15 & $5.56\pm1.05$ & $0.83$ & $145366.93\times10^{-9}$\\
Sloan \textit{i}--MIPS 24~\micron & 23 & $6.7\pm1.05$ & $0.81$ & $2.6\times10^{-6}$\\
Sloan \textit{i}--MIPS 70~\micron & 15 & $5.64\pm1.07$ & $0.83$ & $0.15\times10^{-3}$\\
Sloan \textit{z}--MIPS 24~\micron & 23 & $7.01\pm1.04$ & $0.83$ & $1.16\times10^{-6}$\\
Sloan \textit{z}--PACS 70~\micron & 23 & $6.35\pm0.99$ & $0.81$ & $2.45\times10^{-6}$\\
Sloan \textit{z}--MIPS 70~\micron & 15 & $6.01\pm1.04$ & $0.85$ & $6.46\times10^{-5}$\\
Sloan \textit{z}--PACS 100~\micron & 18 & $5.16\pm0.93$ & $0.81$ & $4.56\times10^{-5}$\\
2MASS J--MIPS 24~\micron & 23 & $6.98\pm1.01$ & $0.83$ & $8.44\times10^{-7}$\\
2MASS J--PACS 70~\micron & 23 & $6.36\pm0.97$ & $0.82$ & $1.68\times10^{-6}$\\
2MASS J--MIPS 70~\micron & 15 & $6.09\pm1.07$ & $0.85$ & $7.12\times10^{-5}$\\
2MASS J--\textit{IRAS} 100~\micron & 12 & $8.6\pm1.81$ & $0.83$ & $77.09\times10^{-5}$\\
2MASS J--PACS 100~\micron & 18 & $4.97\pm0.93$ & $0.8$ & $6.61\times10^{-5}$\\
2MASS H--IRAC 4.5~\micron & 24 & $3.85\pm0.61$ & $0.8$ & $2.52\times10^{-6}$\\
2MASS H--IRAC 5.8~\micron & 24 & $4.44\pm0.7$ & $0.8$ & $2.2\times10^{-6}$\\
2MASS H--MIPS 24~\micron & 23 & $6.86\pm1.04$ & $0.82$ & $1.49\times10^{-6}$\\
2MASS H--PACS 70~\micron & 23 & $6.33\pm0.94$ & $0.83$ & $1.21\times10^{-6}$\\
2MASS H--MIPS 70~\micron & 15 & $5.95\pm1.14$ & $0.82$ & $16.16\times10^{-5}$\\
2MASS H--\textit{IRAS} 100~\micron & 12 & $8.78\pm1.7$ & $0.85$ & $431.21\times10^{-6}$\\
2MASS H--PACS 100~\micron & 18 & $5.12\pm0.87$ & $0.83$ & $2.26\times10^{-5}$\\
2MASS Ks--PACS 70~\micron & 23 & $5.81\pm0.87$ & $0.82$ & $1.39\times10^{-6}$\\
2MASS Ks--\textit{IRAS} 100~\micron & 12 & $7.96\pm1.67$ & $0.83$ & $7.79\times10^{-4}$\\
2MASS Ks--PACS 100~\micron & 18 & $4.61\pm0.76$ & $0.83$ & $1.65\times10^{-5}$\\
IRAC 3.6~\micron--IRAC 4.5~\micron & 24 & $1.38\pm0.19$ & $0.84$ & $2.89\times10^{-7}$\\
IRAC 3.6~\micron--\textit{IRAS} 100~\micron & 12 & $7.07\pm1.36$ & $0.85$ & $0.0\times10^{-1}$\\
IRAC 4.5~\micron--\textit{IRAS} 100~\micron & 12 & $5.64\pm1.21$ & $0.83$ & $89.97\times10^{-5}$\\
IRAC 5.8~\micron--\textit{IRAS} 100~\micron & 12 & $3.68\pm0.79$ & $0.83$ & $87.77\times10^{-5}$\\
\textit{IRAS} 60~\micron--\textit{IRAS} 100~\micron & 12 & $-2.81\pm0.46$ & $-0.89$ & $1069.02\times10^{-7}$\\
\textit{IRAS} 60~\micron--MIPS 160~\micron & 14 & $-3.25\pm0.59$ & $-0.85$ & $139.66\times10^{-6}$\\
PACS 70~\micron--MIPS 160~\micron & 14 & $-2.46\pm0.44$ & $-0.85$ & $1.19\times10^{-4}$\\
MIPS 70~\micron--\textit{IRAS} 100~\micron & 7 & $-2.13\pm0.37$ & $-0.93$ & $222578.79\times10^{-8}$\\
MIPS 70~\micron--MIPS 160~\micron & 13 & $-2.72\pm0.47$ & $-0.87$ & $114097.59\times10^{-9}$\\
MIPS 70~\micron--PACS 160~\micron & 14 & $-2.74\pm0.49$ & $-0.85$ & $12987.26\times10^{-8}$\\
MIPS 70~\micron--SPIRE 250~\micron & 15 & $-3.4\pm0.54$ & $-0.87$ & $2.56\times10^{-5}$\\
MIPS 70~\micron--SPIRE 500~\micron & 15 & $-4.23\pm0.68$ & $-0.87$ & $3.08\times10^{-5}$\\
MIPS 70~\micron--SPIRE 350~\micron & 15 & $-3.75\pm0.61$ & $-0.86$ & $3.46\times10^{-5}$\\
\textit{IRAS} 100~\micron--PACS 160~\micron & 12 & $-3.1\pm0.35$ & $-0.94$ & $5.44\times10^{-6}$\\
\textit{IRAS} 100~\micron--SPIRE 500~\micron & 12 & $-5.69\pm1.03$ & $-0.87$ & $248.74\times10^{-6}$\\
\textit{IRAS} 100~\micron--SPIRE 350~\micron & 12 & $-5.09\pm0.81$ & $-0.89$ & $9.02\times10^{-5}$\\
\textit{IRAS} 100~\micron--SPIRE 250~\micron & 12 & $-4.5\pm0.65$ & $-0.91$ & $3.99\times10^{-5}$\\
PACS 100~\micron--PACS 160~\micron & 18 & $-1.35\pm0.25$ & $-0.8$ & $5.9\times10^{-5}$\\
\hline
\hline
\end{tabular}
\label{flux_ratios}
\end{table*}


\bsp	
\label{lastpage}
\end{document}